\documentclass{article}
\usepackage[utf8]{inputenc}
\usepackage{settings}

\newcommand{\qq}[1]{[\![{#1}]\!]}

\title{\textbf{Post-COVID Inflation \& \\ the Monetary Policy Dilemma: \\ An Agent-Based Scenario Analysis}}

\author{\small Max Sina Knicker$^1$\thanks{knicker@ladhyx.polytechnique.fr}, Karl Naumann-Woleske$^1$, Jean-Philippe Bouchaud$^{1,2,3}$, Francesco Zamponi$^{4,5}$}

\date{
\small
\textit{
$^1$ Chair of EconophysiX and Complex Systems, LadHyX, UMR CNRS 7646, Ecole Polytechnique Paris\\%
$^2$ Capital Fund Management, 75007 Paris, France\\%
$^3$ Acad\'emie des Sciences, Quai de Conti, 75006 Paris, France\\%
$^4$ Dipartimento di Fisica, Sapienza Universit\`a di Roma, Piazzale Aldo Moro 2, 00185 Rome, Italy \\
$^5$ Laboratoire de Physique de l'Ecole Normale Sup\'erieure, ENS, Universit\'e PSL, CNRS, Sorbonne Universit\'e, Universit\'e de Paris, F-75005 Paris, France\\[2ex]
}
}

\begin{document}
\maketitle

\captionsetup[figure]{margin=10pt,font=footnotesize,labelfont=bf,labelsep=endash,justification=centerlast, name={Fig.}}

\begin{abstract}
\noindent
The economic shocks that followed the COVID-19 pandemic have brought to light the difficulty, both for academics and policy makers, of describing and predicting the dynamics of inflation. This paper offers an alternative modelling approach. We study the 2020-2023 period within the well-studied Mark-0 Agent-Based Model, in which economic agents act and react according to plausible behavioural rules.
We include a mechanism through which trust of economic agents in the Central Bank can de-anchor. We investigate the influence of regulatory policies on inflationary dynamics resulting from three exogenous shocks, calibrated on those that followed the COVID-19 pandemic: a production/consumption shock due to COVID-related lockdowns, a supply-chain shock, and an energy price shock exacerbated by the Russian invasion of Ukraine. By exploring the impact of these shocks under different assumptions about monetary policy efficacy and transmission channels, we review various explanations for the resurgence of inflation in the United States, including demand-pull, cost-push, and profit-driven factors.
Our main results are four-fold: (i) without appropriate fiscal policy, the shocked economy can take years to recover, or even tip over into a deep recession; {(ii) the success of monetary policy in curbing inflation is primarily due to expectation anchoring, rather than to the direct economic impact of interest rate hikes; (iii) however, perhaps paradoxically, strong inflation anchoring is detrimental to consumption and unemployment, leading to a narrow window of ``optimal'' policy responses due to the trade-off between inflation and unemployment;} (iv) the two most sensitive model parameters are those describing wage and price indexation. The results of our study have implications for Central Bank decision-making, and offers an easy-to-use tool that may help anticipate the consequences of different monetary and fiscal policies.
\end{abstract}

\newpage
\tableofcontents

\section{Introduction}
Inflation has captured global attention since the onset of the COVID-19 pandemic (from now on, simply ``COVID'') in 2020. In the United States annual inflation reached 4.8\% in 2021 and peaked at 9.1\% in June 2022,\footnote{see e.g. U.S. Bureau of Labor Statistics, Consumer Price Index for All Urban Consumers: All Items in U.S. City Average [CPIAUCSL], retrieved from FRED, Federal Reserve Bank of St. Louis; May 24, 2023.} while Europe experienced highs of 11.5\% in October 2022.\footnote{see e.g.
Eurostat EuroIndicators, report n.~31/2023, https://ec.europa.eu/eurostat/documents/2995521/16310161/2-17032023-AP-EN.pdf} Competing narratives have emerged to explain the mechanisms driving this inflationary surge, which has persisted longer than expected. Policymakers have been blindsided by inadequate models, with the Bank of England admitting it had ``big lessons to learn'' from failure to forecast inflation using existing models.\footnote{see for example the Financial Times, https://www.ft.com/content/b972f5e3-4f03-4986-890d-5443878424ac } 
This paper aims to explore the influence of fiscal and monetary policies on prevailing inflationary dynamics within a complex macroeconomic environment, modeled using the Mark-0 Agent-based Model that is based on alternative modeling foundations.

During the period of high inflation in 2021 and 2022, several theories emerged to explain the underlying causes of persisting inflationary trends. These interpretations led to differing views on appropriate policy responses, ranging from monetary policy interventions such as interest rate hikes to more targeted fiscal policies and price controls. Our analysis contributes to the debate on the appropriate policy responses to post-COVID inflation by providing a flexible framework to assess different policy options in the context of various inflation drivers, including demand-pull, cost-push, and profit-driven inflation. Our framework can accommodate varying behavioral foundations within a complex economy, including agents' trust in the Central Bank's clout and the anchoring of inflation expectations. Our main conclusions are that (i) the economic recovery after the shocks, especially in absence of mitigating fiscal policies, can be much more sluggish than expected -- or even fall into deep recessions beyond dangerous tipping points or ``dark corners''\citep{Blanchard2014WhereDangerLurksa}; (ii) the policy response (both from the government, the Central Bank, and other public authorities) has to navigate a narrow path, facing the trade-off between high inflation (with the risk of a runaway scenario) and high unemployment (with the risk of an economic collapse). In particular too weak a fiscal stimulus is ineffective, and too large a stimulus fuels high inflation; (iii) the success of monetary policy in curbing inflation is primarily due to expectation anchoring, rather than to direct impact of interest rate hikes; (iv) the two most sensitive model parameters (in terms of the inflation outcome) are those describing wage and price indexation, or in other words the bargaining power of workers and the market power of firms. 

An initially dominant view of the post-COVID inflation was based on the ``too much money chasing too few goods'' theory of inflation \citep[see e.g.][]{SoyresEtAl2022FiscalPolicyExcess, SoyresEtAl2022DemandSupplyImbalanceCovid19}, also known as demand-pull inflation. This narrative focuses on the large amount of fiscal stimulus, such as the Coronavirus Aid, Relief and Economy Security Act (the ``CARES Act'') or the American Rescue Plan in the U.S., that have led to excess demand due to increases in disposable income while supply has not adjusted, thus pushing up prices. Some, such as \cite{FergusonStorm2023MythRealityGreat} agree with the demand-pull inflation analysis but contend that ``the final cause of the inflationary surge in the U.S. [...] was in large measure the unequal (wealth) effects of ultra-loose monetary policy during 2020-2021''. 

In contrast to the excess demand interpretation, scholars such as \citet{StiglitzRegmi2022CausesResponsesToday} conclude that ``today’s inflation is largely driven by supply shocks and sectoral demand shifts, not by excess aggregate demand''. This claim is supported by the study of \cite{cavallo2023can}, who present evidence spanning the years 2020 to 2022 illustrating the significant inflationary consequences of unexpected disruptions in product availability and stockouts across various sectors.
{Taking this one step further, \citet{GuerrieriEtAl2022MacroeconomicImplicationsCOVID19} and \citet{FerranteEtAl2023InflationaryEffectsSectoral}  consider the case wherein a negative supply shock on one (or several) sectors can actually lead to a larger drop in demand than the original shock, and consequently a recession.
In the New Keynesian model of \citet{FerranteEtAl2023InflationaryEffectsSectoral} demand reallocation explains a large share of the post-COVID U.S. inflation.
The energy and food price shocks following COVID and the Russian invasion of Ukraine in 2022 were major causes of inflation, but this time from a cost-push inflation perspective wherein firms pass on increases in costs to consumers through prices.}
The strength of the inflationary surge was in large part due to the systemically important nature of the sectors where inflation occurred: energy and food \citep{WeberEtAl2022InflationTimesOverlappinga}. 
These observations suggest an alternative cost-push scenario that is outside the realm of Central Bank's policy toolkit. This is also the conclusion of \citet{BernankeBlanchard2023}, where a granular decomposition of inflation over different factors is proposed.\footnote{We became aware of this paper, dated May 23, 2023, in the last week before finalizing our own work. Similarly to our own approach, \citet{BernankeBlanchard2023} include wage bargaining, labor tightness and trust anchoring in their framework. Among the most important differences, however, are the absence of (i) supply-demand imbalances in the price setting mechanism and (ii) possible price gouging effects. Furthermore, the multiple equilibria and corresponding ``tipping points'' found within our model do not exist in their simplified specification.}

Finally, a recent debate has emerged around profit-driven inflation, wherein sellers have increased prices beyond the increase in costs they face, thus expanding their profit margins. This is based on the observation that profits have increased sharply in the current inflationary climate, as compared to the observed amount of cost increases, even when including concerns of a wage-price spiral, and the excess demand cited by a monetarist perspective \citep{GloverEtAl2023HowMuchHave, StiglitzRegmi2022CausesResponsesToday}.
\citet{WeberWasner2023SellersInflationProfitsa} analysed firms' earnings calls and posit an interpretation based on imperfect competition and market power. Specifically, in an environment of cost increases and supply bottlenecks, firms' market power temporarily increases. This may be enhanced by the fact that when inflation is high, uncertainty about prices increases, such that consumers may be prone to accepting unreasonably high prices (an effect sometimes called ``consumer discombobulation''), thus allowing companies to enlarge their margins.

All of these different interpretations lead to different guidelines for how policy should respond, and whether it should at all.
From a standard monetarist perspective, the Central Bank should raise interest rates in order to push down demand, thus solving the excess demand scenario. On the other hand, in a cost-push scenario, the Central Bank's rate has no effect on the external cost increases and may actually harm the situation if firms choose to pass on the increased costs of debt to consumers. On this, \citet{StiglitzRegmi2022CausesResponsesToday} suggest ``monetary policy, then, is too blunt an instrument because it will greatly reduce inflation only at the cost of unnecessarily high unemployment, with severe adverse distributive consequences'', as in their view inflation is not due to excess demand. In reality, the U.S. Federal Reserve has hiked rates at the fastest rate since Paul Volcker was chairman, in line with a monetarist view. Simultaneously, in 2023 inflation has begun easing, which begs the question of whether the reduction in inflation throughout 2022 is due to monetary policy or to external factors such as the easing of energy prices as with the oil crises of the 1970s \citep{Blinder1982AnatomyDoubleDigitInflation,BlinderRudd2013SupplyShockExplanationGreat}, or else to a spontaneous tendency of inflation to self-heal due to economic forces. But since the Central Bank's mandate to keep inflation low and stable, they are expected to act, generally through interest rate mechanisms and communication.\footnote{There is an ongoing impression that monetary policy is the ``only game in town'' when it comes to inflation response. {See e.g. ECB economist Isabel Schnabel's recent speech about the ``Risk of Stubborn Inflation'', https://www.ecb.europa.eu/press/key/date/2023/html/ecb.sp230619$\_$1$\sim$2c0bdf2422.en.html}
} 
The question remains, to what effect? Are dips in inflation due to exogenous or endogenous factors or to Central Bank policy? What are the consequences of raising rates in this environment? Are the consequences of not acting on inflation with monetary policy greater than those when one does? After all, \citet{BrunoEasterly1998InflationCrisesLongrun} found that ``countries can manage to live with relatively high -- around 15-30 percent -- inflation for long periods''. It is these questions upon which we aim to shed some light by considering various scenarios generated through an Agent-based Model, that allows us to run various counterfactual scenarios. 

To conduct our analysis, we use the Mark-0 Agent-based Model (ABM) originally proposed by \citet{GualdiEtAl2015TippingPointsMacroeconomica}, and extended in \citet{GualdiEtAl2017MonetaryPolicyDarka, BouchaudEtAl2018OptimalInflationTargeta} to study monetary policy and inflation, with an early application to the effects of the COVID pandemic by \citet{SharmaEtAl2020WshapedRecoveryCOVIDa} and a model-exploration study by \citet{NaumannWoleskeEtAl2023ExplorationParameterSpace}. Agent-based Models are computational scenario-generators in which a large number of individual agents, in our case firms, interact based on a set of behavioral rules \citep[see][for more detailed introductions]{DawidDelliGatti2018ChapterAgentBasedMacroeconomicsa,HaldaneTurrell2019DrawingDifferentDisciplinesa, DosiRoventini2019MoreDifferentComplexa}.
These models allow us to study (possibly) out-of-equilibrium systems by generating emergent macroeconomic dynamics, many of which have been successfully replicated \citep[see][for instance]{DosiEtAl2017MicroMacroPoliciesa}. Agent-Based Models also allow one, with minimal cost, to study the impact of several effects, such as the (de-)anchoring of trust in the Central Bank policies, which could be highly relevant in practice but are very hard to take into account within standard rational equilibrium models.  

The Mark-0 ABM is a simplified model of a closed macroeconomy that nonetheless generates a wide variety of phenomena, from stable low unemployment and inflation, to endogenous crises that may oscillate regularly or punctuate long periods of recovery, or even to runaway inflation. The philosophy behind the model is to generate qualitatively plausible scenarios. 
In this context, \citet{GualdiEtAl2017MonetaryPolicyDarka} studied the efficacy of monetary policy in maintaining low unemployment and inflation, and ``find that provided the economy is far from phase boundaries (or `dark corners' \citep{Blanchard2014WhereDangerLurksa}) such policies can be successful, whereas too aggressive policies may in fact, unwillingly, drive the economy to an unstable state, where large swings of inflation and unemployment occur.''
An analysis using the Mark-0 model in fact predicted, as early as June 2020, that the post-COVID recovery could be more sluggish than expected and lead to a period of sustained inflation~\citep{SharmaEtAl2020WshapedRecoveryCOVIDa}.

In this paper we present a more detailed study of the post-COVID recovery by considering three distinct shocks occurring in the Mark-0 model: (1) a COVID-shock that negatively impacts firms productivity and consumer demand for the period of lockdowns, (2) a supply-chain shock on firms' productivity that mimics the after-effects of COVID on global value chains, and (3) an energy price shock that is exacerbated by the Russian invasion of Ukraine. 
Each of these shocks is calibrated to macroeconomic time-series for the United States, such that the magnitude and duration match the observed data.
We then study the effects of these shocks under different assumptions about the activity and efficacy of monetary policy, including the strength of monetary policy reactions to inflation, the ability of the Central Bank to influence firms' expectations, and the structure of firm decision-making on prices and wages. 

Our first results confirm that in the presence of properly calibrated shocks, the economic recovery in absence of any mitigating policy is extremely sluggish, taking at least several years (if not much more) for the economy to return to the pre-crisis equilibrium. We note that this happens with shocks that are calibrated on the observed post-COVID macroeconomic time series, which already incorporate the effect of actually implemented policies; we expect that in absence of any policies the impact would have been even greater.
{This result thus confirms the necessity of some sort of fiscal policy to avoid the economy spiraling into a crisis or a very sluggish recovery.}

A second set of results concerns the impact of such policies. We find that the model properly accounts for the inflation-unemployment trade-off, which leads to a narrow window in which policy can be efficient. For example, a disproportionate response of the Central Bank to inflation leads to unnecessary high unemployment, and an oversized injection of Helicopter Money leads to an unnecessary high inflation. Finally, we find that the anchoring (or de-anchoring) of economic agents' trust in the Central Bank strongly influences the dynamics of inflation, much more than the direct influence of interest rates on economic activity. This is due to the fact that the decision of households to consume, and of firms to increase wages and economic activity is based on the gap between interest rates and expected inflation, which is either close to, or very different from the target inflation, depending on expectation anchoring. This confirms the necessity for the Central Bank to manage expectations. However, keeping expectations strongly anchored to the Central Bank's target may come at a cost: if households believe that inflation is under control while rates are going up, consumption might be hobbled. Similarly, if firms believe that inflation is under control when it is not, high interest rates will hurt employment while wages will lag behind, again leading to a drop in consumption. These effects, perhaps paradoxically, lead to increased unemployment compared to the case where expectations are floating. We conclude that each policy must be carefully tuned to achieve the desired result, and we believe that our modelling tool can guide policy makers in this respect.

Our primary objective is to provide possible scenarios and counterfactuals. As discussed at length in our previous papers \citep{GualdiEtAl2015TippingPointsMacroeconomica, BouchaudEtAl2018OptimalInflationTargeta, SharmaEtAl2020WshapedRecoveryCOVIDa}, our ambition is not to provide precise predictions based on a fully calibrated model, but rather a tool for decision makers to help them apprehend different possible outcomes and anticipate unintended consequences and potential counter-intuitive impacts of their policies. 
We hope that Mark-0 can be usefully added to the policymakers toolbox and help them navigate a radically complex world (see e.g. \cite{king2020radical,bouchaud2021radical}). 

The remainder of the paper is structured as follows: Section \ref{sec:model} gives an overview of the Mark-0 model and the adaptations made for this paper. Pursuing this, Section \ref{sec:mechanisms} then outlines the policy channels that the model contains: interest rate, expectation management, access to credit, Helicopter Money and Windfall Tax. Section~\ref{sec:4} discusses the model's dynamics in absence of exogenous shocks.
In Section \ref{sec:shocks} we calibrate our three shocks, and show how they affect the macroeconomic dynamics with or without an easy credit policy, but without any monetary policy interventions by the Central Bank.  In Section \ref{sec:6} we introduce and discuss monetary policy through interest rates and expectations management, and highlight its effects. 
In Section~\ref{sec:7} we add fiscal policy by examining two distinct kind of stimuli, Helicopter Money and a Windfall Tax.
Section~\ref{sec:sloppy} presents a discussion of the robustness of our model to variations of its parameters, and highlights the risk of hyperinflation.
Finally, Section \ref{sec:conclusion} summarizes our results and lays out some perspectives and ideas for future work.

\section{The Mark-0 Model}\label{sec:model}
The Mark-0 Model is a stylized macroeconomic agent-based model that is qualitatively very rich, as many plausible -- and sometimes un-intuitive -- economic phenomena can emerge at the macro scale. The model was created as a minimalist reduction of the Mark family of Agent-Based Models originally developed by \citet{GaffeoEtAl2008AdaptiveMicrofoundationsEmergenta} and  \citet{GattiEtAl2011MacroeconomicsBottomupa}. Mark-0 was proposed by \citet{GualdiEtAl2015TippingPointsMacroeconomica}, and expanded to monetary policy questions by \citet{GualdiEtAl2017MonetaryPolicyDarka} and \citet{BouchaudEtAl2018OptimalInflationTargeta}. More recently, an analysis of possible economic responses to the COVID-pandemic were studied using Mark-0 by \citet{SharmaEtAl2020WshapedRecoveryCOVIDa}. Unless stated otherwise, we use the model description and formalism of \citet{SharmaEtAl2020WshapedRecoveryCOVIDa}. In this section, we outline the relevant equations for the model and explain the economic intuition behind all of them. 

\subsection{Model Overview}

The model comprises a large set of firms producing a homogeneous consumption good that is purchased by a representative household sector. The household sector goes to the market with a consumption budget $C_B$ computed as a fraction of savings, $S(t)$, wages $W(t)$, payouts from the energy sector $\delta_e \mathcal{E}_e(t)$, and interest on deposits $\rho_d(t)S(t)$, 
\begin{equation}\label{eq:hh_consumption_budget}
    C_B(t) = c(t)\left[S(t) + W(t) + \delta_e \mathcal{E}_e(t) + \rho_d(t)S(t)\right],
\end{equation}
where $c(t) \in (0,1)$ is the consumption propensity out of wealth and income, defined as
\begin{equation}\label{eq:consumption_propensity}
    c(t) = c_0 \left[1+\alpha_c\left(\hat{\pi}(t)-\rho_d(t)\right)\right],\qquad \alpha_c \geq 0,
\end{equation}
where $\hat{\pi}(t)$, specified by Eq.~\eqref{eq:inflation_expectation} below, is the expected future inflation. Eq.~\eqref{eq:consumption_propensity} mimics the behaviour of the classical Euler equation, as the household consumes more when real interest rates, $\rho_d(t) -\hat{\pi}(t)$, are lower. 
The savings of households over time can then be written as:
\begin{equation}\label{eq:savings}
    S(t+1) = S(t) + W(t) + \rho_d(t)S(t) - C(t) + \Delta(t) + \delta_e \mathcal{E}_e(t)
\end{equation}
with dividends received from firms profits $\Delta(t)$ and actual consumption $C(t) \leq C_{B}(t)$ that is given by the matching of demand and production.

Households choose to split their consumption budget between {$N_F$ different firms (labeled by an index $i=1,2,\cdots,N_F$)} based on an intensity of choice model
\begin{equation}\label{eq:hh_demand}
    D_i(t) = \frac{C_B(t)}{p_i(t)} \frac{\exp(-\beta p_i(t))}{\sum_j \exp(-\beta p_j(t))}, \quad \text{s.t.}\quad \sum_i p_i(t) D_i(t) \equiv C_B(t),
\end{equation}
with a price sensitivity $\beta$.\footnote{Good differentiation could easily be included at this stage by replacing $p_i$ by $p_i/\psi_i$, where $\psi_i$ is the preference for good $i$.}

Because the model considers out-of-equilibrium situations, demand $D_i$ for good $i$ and production $Y_i$ may not match, leading to a realized consumption $c_i^R$ given by $\min(D_i,Y_i)$. 

Meanwhile, firms produce consumption goods based on a linear production function, 
\begin{equation} \label{eq:prod_f} 
Y_i(t) = \zeta(t) N_i(t),
\end{equation} 
dependent only on the firm's employed labour force $N_i$ and time-dependent labour-productivity $\zeta$. {The unemployment rate $u(t)$ is then given by $u(t) = 1 - {\sum_i N_i(t)}/{N}$, where $N$ is the number of workers.} At each time step, corresponding to one month throughout this paper, firms can adapt to the current economic environment by choosing three firm-specific variables: the target production $Y_i(t)$, price $p_i(t)$, and wage offered to their employees, $w_i(t)$. We describe these in turn, but note that the time step of one month and the different update parameters $\gamma, g$ are assumed to be small such that production, prices and wages evolve slowly between $t$ and $t+1$, barring the role of external shocks of the COVID type, as discussed in section \ref{sec:shocks} below. 
\begin{enumerate}
\item {\it Production update.} Production adapts to the observed gap between supply and demand at the previous time step
as follows:
 \begin{align}
 \label{y_update}
\begin{split}
    \text{If } Y_i(t) < D_i(t)  &\hskip10pt \Rightarrow \hskip10pt 
     Y_i(t+1)=Y_i(t)+ \min\{ \eta^+_i ( D_i(t)-Y_i(t)), \zeta u^\star_i(t) \} \\
    \text{If }   Y_i(t) > D_i(t)  &\hskip10pt  \Rightarrow \hskip10pt
    Y_i(t+1)= Y_i(t) - \eta^-_i [Y_i(t)-D_i(t)]  \\
\end{split}
\end{align}
where $u^\star_i(t)$ is the maximum number of unemployed workers available to
the firm $i$ at time $t$, {which depends on the wage the firm pays relative to the production-weighted average wage ${W_i(t)}/{\overline{w}(t)}$\footnote{The production weighted average wage is defined as $\overline{w}(t)=\frac{\sum_i W_i(t) Y_i(t)}{\sum_i Y_i(t)}$.} 
\begin{equation}
    u^\star_i(t) = \frac{\mathrm{e}^{\beta W_i(t) / \overline{w}(t)}}{\sum_i \mathrm{e}^{\beta W_i(t) / \overline{w}(t)}}.
\end{equation}}
The speed at which firms hire and fire workers depends on their level of financial fragility $\Phi_i$, defined as the debt-to-sales ratio, where debt is $\mathcal{D}_i(t)$:\footnote{Note that this definition of fragility $\Phi$ slightly differs from that used in our previous publication, in particular the normalisation by the bankruptcy threshold $\Theta$.}    
\begin{equation}\label{eq:fragility}
    \Phi_i(t) = \frac{1}{\Theta}  \, \frac{\mathcal{D}_i(t)}{\min{(p_i(t)D_i(t), p_i(t)Y_i(t))}}.
\end{equation}
Non-indebted firms have zero fragility and $\Theta$ is the maximum debt-to-sales ratio allowed by firms creditors, beyond which firms are declared bankrupt.\footnote{For the detailed bankruptcy settlement, see \cite{GualdiEtAl2017MonetaryPolicyDarka, BouchaudEtAl2018OptimalInflationTargeta}}
Firms that are close to bankruptcy (i.e. $\Phi \approx 1$) are arguably faster to fire and slower to hire, and vice-versa for healthy firms. The coefficients of the hiring and firing rates $\eta^\pm_i$ for firm $i$ (belonging to $[0,1]$)
are given by: 
\begin{equation}\label{eq:etapm}
\eta^\pm_i = \qq{\eta_0^\pm \, (1 \mp \Gamma(t) \Phi_i(t))}, 
\end{equation}
where $\eta_0^\pm$ are fixed coefficients, identical for all firms, and $\qq{x}=x$ when $x \in (0,1)$, $\qq{x} = 1$ for $x \geq 1$ and $\qq{x}=0$ when $x \leq 0$.
The factor $\Gamma > 0$ measures how the financial fragility of firms influences their hiring/firing policy, 
since a larger value of $\Phi_i$ then leads to a faster downward adjustment of
the workforce when the firm is over-producing, and a slower (more cautious) upward adjustment
when the firm is under-producing. $\Gamma$ itself depends on the inflation-adjusted interest rate and takes the following form:
\begin{align}
\label{eq:gamma-coupling}
\Gamma(t) = \max \Big[ \alpha_{\Gamma}\left(\rho_\ell(t) - \hat \pi(t)\right), \Gamma_{0}\Big], \qquad \Gamma_0,\alpha_\Gamma \geq 0,
\end{align}
where $\rho_\ell(t)$ is the rate at which firms can borrow,  $\Gamma_0$ and $\alpha_{\Gamma}$ are free, non negative parameters, the latter being similar to $\alpha_{c}$ that captures the influence of the real interest rate on the hiring/firing policy of firms.

\item {\it Price update}. 
Compared to the previous versions of the Mark-0 model, in this paper the firms' price update, $\Delta p_i(t+1)=({p_i(t+1)-p_i(t)})/{p_i(t)}$, explicitly takes into account the demand-output gap, and a contribution for the change in exogenous (e.g. energy) price $\Delta p_{e, \text{ema}}$ (in \%), to with:
\begin{align}
     \Delta p_i(t+1)&= \gamma \xi_i(t) \cdot \frac{D_i(t)}{Y_i(t)} + g_p\hat{\pi}(t) + g_e \Delta p_{e, \text{ema}}(t) \quad &\mathrm{if} \quad&\begin{cases} D_i(t) > Y_i(t)\\ p_i(t)<\overline{p}(t)\end{cases}\label{eq:price_setting_up}\\
    \Delta p_i(t+1)&= -\gamma \xi_i(t) \cdot \frac{Y_i(t)}{D_i(t)} + g_p\hat{\pi}(t) + g_e \Delta p_{e, \text{ema}}(t)
    \quad &\mathrm{if} \quad&\begin{cases}D_i(t) < Y_i(t)\\ p_i(t)>\overline{p}(t)\end{cases}\label{eq:price_setting_down}\\
    \Delta p_i(t+1)&= g_p\hat{\pi}(t) + g_e \Delta p_{e, \text{ema}}(t) \quad &&\mathrm{otherwise},\label{eq:price_setting_otherwise}
\end{align}

where 
\begin{itemize}
    \item $\gamma$ is a parameter and $\xi_i(t)$ are independent uniform $U[0,1]$ random variables scaled up by the actual demand-output ratio ${D_i(t)}/{Y_i(t)}$ (or its inverse), in order to mimic an increased pressure on prices in case of supply (or demand) gluts. When ${D_i(t)} \approx {Y_i(t)}$, this reduces to the rule used in previous papers, see \cite{GualdiEtAl2015TippingPointsMacroeconomica};
    \item $\hat{\pi}(t)$ is the expected next-period inflation in \% per month, that firms take into account in their price setting mechanism. {This term also accounts for changes in price due to expected changes in wages and input costs, which are part of the expected inflation. A factor $g_p>1$ models an inflation gouging behaviour, while $g_p<1$ means that only part of inflationary costs are passed to the customers.}\footnote{Note that in principle, $g_p$ (and $g_w$) might depend on the firm or on the sector. We have not considered this possibility here. }
    \item $\Delta p_{e, \text{ema}}(t)$ is the exponentially weighted moving average (with parameter $\omega$, as in Eq.~\eqref{eq:inflation_ema} below) of the exogenous price variations that are partially or fully passed on to final customers. The weighted average reflects that changes in energy price are not instantaneously transmitted to customers but rather distributed gradually over several months.
\end{itemize} 
The magnitudes of the corresponding price adjustments are determined by parameters: $\gamma$ for supply-demand imbalance, $g_p$ for inflation expectations, and $g_e$ for changes in the exogenous price. In this regard, $g_e$ can be thought of as the effective energy-share of production that firms want to pass on to customers, whereas $g_p > 1$ would correspond to the much discussed concept of ``greedflation''. 

Note that $\overline{p}$ is the consumer price index, defined as
\begin{equation}\label{eq:cpi} 
    \overline{p} = \frac{\sum_i c^R_i p_i}{\sum_i c^R_i} \ , \qquad c^R_i = \min(D_i, Y_i) \ ,
\end{equation}
with $c^R_i$ the realized consumption of product $i$.\footnote{Here again, our present definition of $\overline{p}$ slightly differs from that used in our previous papers, where we used a production weighted index instead of a consumption based index. We conform in this paper with the more standard definition of inflation based on the Consumer Price Index (CPI).}

\item{\it Wage update.} 
In similar fashion, firms update wages as
\begin{align}
    \Delta w_i(t+1) &= \gamma(1 - u(t))\left(1-\Gamma(t)\Phi_i(t)\right)\,\xi'_i(t) + g_w\hat{\pi}(t) \quad &\mathrm{if} \quad\begin{cases} D_i(t) > Y_i(t)\\ \mathcal{P}_i(t)>0\end{cases}\label{eq:wage_up} \\
  \Delta w_i(t+1)&= -\gamma u(t)\left(1+\Gamma(t)\Phi_i(t)\right)\,\xi'_i(t) + g_w\hat{\pi}(t) \quad &\mathrm{if} \quad \begin{cases} D_i(t) < Y_i(t)\\ \mathcal{P}_i(t)<0\end{cases}\label{eq:wage_down}\\
  \Delta w_i(t+1)&= g_w\hat{\pi}(t)&\mathrm{otherwise}
\end{align}
with $\Delta w_i(t+1) = ({w_i(t+1) - w_i(t)})/w_i(t)$,  and
where $u(t)$ represents the unemployment rate, $\xi'_i(t)$ is a $U[0,1]$ random realisation, and $g_w$ is the wage sensitivity to expected inflation, which could be seen as worker's bargaining power. Firm profits $\mathcal{P}_i$ include the cost of debt $\mathcal{D}_i$, the revenue on cash $\mathcal{E}_i$ and the cost of energy $g_e p_e Y_i$:
\begin{align}
\label{eq:firm-profits}
 \mathcal{P}_i = p_i(t) \min \{Y_i(t), D_i(t) \}  - w_i(t) Y_i(t)  + \rho_d \mathcal{E}_i(t) - \rho_\ell \mathcal{D}_i(t) - g_e p_e(t) Y_i(t) \ . 
\end{align}

The conditions of wage change therefore depend on the demand-supply imbalance, current firm profit $\mathcal{P}_i(t)$ and firm financial health, and also on the tension on the labour market, since lower unemployment leads to higher wage increase. This allows the model to reproduce Phillips curve effects.
{When firms experience positive profits and maintain a positive cash balance, they distribute dividends to households as a fraction $\delta$ of their cash balance, as in Eq.~\eqref{eq:savings}:
\begin{equation}\label{eq:dividends}
    \Delta(t) = \delta \sum_i \mathcal{E}_i(t)
    \, \mathbb{I}\Big(\mathcal{P}_i(t)>0\Big) \mathbb{I}\Big(\mathcal{E}_i(t)>0\Big) \ ,
\end{equation}
where $\mathbb{I}(E)$ is the indicator function of event $E$.}

\item{\it Energy Sector.}
In this paper, we consider an exogenously varying energy price, $p_e(t)$.  Before the explicit shock introduced in Section \ref{sec:shocks}, we consider the real energy price to be constant and equal to the average price of goods, i.e. $p_e(t) = \overline p(t)$. In each period, firms pay a total amount $g_e p_e(t) \sum_i Y_i(t)$ to the energy sector. Subsequently, the energy sector pays a fraction of its accumulated profits to households at a rate $\delta_e$. Therefore the cash balance $\mathcal{E}_e$ of the energy sector writes as 
\begin{equation} \label{eq:energy_sector}
    \mathcal{E}_e(t+1)= \mathcal{E}_e(t) + g_e p_e(t) \sum_i Y_i(t) - \delta_e \mathcal{E}_e(t)
\end{equation}
with the fraction $\delta_e$ of the energy profits that is paid out to households as income. 

Here, $\delta_e$ can be understood as dividends, share sales, or other channels through which energy sector profits circulate back into the economy. The way we introduce the energy sector in this paper as an accounting identity with exogenous prices is arguably simplistic, and can certainly be improved in a future version of the model.\footnote{Note that in the model there is a resource constraint such that the total money $M$ (that is created by the Central Bank) is kept fixed during the simulation The balance sheet of the banking sector can then be written as $M = S(t) + \mathcal{E}^+ - \mathcal{E}^- + \mathcal{E}_e$, with the household savings $S$ and the cash balance of firms $\mathcal{E}^{+/-}$.}
\end{enumerate}
\subsection{Inflation Expectations and Monetary Policy}
The equations above depend on inflation rate expectations $\hat{\pi}(t)$, that we now define. 
We consider that the measure of realized inflation is given by the change in consumption-weighted average price,
\begin{equation}\label{eq:inflation_realized}
    \pi(t) = \frac{\overline{p}(t) - \overline{p}(t-1)}{\overline{p}(t-1)},
\end{equation}
where $\overline{p}$ is defined in Eq. \eqref{eq:cpi}. 

In the model, agents form expectations of future inflation partly on the basis of past realisations and partly on the basis of their trust in the Central Bank ability to enforce its inflation target. More precisely, they use an exponentially weighted moving average over realised inflation,
\begin{equation}\label{eq:inflation_ema}
    \pi_{\text{ema}}(t) = (1-\omega)\pi_{\text{ema}}(t-1) + \omega \pi(t),
\end{equation}
where $\omega$ sets a memory time over which agents perceive realized inflation, equal to $-[\log(1-\omega)]^{-1} \approx \omega^{-1}$ for small $\omega$. Together with the Central Bank's communicated inflation target, $\pi^\star$, all agents form the same inflation expectation, given by a weighted average of $\pi_{\text{ema}}(t)$ and $\pi^\star$:
\begin{equation}\label{eq:inflation_expectation}
    \hat{\pi}(t) = (1-\tau^T(t))\pi_{\text{ema}}(t) + \tau^T(t) \pi^\star,
\end{equation}
where $\tau^T(t)$ is the degree to which expectations are anchored around the Central Bank's target.\footnote{For empirical work on the question of trust, see e.g. \cite{christelis2020trust}.} Consistently with the definition of $\pi_{\text{ema}}$ and the long-term nature of the inflation target, we interpret $\hat{\pi}(t)$ as the expectation of {\it long-term} inflation.\footnote{This is in contrast to the common formulations in DSGE models that operate with one-period-ahead expectations ($t \to t+1$)}

As far as monetary policy is concerned, we assume that the Central Bank sets the baseline interest rate, $\rho_0$, via a classical Taylor rule based on observed (realized) inflation $\pi_{\text{ema}}$:
\begin{equation}
    \rho_0 = \rho^\star+ \phi_\pi\left(\pi_{\text{ema}}(t)-\pi^\star\right),
\end{equation}
with reaction strength $\phi_\pi$. The baseline rate $\rho_0$ is then translated into a time-dependent rate on loans, $\rho_\ell(t)$, and deposits, $\rho_d(t)$, adjusting for the cost of bankruptcies.{These interest rates are defined as 
\begin{align}
    \rho_\ell(t) &= \rho_0(t) + f \frac{\mathcal{D}(t)}{\mathcal{E}^-(t)}, \\
    \rho_d(t) &= \frac{\rho_0(t) \mathcal{E}^-(t) - (1-f) \mathcal{D}(t)}{S + \mathcal{E}^+(t)}
\end{align}
with the positive cash balance $\mathcal{E}^+(t) = \sum_i \max(\mathcal{E}_i, 0)$ and firms total debt $\mathcal{E}^-(t) = -\sum_i \min(\mathcal{E}_i, 0)$. The parameter $f$ determines how the consequences of defaults are allocated to lenders and depositors, interpolating between the costs borne entirely by borrowers ($f$ = 1) and those shouldered entirely by depositors ($f$ = 0).}
In this paper, we do not consider a double mandate for the Central Bank as in \cite{GualdiEtAl2017MonetaryPolicyDarka}, instead focusing only on inflation. We will comment about this below, as strict inflation control may turn out to be highly detrimental to unemployment.

Finally, we allow inflation expectation anchoring to evolve dynamically via\footnote{Note that in our previous paper (\cite{BouchaudEtAl2018OptimalInflationTargeta}), the anchoring parameter $\tau^T$ was assumed to be time independent, a case we will call ``Anchored Expectation'' henceforth.}
\begin{equation}\label{eq:tau_anchoring}
    \tau^T(t+1) =  (1-\omega)  \tau^T(t)+ \omega  \exp{\left[- \alpha_I \frac{|\pi(t)-\pi^\star|}{\pi^\star \phi_\pi} \right]}.
\end{equation}
This equation aims at capturing the fact that the degree of expectation anchoring depends on how closely the realised inflation actually matches the Central Bank target. This is factored into an exponentially weighted moving average with memory time $\approx \omega^{-1}$: realized and target inflation must differ significantly, and for sufficiently long times, for agents to lose trust in the Central Bank.  A larger $\alpha_I$ means economic agents lose trust in the Central Bank more abruptly as the gap between realized inflation and inflation target becomes  significant. We have included the factor $\phi_\pi$ in the denominator to emphasize the fact that stronger commitment of the Central Bank should decrease the sensitivity of anchoring on realized inflation, but of course this extra factor can be reabsorbed into $\alpha_I$.

The calibration of the parameters we choose for the model is based on the results and studies done in the previous work around Mark0 (\cite{GualdiEtAl2015TippingPointsMacroeconomica, GualdiEtAl2017MonetaryPolicyDarka, BouchaudEtAl2018OptimalInflationTargeta, SharmaEtAl2020WshapedRecoveryCOVIDa}). Here, we discuss only the parameters that are relevant to the current work, in particular those identified by the sensitivity analysis in Section \ref{sec:sloppy}. The parameters used in this study can be found in the Table \ref{tablenocb}.

\section{Policy Channels in Mark-0}
\label{sec:mechanisms}

In the augmented Mark-0 model, the central authorities can influence macroeconomic dynamics through (i) the manipulation of interest rates, (ii) anchoring inflation expectations, and (iii) regulating the amount of debt firms can accumulate. Additionally, there is the possibility of fiscal policy in the form of direct injection of cash in the economy (``Helicopter Money'') and Windfall Tax on the energy sector. These manipulations take effect through firms' financial fragility, which directly determines the probability of bankruptcies and indirectly changes the propensity of firms to hire and fire workers, as well as the wage setting behavior. Interest rates and cash injection also have effects on the willingness of consumers to spend. In addition, there is an expectations channel whereby firms and households may or may not trust the Central Bank's communicated inflation target depending on how close actual inflation is to the target. We now consider these mechanics in turn, as these will turn out to be important in response to the COVID related shocks that we will introduce in the next section.

\subsection{The Interest Rate Channel}

The Central Bank's baseline interest rate $\rho_0$ additively influences both deposit ($\rho_d$) and lending ($\rho_\ell$) rates. Through this channel, the interest rate affects firms' wages and hiring strategies, as well as households' spending. 

The impact on households is straightforward and parallels that of the standard Euler equation of intertemporal substitution: higher interest rates, all else being equal, reduce the propensity to consume out of income and wealth through parameter $\alpha_c$ (Eq.~\eqref{eq:consumption_propensity}), effectively decreasing current demand in favor of later consumption. This decrease leads firms to reduce or maintain prices due to excess supply (see Eq. \eqref{eq:price_setting_down}), thereby lowering inflation.

The impact of interest rates on firm behavior occurs through the coefficient $\alpha_\Gamma$, defined in Eq.~\eqref{eq:gamma-coupling}, which influences firms' production target and wage adjustment.  In all cases $\alpha_\Gamma$ affects firm behavior in combination with the firm's fragility $\Phi$ (where $\Phi=1$ implies bankruptcy, see Eq.~\eqref{eq:fragility}). An increase in the baseline rate will increase $\Gamma$, which implies that firms have stronger reactions to excess demand or excess supply. Unhealthy firms ($\Phi>0$)  will therefore more cautiously expand production when needed ($\eta^+$), but more abruptly fire staff in the case of excess supply ($\eta^-$). Hence, an increase of the baseline rate $\rho_0$ will tend to decrease production and increase unemployment as fragile firms feel the brunt of the cost of debt. For similar reasons, an increase of $\rho_0$ implies a stronger downward pressure on wages.    

Therefore, in Mark-0 an increase of the baseline rate ceteris paribus induces a reduction of consumption and an increase of unemployment. The success of a rate hike in curbing inflation depends on the strength of the response of households and of firms to interest rates, and hence on the value of the parameters $\alpha_c$ and $\alpha_\Gamma$. If, as some authors argue \citep{reis2022burst}, the reaction of households and of firms to interest rates are subdued, large hikes would be necessary for monetary policy to have a significant effect on inflation, in part by bringing about a recession \citep{StiglitzRegmi2022CausesResponsesToday}. 

\subsection{The Expectation Channel}\label{sec:expectations}
With rising global inflation, the debate about the importance of inflation expectations for macroeconomic dynamics has been rekindled. Some, such as \citet{Rudd2022WhyWeThinka}, argue that short-term inflation expectations play no role in macroeconomic dynamics, while others, like \citet{Reis2021LosingInflationAnchor}, maintain their importance but question whether they are “de-anchored” from Central Bank’s targets. In the present version of the Mark-0, long-term inflation expectations $\hat \pi$ play a role in price and wage settings: formally, the factor $g_p$ in Eqs. \eqref{eq:price_setting_up} and \eqref{eq:price_setting_down} determines the fraction of expected inflation reflected in firms' next-period prices. In a similar style, the factor $g_w$ in Eqs. \eqref{eq:wage_up} and \eqref{eq:wage_down} controls the rate of wage adaptation to inflation in response to firms’ expected inflation, and is thus adjacent to the idea of labour-bargaining power.

The long term inflation expectation, $\hat{\pi}$, is determined as a mixture of past realized inflation and the Central Bank target, with the Central Bank's effectiveness in reaching its target dictating the level of ``anchoring'' $\tau^T$ (see Eq. \eqref{eq:inflation_expectation} and \eqref{eq:tau_anchoring}).
With this formulation, economic agents lose confidence in Central Bank actions when inflation significantly and persistently deviates from the Central Bank’s inflation target. The parameter $\alpha_I$ represents the sensitivity of trust to the Central Bank's ability to control inflation.

De-anchored beliefs imply firms' inflation expectation is based solely on past observation, running a risk of a wage-price spiral or a hyper-inflation runaway, depending on the imbalance between the parameters $g_p$ and $g_w$. 

The simplest case is equality in bargaining and market power, $g_w=g_p$, which implies that on average both prices and wages are raised in the same proportion to expected inflation, thus having no real effects on the economy.
One then faces either stable inflation when $g_w=g_p<1$ or possible hyper-inflation when $g_w=g_p>1$, see section \ref{sec:sloppy}. 

A more interesting case is $g_p\neq g_w$, implying a differentiation in bargaining vs. pricing power. For $g_p > g_w$, labour bargaining power is lower than firms pricing power. In this case, firms raise prices by more than wages, eroding the purchasing power of workers. This results in lower demand that makes price hikes less likely until firms' excess demand becomes excess supply, reversing the cycle. For $g_w > g_p$, on the other hand, an unstable wage spiral may set in, with higher wages driving demand up, leading to further increases of prices as firms face excess demand. The situation $g_p > 1$, at least for firms with a healthy balance sheet, would correspond to ``greedflation'', i.e. firms trying to use inflation to hide increases in their profits.  

\subsection{Credit Regulatory Policy}\label{sec:regpolicy}

The last channel for the monetary authority to affect the economy within the framework of Mark-0 is by means of the default threshold, $\Theta$. Earlier work on the Mark-0 model found that the default threshold is a key parameter causing a phase transition between a regime of high unemployment and/or endogenous crises (low $\Theta$), and of stable full employment (higher $\Theta$ conditional on low interest rates) \citep[see][]{GualdiEtAl2015TippingPointsMacroeconomica}. This is because higher thresholds give firms more time to adjust their production and price strategy to return to a profitable state, assuming interest payments are low. Empirically, both in the CARES act and the CAA there were relaxations of the bankruptcy laws in the United States.\footnote{CARES refers to the Coronavirus Aid, Relief, and Economic Security Act, and the CAA refers to the Consolidated Appropriations Act of 2021.} Direct state aid to firms may also prevent bankruptcies from occurring. In the realm of our model these policies are mimicked by increasing the bankruptcy threshold $\Theta$ of firms during the shock and reducing  it again after the shock. The ``Easy-Credit policy'' proposed in \cite{SharmaEtAl2020WshapedRecoveryCOVIDa} amounts to set the value of $\Theta$ such that it remains commensurate to the current average firm fragility, to wit 
\begin{equation}\label{eq:adaptive}
    \Theta(t) = \max(\mu \langle \phi \rangle(t), \Theta_0), 
\end{equation}
where $\Theta_0$ is the bankruptcy threshold before the shock, $\mu$ a multiplier, and $\langle \phi \rangle(t)$ the firm-wide average debt-to-sales ratio at time $t$, meaning that only the firms whose debt-to-sales ratio exceeds $\mu \langle \phi \rangle$ must file for bankruptcy. Throughout the following we will use $\mu=1.3$ and $\Theta_0=3.2$, which represents a compromise between ``too robust'' economies when $\Theta_0$ is large, and ``too fragile'' economies when $\Theta_0$ is small. 

This implies that with easier credit access, firms can accumulate debt during the shock without the fear of bankruptcy. This, in turn, affects the firing and hiring policies of firms as they can continue with their business as usual, leading to a hiring and firing as if they were not fragile (see Eq.~\eqref{eq:fragility}). \cite{SharmaEtAl2020WshapedRecoveryCOVIDa} have shown that such policy is effective in speeding up the recovery to pre-shock levels after severe COVID-like shocks, without such policy long-lasting recessions would otherwise ensue.  

\subsection{Helicopter Money}\label{sec:mechanism_helicopter}
When interest rates are low and access to credit is loose, but shocks nonetheless persist, central authorities may turn to less conventional policies such as ``Helicopter Money'' to address economic downturn. Introduced by \citet{friedman1969optimum}, this implies an injection of money directly into the hands of economic agents to increase spending and thus economic output.\footnote{By contrast, quantitative easing implies that money created by the Central Bank is used to purchase government bonds and distributed through the government, whereas Helicopter Money is directly distributed in the economy \citep{ugai2007effects}.} While some view Helicopter Money as a potential cause of hyperinflation and currency devaluation, it can be an effective response to financial crises and pandemics by stimulating real economic activity \citep{reis2022helicopter}. In this paper we consider Helicopter Money as introduced by Friedman: a distribution of newly printed money by the Central Bank directly to households, resulting in a decrease in the Central Bank's net worth and an increase in the net worth of households. This injection of money is modeled as a one-time increase of savings of households by a factor $\kappa_H>1$, such that $S \rightarrow \kappa_H S$ instantaneously after all shocks. The increased savings of households boost demand through an increased consumption budget $C_B$ (see Eq.~\eqref{eq:hh_demand}) and finally end up on the balance sheet of firms where it is then redistributed to households as wages and dividends.

\subsection{Windfall Tax}\label{sec:mechanism_windfall}

Whereas Helicopter Money stimulates consumption by injecting external funds into households savings, a tax on ``windfall'' profits, which refer to large unforeseen profits \citep{chennells1997windfall}, achieves a similar outcome by redistributing money through taxation. Following the Russian invasion of Ukraine, there has been a significant surge in fossil fuel prices, leading to unexpectedly high returns for utilities and fossil fuel producing companies \citep{Weber2022BigOilProfitsa}, and profit-driven price markups for other firms \citep{WeberWasner2023SellersInflationProfitsa}, leading to strong debate amongst policymakers about implementing a Windfall Tax. The specific design of the Windfall Tax as a rent-sharing fiscal policy may vary depending on the particular circumstances \citep{baunsgaard2022taxing}. 

In this paper, we implement a ``Windfall Tax'' as a temporary increase of the fraction of energy sector cash balance that is re-injected to the savings of households, i.e. $\delta_e \rightarrow \delta_e + \Delta \delta_e$. This tax has a duration of two years, commencing one year prior to the end of the price shock. The primary objective of increasing $\delta_e$ is to boost consumption by enhancing households' savings through the redistribution of the increased cash balance in the energy sector that follow the energy price increases.

\section{Stationary Dynamics: the Role of the Central Bank} 
\label{sec:4}

\subsection{The Economic Equilibrium Without Monetary Policy}\label{sec:baselinedescription}
\label{sec:4.1}

\begin{figure}[t]
    \centering
    \includegraphics[width=\textwidth]{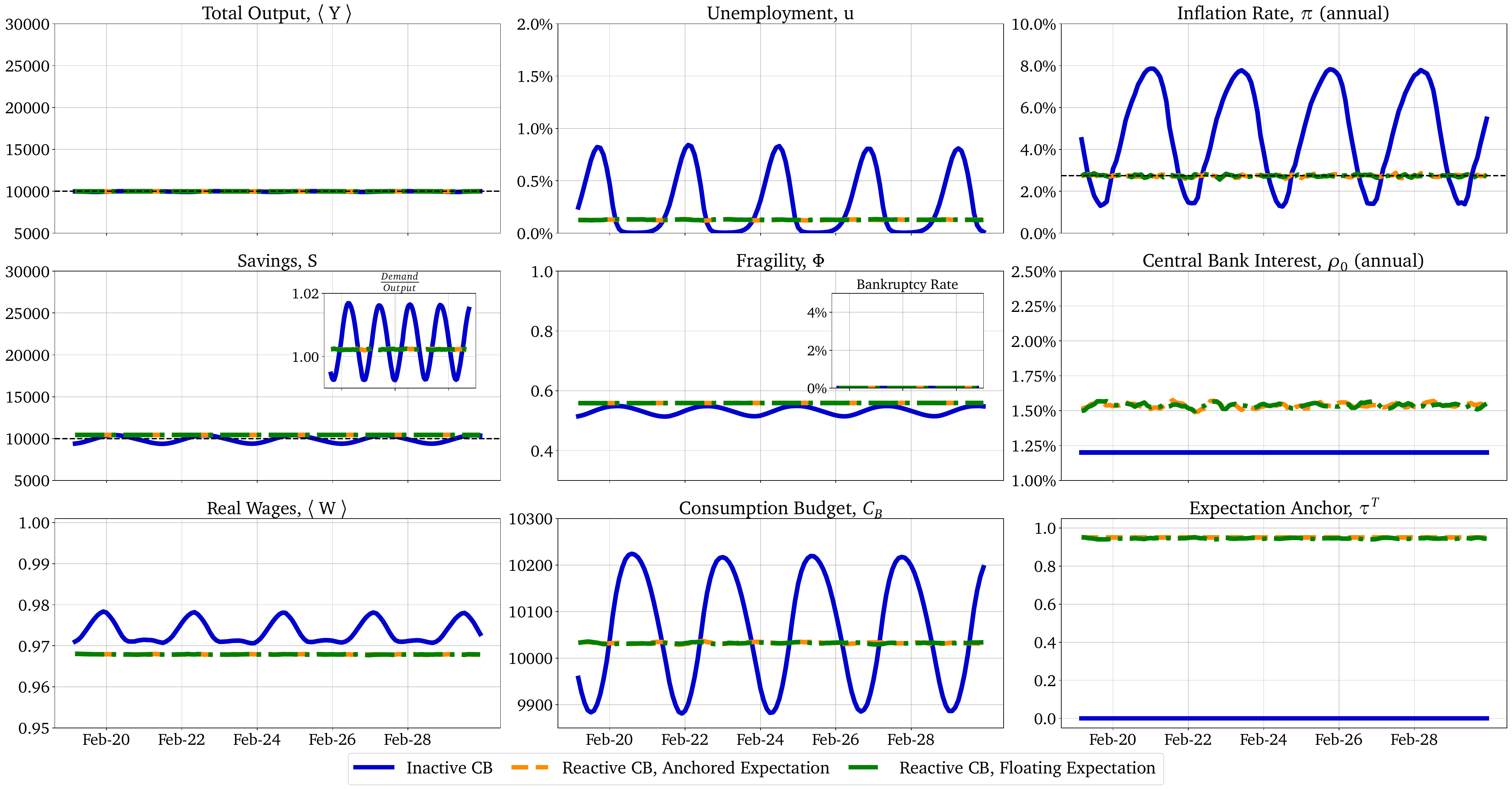}
    \caption{\textbf{Economic dashboard for the three Central Bank scenarios described in sections \ref{sec:baselinedescription}, \ref{sec:policydescriptions} in absence of any policy and shock:} Blue lines: Inactive Central Bank scenario ($\pi^\star=\phi_\pi=0,\; \rho^\star=1.2\%$ p.a.). Orange lines: Reactive Central Bank with Anchored Expectation scenario ($\pi^\star=2.4\% \,\, \text{p.a.},\;\phi_\pi=1,\;\tau^T=0.95$). Green lines: Reactive Central Bank with Floating Expectation scenario. The complete parameter set can be seen in table \ref{tablenocb}. The two insets correspond to the demand/output ratio and the bankruptcy rate.}
    \label{fig:baseline_noshocks}
\end{figure}

To begin our analysis, we consider the economy without shocks. Our first choice of parameters corresponds to a Central Bank that does not react to inflation (i.e. $\pi^\star=\phi_\pi=0$) while maintaining low baseline interest rates ($\rho^\star=1.2\%$ p.a.).
By consequence, agents form inflation expectations based only on their own observations ($\tau^T=0$, see Eq. \eqref{eq:inflation_expectation}). 
In this scenario, called ``Inactive CB'', and for the choice of parameters detailed in Table \ref{tablenocb}, the economy spontaneously evolves into a steady state of full capacity, as evidenced by the low, oscillating unemployment rate around $0.4 \%$ (see Figure \ref{fig:baseline_noshocks}). There are small endogenous output and unemployment fluctuations of roughly two years in duration. Simultaneously, there are stronger oscillations in the demand and supply for goods, and in real wages. These fluctuations cause firms to adjust both their production and prices, thus generating a small ``business cycle'' with periodic imbalance between demand and output due to price adjustments. But in the absence of an active Central Bank target, the amplitude of the resulting inflation oscillations is found to be substantial, ranging between $1.5 \%$ and $8 \%$ p.a., suggesting that monetary policy indeed iron out business cycles. As already found by \cite{GualdiEtAl2015TippingPointsMacroeconomica}, these cycles emerge endogenously from the oscillating feedback loop of prices, demand and savings.
In this scenario, called ``Inactive CB'', and for the choice of parameters detailed in Table \ref{tablenocb}, the economy spontaneously evolves into a steady state of full capacity, as evidenced by the low, oscillating unemployment rate around $0.4 \%$ (see Figure \ref{fig:baseline_noshocks}). There are small endogenous output and unemployment fluctuations of roughly two years in duration. Simultaneously, there are stronger oscillations in the demand and supply for goods, and in real wages. These fluctuations cause firms to adjust both their production and prices, thus generating a small ``business cycle'' with periodic imbalance between demand and output due to price adjustments. But in the absence of an active Central Bank target, the amplitude of the resulting inflation oscillations is found to be substantial, ranging between $1.5 \%$ and $8 \%$ p.a., suggesting that monetary policy may iron out business cycles, as indeed found below. As already discussed in \cite{GualdiEtAl2015TippingPointsMacroeconomica}, these cycles emerge endogenously from the oscillating feedback loop of prices, demand and savings.

Note however that depending on the choice of parameters, the economy may settle into a much less favourable state. In particular, when the hire-to-fire ratio is too small or when the baseline interest rate is too high, the economy collapses, see \cite{GualdiEtAl2015TippingPointsMacroeconomica, BouchaudEtAl2018OptimalInflationTargeta}. Similarly, when the bankruptcy threshold $\Theta_0$ is too low, unemployment remains at a relatively high level. 

\subsection{Stabilizing Inflation through Monetary Policy} 
\label{sec:policydescriptions}

At this point, we introduce an active Central Bank and study its impact on inflation and unemployment. Specifically, we set the Central Bank's inflation target to $\pi^\star=2.4\%$ p.a. and its reaction to inflation variation to $\phi_\pi=1$. In this configuration, we assume that the Central Bank works with a well-anchored inflation expectation fixed at $\tau^T=0.95$. In the pre-COVID period core inflation was quite stable as a result of an arguably successful monetary policy \citep{miles2017and}. Simultaneously, the Central Banks had been acting in a transparent and measurable way despite the historically low interest rates \citep{Reis2021LosingInflationAnchor}, suggesting a high degree of trust. We call this constant $\tau^T$ scenario ``Reactive CB, Anchored Expectation" in the following. In this case, Figure \ref{fig:baseline_noshocks} shows that inflation is reduced to an average of $\langle \pi_{AT} \rangle \approx 2.7 \%$, close to the $2.4 \%$ target, while nearly completely suppressing the business cycle with a very low stable unemployment. On the other hand, there is a slight reduction in real wages and higher interest rates.

Finally, we introduce a third scenario, referred to as ``Reactive CB, Floating Expectation'', where we allow agents' trust in the Central Bank to vary depending on the perceived success of the CB to bring inflation to roost, see $\tau^T(t)$ defined in Eq.~\eqref{eq:tau_anchoring}. We choose as parameter values $\omega=0.2$ (corresponding to a memory time of 5 months) and a sensitivity to off-target inflation of $\alpha_I=0.4$. Again, all other parameters are taken from Table \ref{tablenocb}. In the absence of external shocks, this configuration leads to a similarly high degree of trust in the Central Bank, and is otherwise identical to the case where trust is anchored -- simply because inflation is on target. 

\section{The COVID Shock and its Aftermath in the Absence of Monetary Policy}\label{sec:shocks}

\begin{figure}[htpb!]
    \centering
    \includegraphics[width=0.8\textwidth]{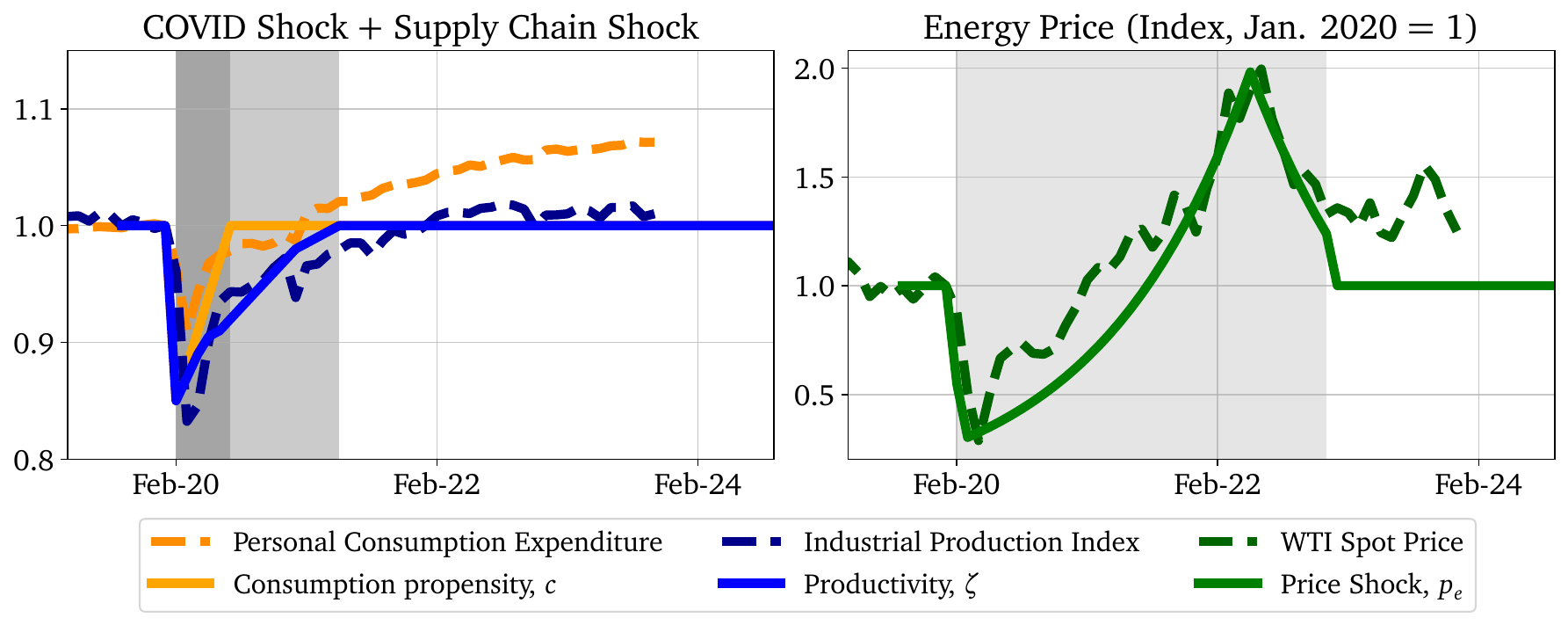}
    \caption{\textbf{Empirical Shocks}: Fitting of shock scenarios to macro data from the US up to end of 2022, when the present analysis was conducted. The empirical series (dashed lines)
    are indexed to January 2020. (Left Panel) Empirical series for the Personal Consumption Expenditure (orange, dashed line) and Industrial Production Index (blue, dashed line). Solid lines represent our model for the shocks to consumption propensity $c$ (orange) and productivity $\zeta$ (blue). Dark grey area corresponds to the COVID Shock and the grey area to the Supply Chain Shock in the shock scenarios that we apply to Mark-0. (Right Panel) WTI Crude oil spot price during the periods of shock (green, dashed line). Modelling of the artificial price series (green, solid) in the Mark-0 model for a period of two and a half years (light grey area). Data retrieved from \cite{FredPCE, FREDCPI, FREDWTI}.
    }
    \label{fig:empirical_shocks}
\end{figure}

\subsection{Modelling the Three COVID Shocks}

In order to assess the effects of recovery and monetary policies in response to the events of 2020-2022, we introduce in Mark-0 three shocks calibrated on U.S. data: (1) a COVID shock, (2) a supply-chain shock, and (3) an energy price shock. The data was retrieved from the Federal Reserve Bank of St. Louis, and are shown in Figure \ref{fig:empirical_shocks} where we have indexed values to January 2020 and calibrated the model until end of 2022.\footnote{This date corresponds to the moment we started our study. We have not tried to re-run the model including more recent data.} We now discuss these shocks in turn:
\begin{enumerate}
    \item \textbf{COVID}: The first major COVID outbreak emerged in the US in February 2020, prompting the US government to declare a public health emergency, followed by the implementation of stay-at-home orders in March of the same year. The COVID outbreak led to a significant reduction in personal consumption expenditure by households, falling by around $15\%$ in February 2020 (Figure~\ref{fig:empirical_shocks}, dashed orange line), which was of the same order of magnitude as the decline in the Industrial Production Index during the same month (dashed blue line). It took about 5 months for aggregate Personal Consumption Expenditure to recover to its pre-pandemic level. In accordance with this, we consider a shock to the consumption propensity $c(t)$ of households by $15\%$, which appears in Eq.~\eqref{eq:consumption_propensity}, recovering linearly over 5 periods (solid orange line in Figure~\ref{fig:empirical_shocks}).
    \item \textbf{Supply Chains}: During the initial COVID outbreak, firms in the US laid off a large number of workers, reducing their production to a similar degree as the personal consumption expenditure. However, while personal consumption recovered within 5 months, it took an additional 10-15 months for industrial production to return to its pre-pandemic state, as can be seen in the industrial production index (dashed blue line, Figure \ref{fig:empirical_shocks}). This was due to a plethora of idiosyncratic supply chain disruptions, such as logistics and transportation difficulties, semiconductor shortages, pandemic-related restrictions on economic activity, and labor shortages that led to the slower recovery of production in the industrial sector \citep{attinasi2022supply}. In the context of this paper, we model this by a shock to firm productivity $\zeta(t)$ (defined in Eq. \eqref{eq:prod_f}) of an initial magnitude of 15\%, with a recovery of 15 months, see solid blue line in Figure~\ref{fig:empirical_shocks}.
    \item \textbf{Energy Prices}: Finally, we consider the energy price shock. The reduction of demand and production throughout the pandemic led to a supply glut in energy markets, which led to a steep decrease in energy prices, such as oil, by up to 70\% for immediate delivery of West Texas Intermediate crude oil (Figure \ref{fig:empirical_shocks}, dashed green line). As the recovery period began, external factors such as extreme weather conditions in various parts of the world and maintenance work that had been postponed during the pandemic caused a surge in demand, and the energy prices thus rebounded quickly \citep{zakeri2022pandemic}. Unfortunately, with the Russian invasion of Ukraine in February 2022 this rebound was further exacerbated to a global energy crisis, due to Russia's position as a major global exporter of natural gas and oil. By June 2022, the WTI crude oil spot price had peaked, rising nearly 100\% compared to pre-pandemic levels. Following this, the global recovery and adjustment of energy markets has led to a sharp easing of energy prices. For Mark-0, these processes are external to the model, such that we introduce here an exogenous price shock to firms' price update Eqs.~\eqref{eq:price_setting_up}-\eqref{eq:price_setting_otherwise}. Specifically, firms' prices change by an additional exogenous factor $g_e \Delta p_{e,\text{ema}}(t)$, where $g_e$ is a constant factor akin to the energy-share in production, and $\Delta p_{e,\text{ema}}(t)$ is an exponentially weighted moving average of the time-dependent monthly percentage change in energy prices. Thus, the transmission of the change in energy prices to firms' product prices is smoothed, as expected to be the case in reality. 
    
    The form of the moving average is the same as for trust in Central Bank Eq.~\eqref{eq:tau_anchoring} and inflation expectations
    Eq.~\eqref{eq:inflation_ema}, with the very same memory time parameter $\omega$. Our artificial energy price series is shown in Figure \ref{fig:empirical_shocks} as the solid green line, and is based on the WTI Spot Price.\footnote{Prior to the shock, we set $p_e(t) = \overline{p}(t)$.} Alternative calibrations with the energy component of the U.S. Consumer Price Index or the Henry Hub Natural Gas price lead to similar shapes and magnitudes. In the following we will choose $g_e = 3.25 \%$, i.e. half of the energy share of the GDP in the US, where the factor 2 accounts for inventories and partial substitutability. We also posit that the profits of the energy sector are  transferred back to households at an effective rate of $\delta_e=4\%$ per time step, unless stated otherwise.
\end{enumerate}

\subsection{The Effect of Shocks Without Stabilisation Policies}

\begin{figure}[H]
    \centering
    \includegraphics[width=\textwidth]{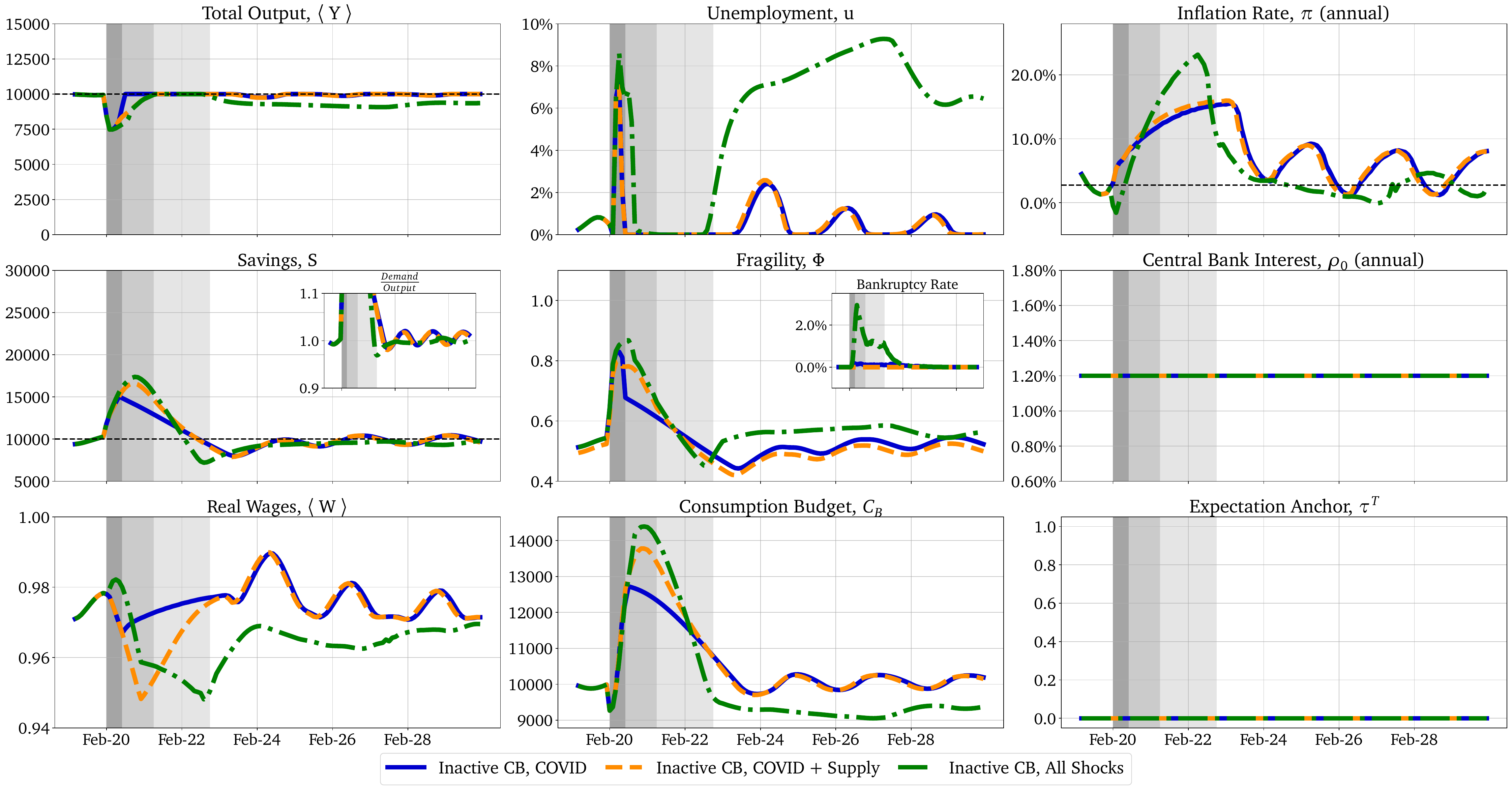}
    \caption{\textbf{Economic dashboard for the three shocks in the Inactive Central Bank scenario and in the absence of any policies:} The dynamics for the three shock scenarios, COVID only (blue), COVID and Supply Chain shock (orange) and all shocks (green) when the Central Bank is inactive and no policy is applied. The areas shaded in grey indicate the duration of the three shocks: the COVID shock lasting until the end of the dark grey area, the supply chain shock until the end of the grey area, and the energy price shock until the end of the light grey area. The dashed black line represents pre-shock averages. In the first two cases, the economy is able to recover on its own (although with significant fluctuations that last for a few years), but significantly gets worse when the energy shock is turned on. 
    }
    \label{fig:baseline_shockeffects_nopolicy}
\end{figure}

To develop some intuition, we now build three counterfactual cases where (i) only the COVID shock, (ii) COVID + supply chain shocks and (iii) COVID + supply chain + energy price shocks hit the Mark-0 economy, without any stabilisation policy put in place. It should be noted that this exercise is somewhat optimistic, as our shocks have been calibrated on an economy where emergency policies were actually implemented. This is discussed further below. Moreover, in the Inactive Central Bank scenario discussed in this section, economic agents form their inflation expectations solely on the basis of realized inflation, which means that there is no anchoring of expectations ($\tau^T=0$, see Eq.~\eqref{eq:inflation_expectation}). All other parameters of this scenario can be found in Table \ref{tablenocb}.

\begin{enumerate}[{Case} (i)]
    \item After the impact of only the COVID shock, the economy remains operational with full capacity but experiences a persistent and high inflation that peaks at 15.5\% and only starts to recede by the end of 2023, with significant fluctuations that persist for a few years (Figure \ref{fig:baseline_shockeffects_nopolicy}, blue). All other observables return to their pre-shock levels by the end of 2023, with small oscillations around the steady state in accordance with the Inactive Central Bank scenario without shocks.
    \item When adding the supply chain shock (Figure \ref{fig:baseline_shockeffects_nopolicy}, orange), the dynamics remain very similar: the economy returns to its steady state without fiscal or monetary policy. Remarkably, inflationary dynamics are almost identical, albeit with a slightly higher peak value (16.0\%) reached slightly earlier in 2023. The biggest difference is the prolongation of the recovery period, which leads to a steeper real wage dip.
    \item Taking now the energy price shock into consideration, the situation abruptly changes. Initially, we observe a short deflation period due to the steep drop in energy prices, increasing real wages, supply-demand imbalance and the fragility of firms due to the decreased prices.\footnote{
    We consider here a direct price-shock transmission. In practice, energy is purchased in various forms and often with inventories and financial contracts to insure against price volatility. In this respect, it might make more sense to define the effective price of energy $p_e(t)$ in the production process as an exponential moving average of the WTI spot price. This would smooth out the initial dip and lead to more realistic inflation time series. We leave this for a later study.} Immediately following this, we observe an explosion in inflation peaking at 23.2\% mid-2023. Firms fragility increases due to the price shock which causes bankruptcies (peaking at 3 \% rate), unemployment and a decrease of wages. In the long run, this disruption is so strong that wages recover only very slow. This starts a feedback cycle of low demand that leads to a decrease of output, which causes a reduction in savings and therefore demand drops. This feedback cycle consistently increases unemployment, and decreases savings and wages, which takes the economy significantly longer than 10 years to recover fully with a high and persistent unemployment between 6\%-8\%. 
\end{enumerate}

In the realm of the Mark-0 model, the COVID and supply chain shocks would thus have had a minor long term impact on output, but would have led to substantial medium term inflation due to excess savings, as predicted using the Mark-0 model as early as June 2020 in \cite{SharmaEtAl2020WshapedRecoveryCOVIDa}. The energy price shock, however, is the last straw on the camel's back and, in the absence of monetary and fiscal policies, has a strongly detrimental long-term impact on the economy. This is a consequence of the existence of discontinuous transition boundaries (a.k.a. {\it tipping points}) in parameter space, as emphasized in \cite{GualdiEtAl2015TippingPointsMacroeconomica} and, within the specific context of the COVID shock, in \cite{SharmaEtAl2020WshapedRecoveryCOVIDa}. As a scenario-generating tool for policymakers, our model demonstrates that in the absence of mitigating policies, the full sequence of COVID, supply chain and energy shocks can trigger a negative feedback cycle, manifested as a downward wage spiral, that results in a collapse of demand and a full blown crisis.  However, we see that the first two shocks, as modelled above, are of sufficiently mild amplitude not to trigger a complete collapse of the economy. In the following section, we study counterfactuals with larger ``bare'' (unmitigated) shock amplitudes, where mitigating policy measures are indeed needed to prevent a catastrophic collapse of output beyond a certain tipping amplitude of the initial COVID shock alone.   

\subsection{Sensitivity to Shock Magnitude and the Role of Easy-Credit Policy}\label{sec:sensitivityshocks}

\begin{figure}[h]
    \centering
    \includegraphics[width=0.8\textwidth]{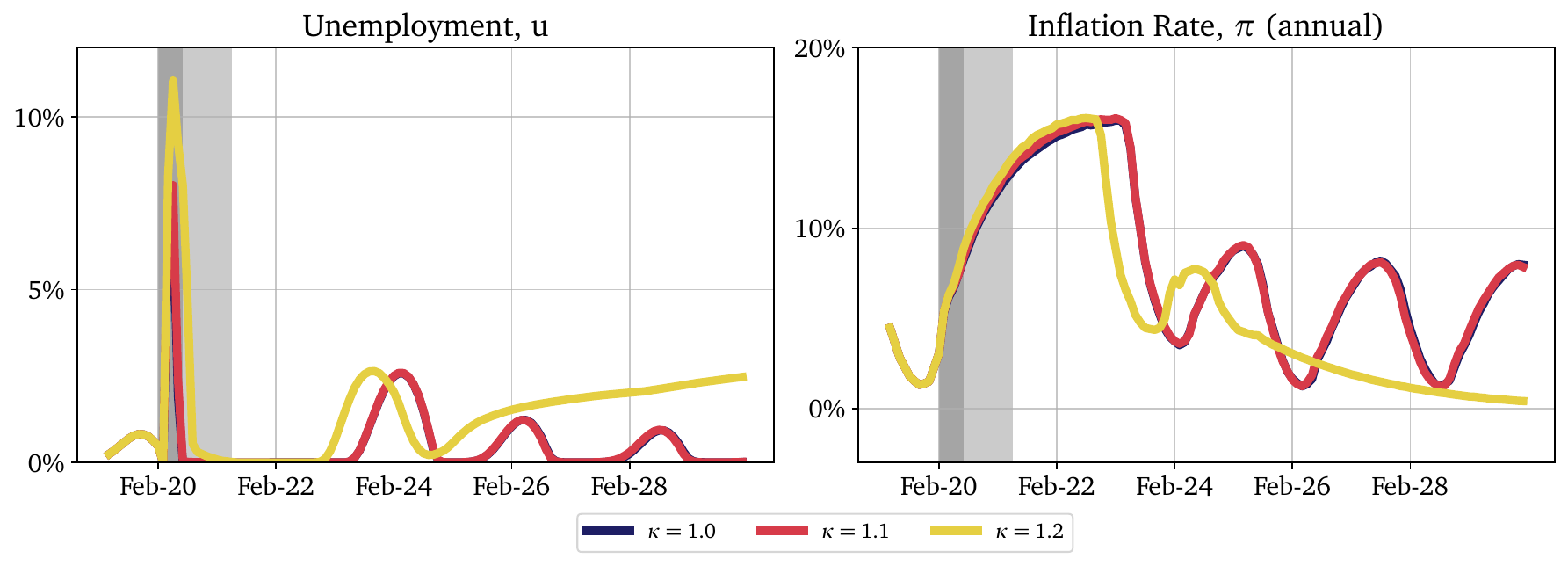}
    \caption{\textbf{Counterfactual COVID shock in the Inactive Central Bank scenario and in the absence of any policies:} Unemployment (left) and inflation (right) in the Inactive Central Bank scenario without any policy, with an amplified COVID shock by a factor $\kappa$ and Supply Chain shock. $\kappa=1$ (blue line) corresponds to the COVID shock considered in the previous section, with a very low level of unemployment. With $\kappa=1.1$ (red line), the unemployment and inflation dynamics change only very slightly, as the shock is mild. As soon as $\kappa \ge 1.2$, the economy is unable to recover spontaneously, and output collapses as demonstrated in the $\kappa=1.2$ case (yellow line).
    }
    \label{fig:counterfactual}
\end{figure}

\begin{figure}[t]
    \centering
    \includegraphics[width=\textwidth]{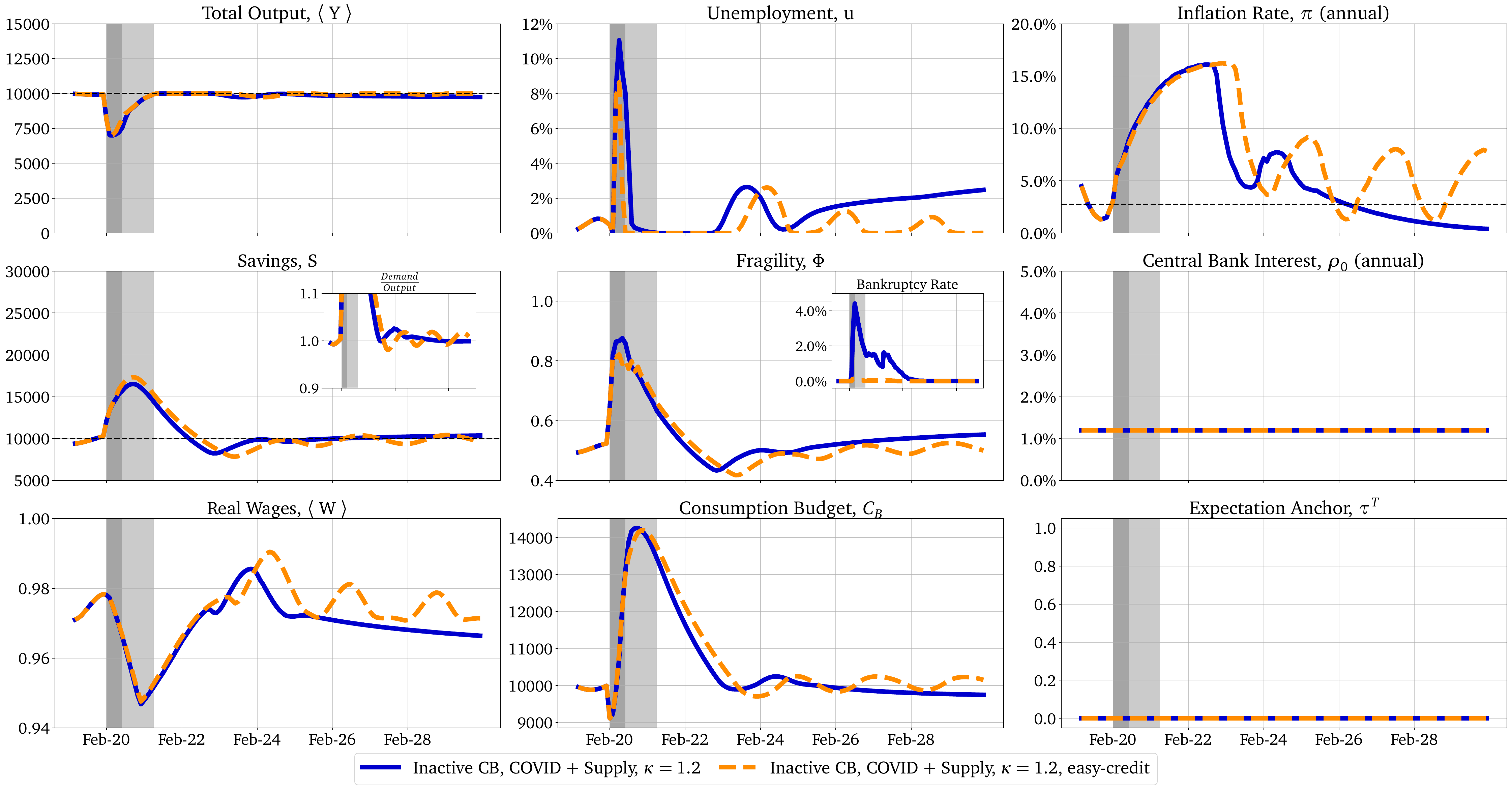}
    \caption{\textbf{Economic dashboard for a stronger COVID shock in the Inactive Central Bank scenario, with and without Easy-Credit policy:} The dynamics of the Inactive CB scenario in case (ii) (COVID and supply chain shock but no energy price shock), for a COVID shock strength of $\kappa=1.2$ without Easy-Credit policy (blue) and $\kappa=1.2$ with Easy-Credit policy, $\mu=1.3$ (orange). We see that policy is effective in preventing the economic collapse in the case of the stronger COVID shock.
    }
    \label{fig:counterfactual_adaptive}
\end{figure}

Our model is based on data that already incorporates policy measures, which were indeed in place in 2020-21 to mitigate shocks. As such, it is possible that the unmitigated shocks were actually more severe than what we observe retrospectively in data. Here, we investigate counterfactual (ii)-scenarios with different shock magnitudes, where we shut down the final energy shock but scale the amplitude of the observed COVID shock by a factor $\kappa$ (see Figure \ref{fig:empirical_shocks}, left panel dark grey area). This stronger shock could have taken place if for example the CARES act and the CAA where not authorized.\footnote{The supply chain shock is left unchanged in the present exercise, but we observe almost the same phenomenon without such additional shock.} As $\kappa$ increases, firms go progressively bankrupt due to their high financial fragility $\Phi$, leading to a higher unemployment rate during the shock. Households save money during the shock, resulting in higher demand, leading to upward pressure on prices and somewhat higher inflation. Therefore in the long run, the scenarios with intensified shocks $\kappa \lesssim 1.2$ very slightly increases inflation and long-term unemployment. However, for larger $\kappa$, a tipping point is reached, as in \cite{SharmaEtAl2020WshapedRecoveryCOVIDa}: more initial bankruptcies and firms' hesitation to increase wages after the shock leads to a collapse of output and deflation, see Figure~\ref{fig:counterfactual} (for a full dashboard see Figure \ref{cf21_covid} in Appendix \ref{appx:counterfactual}). Eventually, the shock is too strong for the economy to recover, and the negative feedback loop of a downward wage spiral causes the economy to collapse.

The default threshold of firms $\Theta$, holds a pivotal role in the economic downturn; if chosen too low an excessive number of firm bankruptcies ensues, which can be detrimental to the overall economy. In this case, allowing firms to accumulate more debt by easing bankruptcy rules is highly effective to keep the economy on an even keel. In fact, as shown in Figure~\ref{fig:counterfactual_adaptive}, the ``Easy-Credit'' policy defined by Eq.~\eqref{eq:adaptive} with $\mu=1.3$ for the bankruptcy threshold $\Theta$ manages to substantially reduce the impact of strong COVID-related shocks, even with an increased initial COVID shock ($\kappa=1.2$) that would collapse the economy on its own. With Easy-Credit policy, firms can maintain wages and do not need to fire employees to remain solvent, allowing the economy to recover. However, unemployment still peaks above 8\% (with a second hump mid 2024) and inflation reaches 16.3\% at the end of 2022. As we will discuss in section~\ref{sec:sloppy} below, the main issue with such a period of high inflation is the risk of de-anchoring inflation expectations, opening the path to possible hyper-inflation.

In order to control inflation, one needs to consider the effect of monetary policy. We thus now turn to the study of the same sequence of three non-amplified shocks, with Easy-Credit policy and with a fully active Central Bank, distinguishing between the case of Anchored Expectation and the case of Floating Expectation in its ability to curb inflation.  

\section{Monetary Policy Response to Inflationary Shocks}
\label{sec:6}

In the previous section, we showed that the loosening of regulatory bankruptcy policy during severe shocks prevents the Mark-0 economy from collapsing, but does not address the issue of high inflation rates (Section \ref{sec:sensitivityshocks}). To study inflation mitigation policies, we now introduce a Taylor-rule based monetary policy into the mix, 
i.e. we combine the situation explored in Section~\ref{sec:shocks} (all shocks and Easy-Credit policy without Central Bank) with the different monetary policy scenarios discussed in
Section~\ref{sec:4}. All numerical experiments henceforth are run with the Easy-Credit policy described in Eq.~\eqref{eq:adaptive} with $\mu=1.3$, as in Section~\ref{sec:shocks}. As the economy returns to its steady state, the Easy-Credit policy becomes equivalent to the fixed default threshold, such that its duration depends on the recovery period.\footnote{Eq. \eqref{eq:adaptive} indeed implies $\Theta(t)=\Theta_0$ in the steady state when the average fragility is low, which is the case for the choice of parameters made in this paper.}
Note that in what follows, we consider the three shocks together, keeping the amplitude of the COVID shock to the one calibrated on the observed data (i.e. $\kappa = 1$ in the language of Section~\ref{sec:shocks}). As already discussed, such calibration may underestimate the severity of the unmitigated shock, since the US government immediately implemented Easy-Credit and Helicopter Money measures to alleviate the COVID shock. Note also that unemployment is not an explicit target of our Central Bank, in the sense that the Taylor-rule is only responsive to realized inflation and not to unemployment.  

\subsection{An Inactive Central Bank}

We begin our study by considering again the situation with an Inactive Central Bank as in Section~\ref{sec:4.1}, but now with all shocks together and with the Easy-Credit policy being implemented
(Figure~\ref{fig:activeCB_ap_short}, full blue line; see Figure~\ref{all_scenarios} in Appendix \ref{appx:dashboards} for all macroeconomic timeseries).
In this case, over the first few months, the combination of shocks causes a spike in unemployment that peaks around 8\% during a deflationary phase, 
but the economy quickly recovers to full employment, yet at the price of a sustained inflationary period that peaks above 20\% before slowly reverting to equilibrium,
consistently with the discussion of Section~\ref{sec:shocks}: the Easy-Credit policy can avoid full collapse, but only at the price of high inflation.

\subsection{Monetary Policy without Anchoring}
{
Before introducing trust dynamics, we consider the case where agents form their inflation expectations without considering the Central Bank's target inflation, but the Central Bank is nonetheless actively pursuing monetary policy.
We thus set $\tau^T=0$ such that inflation expectations $\hat{\pi}(t)$ are simply an exponentially weighted moving average over realized inflation. 
In this case, we find that the Central Bank initially decreases rates before rapidly increases interest rates as inflation increases (see Appendix \ref{appsec:anchoring} for the corresponding dashboard).
However, there is very little resulting effect on inflation, which displays roughly the same dynamics as when the Central Bank is inactive. In other words, the direct impact of higher interest rates on inflation is weak, and expectation anchoring will be dominant in this respect. 

However, the de-Anchored Expectations lead to a doubling in magnitude of the initial spike in the unemployment rate, together with a higher (and oscillating) peak following the end of the shocks, as compared to the case where monetary policy is inactive. This is due to the spike in debt induced by bankruptcies accompanying the initial wave of shocks.

We can thus draw a first conclusion that in the absence of any sort of expectation anchoring the main impact of monetary policy is detrimental to the unemployment rate, which is in fact not within the Central Bank's mandate in our model setup.
}

\subsection{Monetary Policy with Anchored Expectation}\label{sec:interest_only}

We next consider a monetary policy experiment with a responsive Central Bank in the ``Anchored Expectation'' scenario. In the Mark-0 model, as in real life, the Central Bank's aim is to keep inflation on target. Here, following Section~\ref{sec:policydescriptions}, the inflation target is chosen to be $\pi^\star = 2.4 \%$ p.a. and the Taylor rule strength is $\phi_\pi = 1$. Moreover, we optimistically assume that the Central Bank has successfully communicated to all economic actors that it has inflation under control, thereby convincing them to believe that long-term inflation will be close to the Central Bank's inflation target, with a fixed anchor weight $\tau^T = 0.95$. 

\begin{figure}[t]
    \centering
    \includegraphics[width=\textwidth]{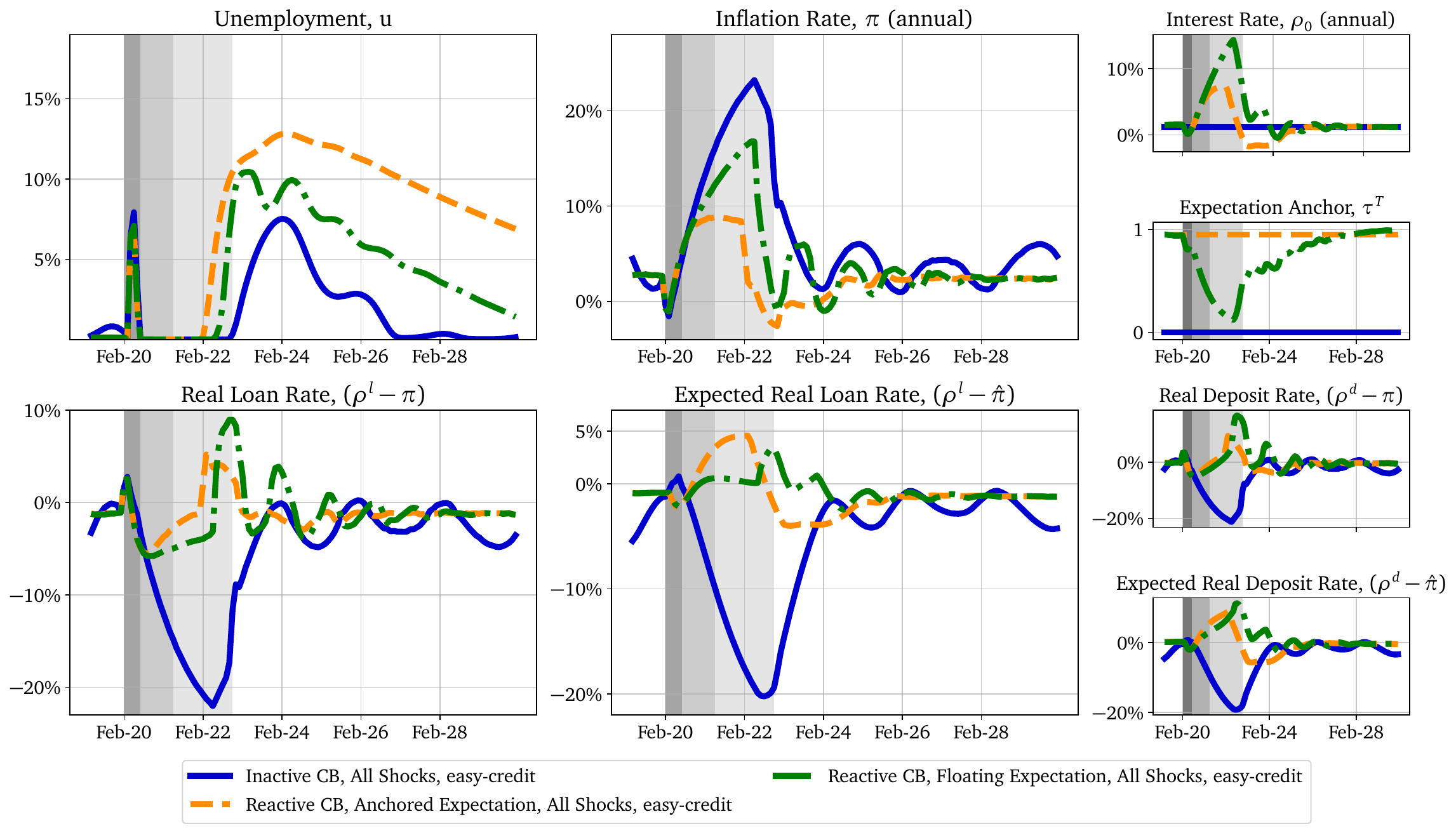}
    \caption{\textbf{Unemployment and Inflation with all shocks, Easy-Credit policy, and three distinct monetary policy scenarios}. (Blue lines) Inactive Central Bank scenario. (Orange lines) Reactive Central Bank with Anchored Expectation, with anchor parameter $\tau^T=0.95$. A Taylor rule policy successfully decreases peak inflation but increases peak unemployment. (Green lines) Reactive Central Bank with Floating Expectation. Here, trust is eroded during the high inflation period. Monetary policy then fails at reducing inflation, with the risk of hyper-inflation lurking (see section \ref{sec:sloppy}); unemployment remains lower because of the higher inflation.
    }
    \label{fig:activeCB_ap_short}
\end{figure}

In the case of Anchored Expectation and variable interest rates with all three shocks (Figure \ref{fig:activeCB_ap_short}, dashed orange line), the Central Bank is successful in taming inflation: peak inflation is lower than in the case without an active central bank (peak at 8.8\% vs 23.2\% in the Inactive Central Bank scenario) and quickly tapers off. Although the inflation rate is not in line with the Central Bank's target throughout the crisis, it appears that monetary policy is able to reign in inflation. The reduction in inflation is not primarily due to the impact of interest rate policy but rather to the strong anchoring of expectations, which significantly dampens expected (and thus realized) inflation, as upward pressure on both prices and wages are reduced. Such anchoring moderates the price increases after the initial COVID shock, but does not dampen the deflationary dynamics towards the end of the energy price shock, as this is driven by a strongly increasing unemployment that puts downward pressure on wages (see Figure \ref{all_scenarios} in Appendix \ref{appx:dashboards} for all macroeconomic timeseries).

The initial price shock in the first few months causes bankruptcies that are less numerous with a reactive Central Bank, hence also reducing the initial unemployment spike slightly (peak at 8.1\% in the Inactive Central Bank scenario and 7.4\% with Active Central Bank with Anchored Expectation). However, at the end of the shock the Central Bank effort to control inflation results, within an Anchored Expectation scenario, in a peak unemployment of 12.8 \% in Feb 2024, 5\% higher than in the Inactive Central Bank scenario. {Despite realized inflation surging during shocks, firms inflation expectations stay anchored to the central bank's target. This results in a substantial decline in real wages during the shock and a sluggish recovery thereafter.
Furthermore, firms' wariness of the cost of debt is increased (through parameter $\Gamma$, Eq. \eqref{eq:gamma-coupling}), leading to an increase of the firing rate. Similarly, households expect a controlled level of inflation and hence reduce consumption in the face of high interest rates that favour savings. 

All these effects lead to higher unemployment, impeding the overall post-shock recovery.} We thus see that in the Mark-0 model, the Central Bank is stuck between a rock and a hard place with its inflation-unemployment trade-off, meaning that in the context of these shocks, monetary policy is not a panacea: some form of Keynesian stimulus or ``Helicopter Money'' is needed in conjunction with monetary policy to restore the economy to its pre-crisis steady-state -- see section \ref{sec:helicopter} below. In this regard, a Central Bank with an explicit dual-mandate would potentially navigate the trade-off better.
 
\subsection{Monetary Policy with Floating Expectation}\label{sec:floatingtrust}

The scenario of high and Anchored Expectation in monetary policy for long-term inflation expectations might be a good approximation during long periods of stability, but in periods of shocks, the anchoring $\tau^T$ may decrease when inflation deviates strongly from the Central Bank's target \citep{Reis2021LosingInflationAnchor}. We model such a Floating Expectation effect through Eq.~\eqref{eq:tau_anchoring}, where the anchoring parameter decreases when the spread between observed and target inflation increases. We set $\alpha_I=0.4$ and a memory time of $\sim5$ months ($\omega=0.2$). These values imply that when inflation reaches $4$ times the Central Bank's target $\pi^\star$, trust in the Central Bank falls to approximately a third of its initial value after approximately one year. Larger values of $\alpha_I$ would lead to an even steeper de-anchoring of inflation expectations. Such a loss of faith in monetary authorities further increases realized inflation as economic agents expect higher inflation in the future and take this into account when setting prices and wages. This may lead to a self-fulfilling feedback loop between expected inflation and real inflation through a wage-price spiral (see section \ref{sec:sloppy}). 

Compared to the Anchored Expectation case, the surge of inflation after all COVID shocks leads to a loss in trust that almost vanishes (i.e. $\tau^T \to 0$) during the energy price shock. The consequence of this is a higher realized inflation rate, that reaches a peak value only slightly below the Inactive Central Bank scenario (Figure~\ref{fig:activeCB_ap_short}, right).\footnote{For the full economic dashboard in the case of dynamic expectations, see Appendix \ref{appx:dashboards}, Fig. \ref{activeCB_dyn_ap}, and compare with  Fig. \ref{activeCB_ap}.} Unemployment, on the other hand, turns out to be {\it higher} than with an Inactive Central Bank (see Figure~\ref{fig:activeCB_ap_short}, left), {but {\it lower} than when trust is well anchored. As noted above, low expected inflation is detrimental to economic activity when interest rates are high. Therefore, perhaps paradoxically, loss of trust improves the situation in terms of unemployment, but is of course detrimental to inflation, with the lurking possibility of runaway situations. This again illustrates the quandary faced by Central Banks, and the importance of endowing them with a double mandate.} 

A strong factor of the loss in trust is the exogenous nature of the energy price movements. Despite the strong hike up to a real interest rate of 3.4\% per year (16.8\% nominal target rate), inflation remains far above target. Once these shocks have passed, the Central Bank is arguably quite successful in keeping inflation closer to its steady state value. However, similarly to the Anchored Expectation case, the economy now contends with a lower, but persistent unemployment problem. The unemployment here is driven by the energy price shock that causes a strong surge in inflation, which erodes household savings and reduces real wages by 4\% compared to its steady state value, thus resulting in a drop in demand (see Figure~\ref{activeCB_dyn_ap} in Appendix \ref{appx:dashboards}). 
In contrast to the Anchored Expectation scenario, unemployment in the Floating Expectation scenario recovers to full employment faster. As noted above, this paradoxical result is driven by the comparatively higher real wages and consumption propensity, which allows for a faster recovery of the consumption budget and consequently output.

As with both the Inactive Central Bank and Anchored Expectation scenario, strong exogenous inflation outside of the Central Bank's sphere of influence leads to a severe contraction of the economy, which recovers only slowly. The convergence to pre-crisis equilibrium takes more than 18 years in the Anchored Expectation scenario and 11 years in the Floating Expectation scenario. This suggests that in the face of exogenous price shocks, fiscal policy shoring up consumer's budget or reducing energy price effects on firms and households can be a more effective tool than monetary policy. 

\subsection{Sensitivity to the Strength of the Monetary Policy}

In circumstances of crises, as with our shocks, Central Banks should possibly react more strongly to inflation and keep expectations anchored. We thus consider a larger Central Bank Taylor reactivity parameter  $\phi_\pi = 2.0$ (see Appendix \ref{appx:strongerCB}, Figure \ref{activeCB_stronger_ap} and \ref{activeCB_dyn_stronger_ap} for details).

In the case of Anchored Expectation, an increase in $\phi_\pi$ hardly changes inflation, but stronger interest rate hikes result in significantly higher unemployment rates. 
Hence, stronger monetary policy is clearly detrimental in this case. By contrast, in the Floating Expectation scenario with dynamic trust, an increase in $\phi_\pi$ signals a stronger commitment of the Central Bank to control inflation. Therefore, all else being the same, inflation expectations are lower which, in a self-fulfilling manner, cuts realized inflation from 16.8\% to 11.2\% and keeps trust better anchored. However, as in the Anchored Expectation case, the sharper interest rate hikes by the Central Bank lead to a significant increase in the unemployment rate (from 12.8\% to 20.5\%) and a much longer recovery time for the economy as a whole. Again, our model confirms and quantifies the observation made by \citet{StiglitzRegmi2022CausesResponsesToday}, quoted in the introduction: monetary policy interest tools tend to be too blunt, curbing inflation at the cost of unnecessarily high unemployment.

\subsection{Sensitivity to Transmission Channels}

Within our modelling framework, the impact of monetary policy relies on the efficiency of three transmission channels: (a) expectation anchoring, discussed above; (b) sensitivity of consumption on real interest rate, through parameter $\alpha_c$; (c) sensitivity of firms' hiring and wage policies on real interest rate, through parameter $\alpha_\Gamma$. We have run some simulations to check the dependence of the state of the economy on these last two parameters in the Anchored Expectation case, see Appendix \ref{appx:alphacgamma}. The conclusion of this exercise is that increasing $\alpha_c$ and $\alpha_\Gamma$ has a minor direct effect on inflation but a significant effect on unemployment, depending on the sign of the real interest rate $\rho - \hat{\pi}$. As expected, a raise of interest rates degrades economic activity, all the more so when the sensitivity of firms and households to the real rate is higher. As already stated, the main transmission channel through which monetary policy impacts inflation is expectation anchoring, rather than directly through cost of loans or income on savings.        

\section{Fiscal Stimulus}
\label{sec:7}

Section~\ref{sec:6} showed that in the presence of exogenous inflation drivers, the Central Bank can reduce inflation only at the cost of high unemployment. The mechanisms responsible for unemployment are the increased burden of debt that weighs on wages and increases the speed at which firms lay off workers. This in turn lowers demand, an effect amplified by the erosion of household's purchasing power by the lingering inflation. One way to tackle this issue is to ensure that demand does not drop as severely by increasing it through fiscal stimulus (see Section \ref{sec:mechanism_helicopter}). We test this policy device here by considering different kinds of fiscal stimuli. 
Unless otherwise indicated, the fiscal stimuli are applied on the scenario of Section~\ref{sec:floatingtrust}, i.e. with all shocks, Easy-Credit policy, and an Active Central Bank with Floating Expectation, which we believe to be the scenario closest to reality.

\subsection{The Effects of Helicopter Money}
\label{sec:helicopter}

\begin{figure}[h]
    \centering
    \includegraphics[width=\textwidth]{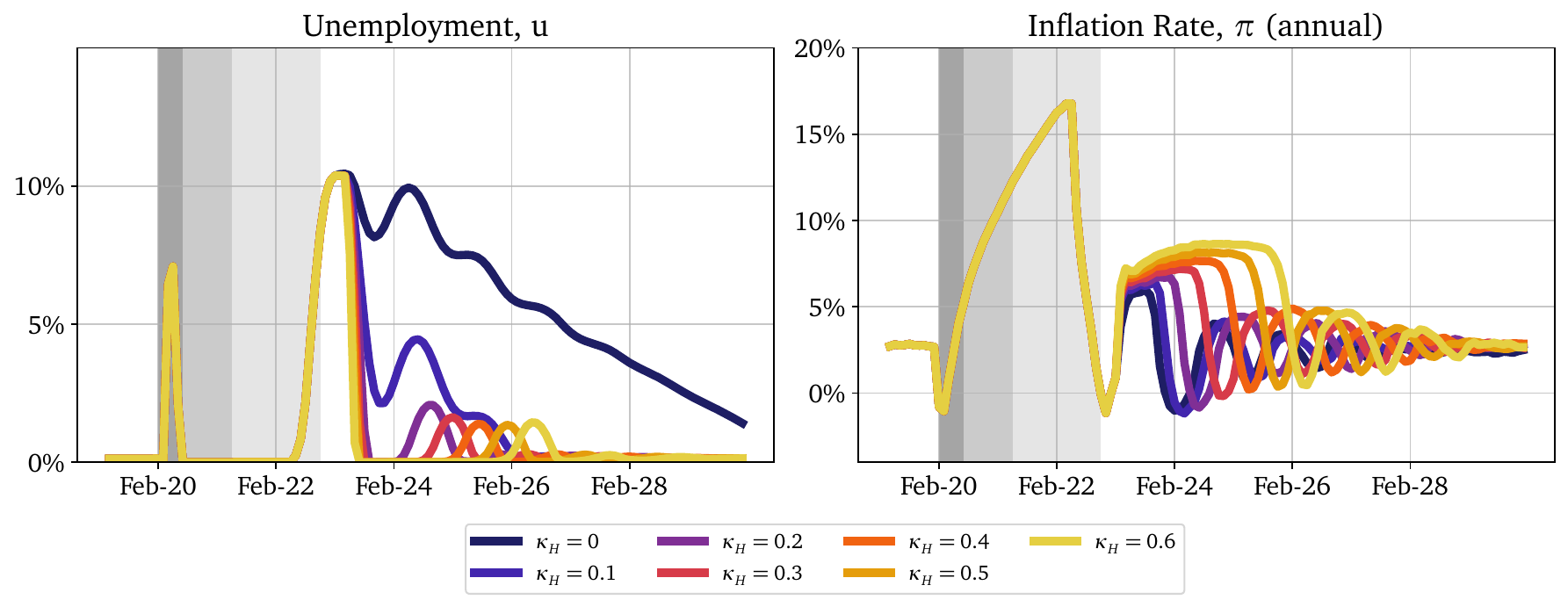}
    \caption{\textbf{Helicopter Money in the Reactive Central Bank with Floating Expectation scenario, Easy-Credit policy, and all shocks.} Unemployment (left) and inflation (right) for a helicopter drop of size $\kappa_H$ times savings one month after the price shock. Already with $\kappa_H\ge 0.2$ unemployment is reduced almost to zero. A further increase of Helicopter Money only increases the duration of the high-inflation period, without further reducing unemployment. Note the small unemployment and inflation ``ripples'' persisting several years after the initial shock.}
    \label{activeCB_dyn_ap_helicopter_spectrum}
\end{figure}

Following~\citet{SharmaEtAl2020WshapedRecoveryCOVIDa}, we begin by considering a one-time increase of the households' budget by a factor $\kappa_H \in [0\%, 60\%]$ one months after the end of the energy price shock (July 2023). The goal is to find an optimal multiplier $\kappa_H$ that reduces unemployment while keeping excess inflation to a minimum
in amplitude and duration. Here we find that this minimum is reached for $\kappa_H \approx 20\%$ (see Figure \ref{activeCB_dyn_ap_helicopter_spectrum} for the Reactive Central Bank with Floating Expectation case).\footnote{See Appendix \ref{appx:helicopter}, Figures \ref{baseline_ap_helicopter_spectrum}, \ref{activeCB_ap_helicopter_spectrum} for the other two monetary policy scenarios, and Figures \ref{baseline_ap_helicopter}, \ref{activeCB_ap_helicopter} and \ref{activeCB_dyn_ap_helicopter} for the complete economic dashboards of all scenarios.}
This choice of $\kappa_H$ matches the magnitude of the ``Emergency Money for the People Act'', which entailed the US government providing direct payments of 2000 USD per month for a maximum of 12 months to support individuals during the COVID pandemic.

In all considered scenarios, the fiscal stimulus package leads to a quick recovery of production to its pre-shock levels and unemployment at near 0\%, thus successfully eliminating the steep recession and long recovery following the energy price shock, as shown in Figure~\ref{activeCB_dyn_ap_helicopter_spectrum}. However, policymakers face an inflation-unemployment trade-off, as the stimulus generates a spurt of inflation. In the absence of monetary authority, inflation due to the injection of money rises to 15.8\% (see Appendix \ref{appx:helicopter}, Figure \ref{baseline_ap_helicopter_spectrum}). However, this is an endogenous inflation, which means that in the Floating Expectation case, the Central Bank can raise rates to curb consumption propensity and keep inflation below 10\% p.a. (Figure \ref{activeCB_dyn_ap_helicopter_spectrum}). Increasing the stimulus leads to a higher inflation peak and a longer duration of high inflation.

The inflationary period after the stimulus is in part due to a short-term de-anchoring of expectations as the demand stimulus leads to higher prices in the context of a tight labour market preventing production increases. Yet, here one can take the position that the economy is in good shape from a macroeconomic perspective, with near-zero unemployment despite an above-target inflation.\footnote{Because Mark-0 treats households at the aggregate level, we do not analyze the distributional consequences of inflation here, though inflation is always and everywhere differential in its effects.} In this case, the Anchored Expectation case would be much more favorable as the high unemployment problem is resolved with only a minimal rise in inflation, around 7\% annualized over 8 months, as soon as $\kappa_H \gtrsim 0.2$. 

\subsection{The Effects of a Windfall Tax}
\label{sec:windfall}

\begin{figure}[H]
    \centering
    \includegraphics[width=\textwidth]{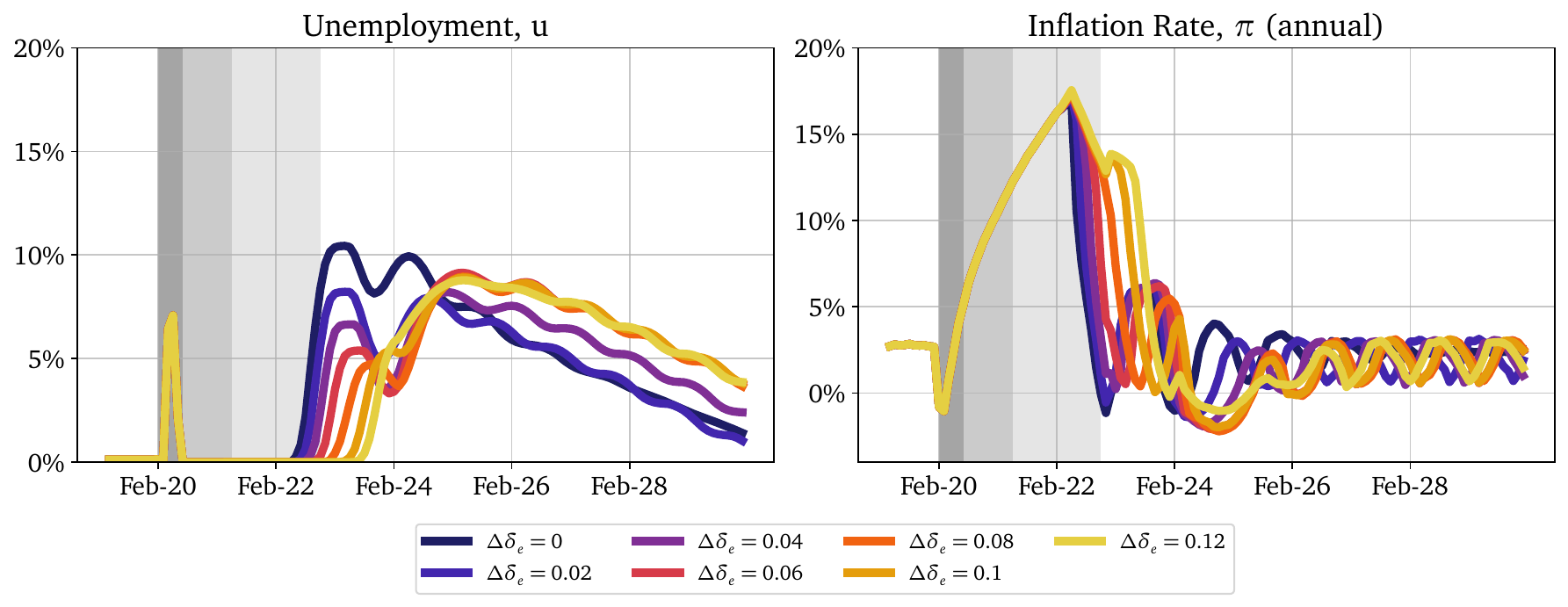}
    \caption{\textbf{Windfall Tax in the Reactive Central Bank with Floating Expectation scenario, Easy-Credit policy, and all shocks.} Unemployment (left) and inflation (right) for a Windfall Tax of $\delta_e + \Delta \delta_e$ one year before the end of the price shock with a duration of two years. With $\Delta \delta_e \approx 4\%$ unemployment is reduced strongly. A further increase of tax does only increase unemployment again. For even larger $\Delta \delta_e$, inflation increases because of increased demand due to increased savings.}
    \label{activeCB_dyn_ap_windfall_spectrum}
\end{figure}

Another stimulus that central authorities can use to alleviate the impact of the price shock is a Windfall Tax. This policy redistributes the excess profits generated by the energy sector as a result of rising energy prices to household savings with the aim to stimulate consumption and consequently reduce unemployment. We test the effectiveness of the policy, which in our modelling consists in an increase of the energy sector payout rate $\delta_e \rightarrow \delta_e + \Delta \delta_e$ one year before the end of the price shock with a duration of two years, with $\Delta \delta_e \in [0\%,12\%]$.\footnote{We remind here that $\delta_e$ is the fraction of the energy sector cash balance redistributed to households at each time step, see Eq. \eqref{eq:energy_sector}.} 

Our aim is to design the fiscal stimulus in a manner that minimizes both unemployment and inflation. We find that for $\Delta\delta_e=2\%-4\%$ the excess profits of the energy sector are redistributed such that unemployment is significantly reduced at all times while inflation remains under control. Further increasing $\Delta\delta_e$ reduces the unemployment in 2023, but at the cost of increasing unemployment in 2024 after the Windfall Tax has ended (see Figure \ref{activeCB_dyn_ap_windfall_spectrum}).\footnote{For more detail, see Appendix \ref{appx:windfall}, Figures \ref{baseline_ap_windfall_spectrum}, \ref{activeCB_ap_windfall_spectrum}.} As long as the tax is active, the increase in $\delta_e$ results in higher savings, leading to an increased consumption budget and therefore higher demand. This has the effect of reducing unemployment as firms expand production, as well as reducing firm fragility due to increased profitability. However, as the Windfall Tax increases in the Floating Trust scenario, inflation also increases and expectations remain de-anchored for longer, compelling the Central Bank to increase interest rates. As the tax ends, the household's consumption budget has been reduced to below the case without a Windfall Tax, and the real interest rate remains high. This leads to a negative spiral of lower demand and output, higher unemployment, leading to lower income, until the system equilibrates again.

The inflationary effects of the increase in savings due to the Windfall Tax remain minor for small values of $\Delta \delta_e$. However, once the Windfall Tax exceeds $\Delta \delta_e \sim 10\%$, a longer period of high inflation after the shock is observed (see Figure \ref{activeCB_dyn_ap_windfall_spectrum} and Appendix \ref{appx:windfall}, Figure \ref{activeCB_dyn_ap_windfall}) as inflation expectations remain de-anchored for a longer period.

The Anchored Expectation scenario presents a different challenge. Again, up to a threshold, the Windfall Tax reduces the initial and long-term unemployment spike, with $\Delta\delta_e=4\%$ still remaining a robust choice (see Appendix \ref{appx:windfall}, Figures \ref{activeCB_ap_windfall_spectrum}, \ref{activeCB_ap_windfall}). However, this time at the cost of inflation (or less deflation) on the way back to the pre-crisis steady state. In this regard, policymakers might opt instead to minimize the volatility of inflation to let it smoothly return to the Central Bank's target rate, which actually suggests a higher Windfall Tax of $\Delta\delta_e=8\%$ for instance.

We conclude that when implementing a Windfall Tax as a fiscal stimulus, excessive tax levels can have the unintended consequence of exacerbating long-term unemployment, while a measured approach can both reduce unemployment and smooth inflation dynamics.

\section{Model Sensitivity: The Dangers of a Wage-Price Spiral} \label{sec:sloppy}

\subsection{``Sloppiness'' Analysis}

To develop a deeper understanding of the reaction to shocks of our model economy, we explore the parameter sensitivity of this configuration of the Mark-0 model using the ``sloppy model'' methodology put forth by Sethna and collaborators (see \cite{TranstrumEtAl2015} for an introduction) and applied to macroeconomic models by \citet{NaumannWoleskeEtAl2023ExplorationParameterSpace}. This method identifies the \textit{stiffest} parameter combinations, along which one observes the largest change in outcomes for small perturbations. In particular, we consider the parameter sensitivity of the dynamics of the inflation rate and of the unemployment rate (for methodological details, see Appendix \ref{appx:sloppy} and \citet{NaumannWoleskeEtAl2023ExplorationParameterSpace}). Overall, this approach provides a more comprehensive understanding of parameter sensitivity by considering the curvature of the model manifold, as compared to one-at-a-time parameter sensitivity analysis.

With respect to both inflation and unemployment, the balance of wage indexation (bargaining power) $g_w$ to price indexation (market power) $g_p$, has by far the strongest influence on the outcome dynamics for the inactive CB and Floating Expectation scenarios (see detailed results in Appendix \ref{appx:sloppy} and \ref{appx:gpgw}). 
Beyond inflation itself, eigenvectors are combinations of various pricing parameters including $\gamma$, $g_p$ and $g_e$ for both unemployment and inflation. Central Bank parameters are present but not significantly.
This aligns with our findings that the Mark-0 economy always maintains a tight connection between unemployment and inflation, indicating that it may not be possible to minimize one without affecting the other (see Appendix \ref{appx:sloppy}, Figures \ref{sloppy_evecs_activecb}, \ref{sloppy_evecs_activecb_dyn}).

\cite{GualdiEtAl2015TippingPointsMacroeconomica} have demonstrated the significance of the default threshold of firms $\Theta$ in determining the economic phase of the Mark-0 model. This parameter exhibits a non-linear behavior, where minimal alterations have negligible effects on the overall dynamics until a critical ``tipping point" is reached. Beyond this threshold, the dynamics undergo substantial transformations. It is important to note that with the ``sloppiness" analysis, we can only identify local sensitivity in parameter combinations, where slight deviations from these combinations result in noticeable changes in observed dynamics. As already found in \cite{NaumannWoleskeEtAl2023ExplorationParameterSpace}, multiple phase transitions occur along the parameter axis of $\Theta$, although minor perturbations do not alter the curvature of the loss function, therefore we do not find $\Theta$ to be crucial in our results. However it is crucial to distinguish between local sensitivity, which can be determined by the sloppiness approach, and global sensitivity to parameters, which can trigger tipping points in the model \cite{GualdiEtAl2015TippingPointsMacroeconomica}.

\begin{figure}[t]
    \centering
    \includegraphics[width=\textwidth]{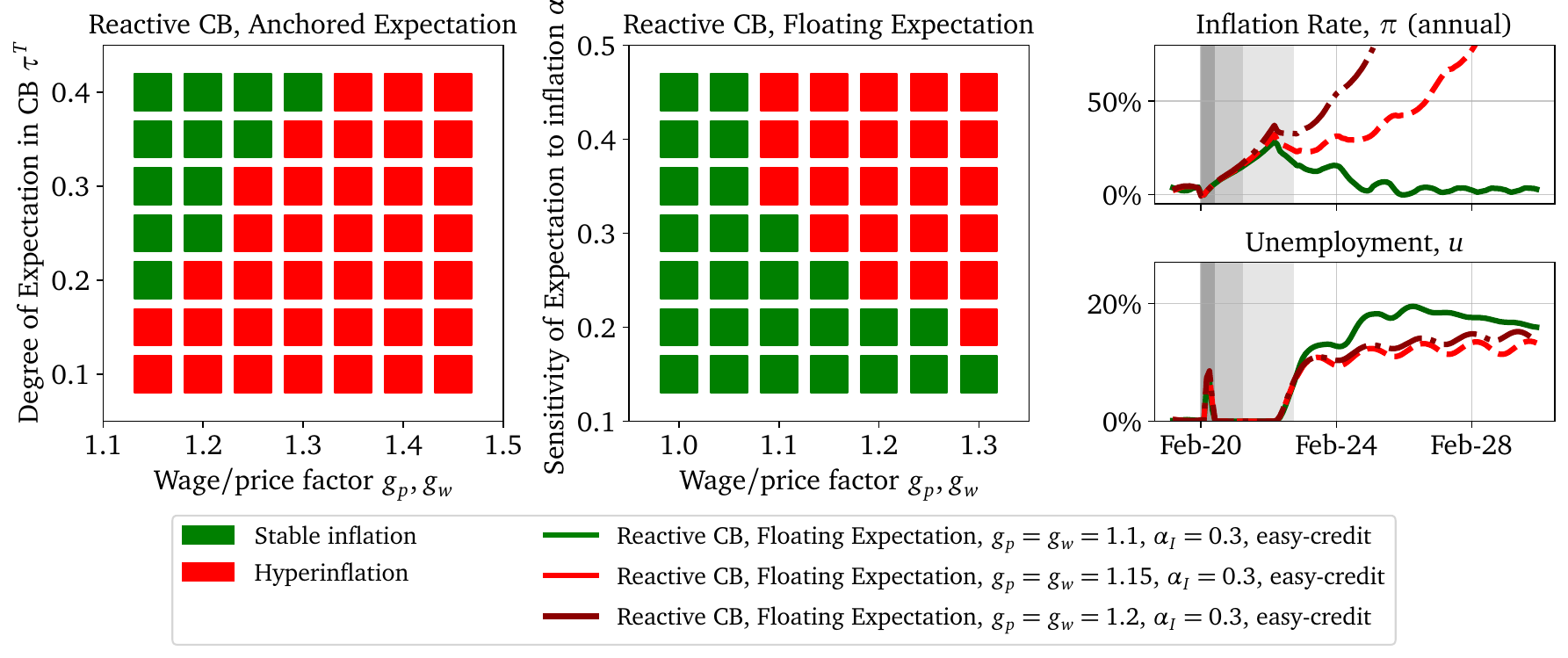}
    \caption{\textbf{Hyperinflation tipping points.} Left: Stable inflation vs. Hyperinflation in the plane ($g_p=g_w$, $\tau^T$) in the Anchored Expectation scenario and in absence of exogenous shocks. Note that inflation can be stable even when $g_p=g_w > 1$ for strong enough anchoring. Middle: Stable inflation vs. Hyperinflation in the plane ($g_p=g_w$, $\alpha_I$) in the Floating Expectation case and in absence of exogenous shocks. Right: inflation rate (top) and unemployment (bottom) as a function of time, in the Floating Expectation case, $\alpha_I=0.3$ and different values of $g_p=g_w$, after the three COVID shocks. When indexation is too strong, an hyperinflation regime sets in at the end of the energy price shock.}
    \label{hyperinflation_phase}
\end{figure}

\subsection{The Risk of a Hyperinflation Spiral}

We can interpret the sensitivity to $g_w$ and $g_p$ in light of the dangers of a hyperinflation episode resulting from a wage-price spiral. \citet{alvarez2022wage} cite a concern that hyperinflation may occur if firms increase wages in response to higher inflation, leading to an increase in purchasing power and ultimately feeding into a wage-price spiral in the current macroeconomic environment. This feedback loop is influenced by the indexation of prices and wages to firms' inflation expectations \citep{holland1988changing}, i.e. by the value of parameters $g_p$ and $g_w$ in our model. In the simplest case where bargaining power and market power are equal ($g_p=g_w$), the economy reaches a stable inflationary state, marked by cyclical fluctuations due to mismatches in supply and demand (see Figure \ref{fig:baseline_noshocks} in Section \ref{sec:model}, for which $g_w=g_p=0.8$). When indexation is weak ($g_p=g_w<1$), neither wages nor prices fully incorporate inflation expectations, and fluctuations in the inflation rate due to mismatches in demand and supply are dampened. Conversely, for strong indexation ($g_p=g_w>1$), the economy may enter a state of hyperinflation, though as wages and prices increase equally fast, there are no real effects (omitting, of course, the cost of inflation itself due to ``menu costs''). Only when $g_p\neq g_w$ does the economy collapse. 

Introducing monetary policy implies that hyperinflation can be staved off due to the anchoring of inflation expectations. In the case of Anchored Expectation, inflation remains stable until a critical point $g_p = g_w < g^\star$ where $g^\star > 1$ depends on the commitment of the Central Bank as well as the strength of expectation anchoring to the its target (see Figure \ref{hyperinflation_phase}, left panel). The same holds true with Floating Expectation (middle panel). However, in this case an interesting range of parameters where inflation remains stable for $g^\star > g_w = g_p > 1$, until a strong enough shock occurs, which triggers a loss of trust in the Central Bank, tipping the economy into a hyperinflation phase (as illustrated in Figure \ref{hyperinflation_phase}, right panel). This is precisely the scenario that Central Bankers want to avoid by doing ``whatever it takes''.

If bargaining and market power differ ($g_p \neq g_w$), the effect on unemployment is significant (see Appendix \ref{appx:gpgw}, Figure \ref{sensitivity_gp_+_gw_const}). In the Anchored Expectation case, when market power is greater than bargaining power -- a.k.a. ``greedflation'' -- ($g_p > g_w$), unemployment increases after the shock and only slowly decreases as wages adjust at a slower pace to pre-crisis levels, leading to sluggish demand recovery. Consequently, firms do not see the need to increase their production faster. In contrast, if bargaining power exceeds market power ($g_w > g_p$), wages recover more quickly, resulting in an earlier recovery of pre-crisis purchasing power, which helps to reduce unemployment. This conclusion is in line with the result of section \ref{sec:helicopter} on the impact of Helicopter Money or, more generally, of Keynesian stimulus: increasing the consumption budget of households is quite efficient at reviving an ailing economy, but only if hyperinflation can be avoided. 

\section{Summary \& Conclusions}\label{sec:conclusion}

In this paper, we have expanded the Mark-0 Agent-based Model to assess the inflationary dynamics following the COVID pandemic and energy crisis of 2020-2023. Our results highlight the narrow path of monetary policy to balance unemployment and inflation, as well as the benefits of joint monetary and fiscal policy packages.

We extended the Mark-0 model used in~\citet{SharmaEtAl2020WshapedRecoveryCOVIDa} in two directions (Section~\ref{sec:model}):
First, a dynamically evolving trust in the Central Bank's ability to control inflation, such that expected inflation remains anchored to the Central Bank's target if trust is high, but persistent off-target inflation realisations lead to a loss of trust in the Central Bank. Second, a simple ``exogenous'' energy sector from which firms buy energy that allows us to introduce an energy price, and slowly re-inject the energy profits into the economy as dividend payouts. This slightly expanded Mark-0 model allows us to investigate several policy channels (Section~\ref{sec:mechanisms}): (a) monetary policy via the management of interest rates and expectations, (b) an Easy-Credit regulatory policy, and fiscal policies such as (c) Helicopter Money and (d) a Windfall Tax. 

The Mark-0 steady state economy is then perturbed by three calibrated shocks (Section~\ref{sec:shocks}): a consumption propensity drop (representing lockdowns), a productivity drop (representing supply chains disruptions), and an energy price shock (first a drop, representing reduced demand during lockdowns, followed by a rise due do demand recovery exacerbated by the Russian invasion of Ukraine). We show in Section~\ref{sec:shocks} that the Easy-Credit policy, which was one of the first emergency responses to the COVID pandemic, is able to alleviate the impact of the shocks, but at the price of high and sustained inflation, as observed in the data and predicted as early as June 2020 by \citet{SharmaEtAl2020WshapedRecoveryCOVIDa} within the Mark-0 framework.

We investigated whether monetary policy alone can control a surge of inflation, and at what cost to unemployment (Section~\ref{sec:6}). 
Our results show that if agents' expectations
are fully de-anchored, i.e. expected inflation coincides with past realized inflation and not to the Central Bank target, monetary policy has very little impact on inflation. If on the contrary expectations
remain fully anchored, inflation remains closer to the Central Bank target, but at the price of a strong recession leading to a wave of unemployment.
Meanwhile, if agents' expectations evolve dynamically, inflation rises well above target during the shock leading to de-Anchored Expectations. Monetary policy is then again essentially ineffective in controlling inflation while causing unemployment to increase in comparison to a baseline case with no monetary policy at all. 

This conclusion is robust to variations in parameters related to the channels of monetary policy. Our framework thus supports the view that the efficacy of monetary policy comes at the cost of unnecessarily high unemployment~\citep{StiglitzRegmi2022CausesResponsesToday}, {in particular, quite paradoxically, if inflation expectations are strongly anchored (see the discussion in Section~\ref{sec:6}).}

Within Mark-0, trust anchoring is thus a crucial determinant of the success of the Central Bank inflation mitigation policy, far more important than the direct economic impact of higher interest rates (noting that we do not model in detail a financial sector). This resonates with Bernanke's statement: ``Expectations matter so much that a Central Bank may be able to help make policy more effective by working to shape those expectations”,\footnote{see Bernanke, B. (2013), ``Communication and Monetary Policy”, speech at “National Economists Club Annual Dinner”. (https://www.federalreserve.gov/newsevents/speech/bernanke20131119a.pdf)} and points to the importance of {\it narratives} in shaping expectations \citep{shiller2020narrative}.  In turn, fine-tuned fiscal policy combined with monetary policy can be successful in controlling inflation while keeping unemployment around acceptable levels (Section~\ref{sec:7}). However, this requires a high degree of precision to be effective: too weak a fiscal stimulus is ineffective, and too large a stimulus leads to further high inflation. 
Finally, in Section~\ref{sec:sloppy} we show that the pricing power of firms and the bargaining power of workers play a crucial role. Depending on their relation, the economy can experience a runaway hyperinflation wage-price spiral. This points to an additional difficulty in properly calibrating monetary and fiscal policy due to a variety of possible tipping points or ``dark corners'' lurking around.

Overall, this study of the Mark-0 economy shows that
(i)~the economic recovery can be very sluggish due to self-fulfilling expectations and other non-linear feedback loops, 
(ii)~there is a tension between inflation and unemployment that is robustly captured by the model, 
(iii)~in the Mark-0 economy, monetary policy is effective at controlling inflation provided trust is anchored (but not because of real economic effects of higher interest rates), and (iv)~fiscal policy can alleviate some of the negative unemployment effects of inflation-focused monetary policy.
While our study did not exhaust all possible cases and parameters, we believe that it illustrates both the realism and the richness of the Mark-0 economy. It offers a versatile tool with which policy makers can easily play in order to forge their intuition about what may happen if they turn this knob or add that policy measure. Indeed, almost all possible narratives that emerged during the recent debate about post-COVID inflation can be captured and reproduced by our model with proper choices of parameters and shock specifications. Furthermore, the procedure proposed in Section~\ref{sec:sloppy} can be used to rigorously establish which predictions of the model are robust to the choice of parameters, most of which are difficult to properly calibrate on data.

Our model is obviously incomplete and improvable on many counts. 
In particular, it could be extended to include (i) a dual mandate central bank as in \citet{BouchaudEtAl2018OptimalInflationTargeta,GualdiEtAl2017MonetaryPolicyDarka} to address negative unemployment side-effects, (ii) a disaggregated household sector (e.g. distinguishing wage and rent earners, and by accumulated wealth) to assess distributional implications of policy, (iii) production networks and energy supply chains with differentiated products and commodities \citep[as in][for instance]{dessertaine2022out}, which introduces firm distribution effects and systemic risk, (iv) a financial sector to more closely model interest rate pass-through, lending choices and financial systemic risk, and (v) more detailed labour market structures to account for the change in occupational structure post-COVID and demographic trends.
However, we do believe, as we already argued in \cite{GualdiEtAl2015TippingPointsMacroeconomica}, that the Mark-0 model should already be part of the toolkit of Central Banks, if only as an inspiring {\it scenario generator}, or ``telescope for the mind'', especially in times of great modelling uncertainty during which it is crucial to be at least ``roughly right'' and avoid being blindsided by spurious Black Swans.\footnote{See for example Mark Buchanan, ``This Economy Does Not Compute'', New York Times, October 2008, \cite{GualdiEtAl2015TippingPointsMacroeconomica,king2020radical,bouchaud2021radical}, or the recent Financial Times piece cited in the introduction, https://www.ft.com/content/b972f5e3-4f03-4986-890d-5443878424ac} 

\section*{Declarations}
\textbf{Competing Interests:} \; The authors have no competing interests to declare that are relevant to the content of this article.
\newline
\textbf{Financial Interest:} \; The authors have no financial or non-financial interests to disclose.

\newpage
\bibliography{Mark0_Inflation_References.bib}

\newpage
\appendix
\section{Appendices} 
\subsection{Table of Parameters}
\begin{table}[H]
\centering
\caption{Inactive Central Bank parameter set for Mark-0 Model (see Section \ref{sec:baselinedescription}) . Parameters, which change for the Reactive Central Bank and Anchored/Floating Expectation scenarios, can be taken from sections \ref{sec:interest_only} and \ref{sec:floatingtrust} respectively and are indicated in the table with a star. All rates are given in monthly time scales.}
\begin{tabular}{clc}
\hline
 &Parameter description & Value   \\
\hline
 $R_0$ & Ratio of hiring-firing rate ($\eta^+/\eta^-$): & 2.0   \\
 $\Theta$& Maximum credit supply available to firms : &  3.2     \\
 $\Gamma_0$& Financial Fragility sensitivity:  &  0.0    \\
 $\rho^\star$& Baseline interest rate: &    0.001*   \\
 $\alpha_c$& Influence of deposit rates on consumption &  12     \\
 $\phi_\pi$& Intensity of interest rate policy of Central Bank: &  0.0*     \\
 $\pi^*$& Central Bank inflation target: &  0.0*     \\
 $\tau^T$& Inflation target trust parameter: &  0.0*     \\
 $g_w$& Factor to adjust wages to inflation expectations: &  0.8     \\
 $g_p$& Factor to adjust prices to inflation expectations: &  0.8     \\
 $y_0$& Initial production: &  0.7     \\
 $\gamma$& Parameter to set adjustment of prices: &  0.01     \\
  $\eta_{0}^-$& Firing propensity: & 0.2  \\
 $\alpha_\Gamma$& Influence of loans interest rate on hiring-firing policy: &  450     \\
  $c_0$& Fraction of savings in consumption budget: & 0.5      \\
  $\beta$& Price sensitivity parameter: & 2      \\
  $\phi$& Revival probability per unit time: & 0.1      \\
  $\omega$& Moving average parameter: & 0.2      \\
 $\delta$& Dividend  rate: & 0.02  
 \\
 $\delta_e$& Fraction of energy sector's equity redistributed: & 0.04   \\
 $g_e$& Share of Energy Price share in GDP: & 0.0325      \\
 $f$ & Interpolation parameter of cost of default: & 0.5 \\
 $\mu$& Easy-Credit policy multiplier: & 1.3      \\
  $N_{F}$& Total number of firms: & 10000      \\
 \hline
\end{tabular}
\label{tablenocb}
\end{table}

\subsection{Counterfactuals} \label{appx:counterfactual}

\begin{figure}[H]
    \centering
    \includegraphics[width=\textwidth]{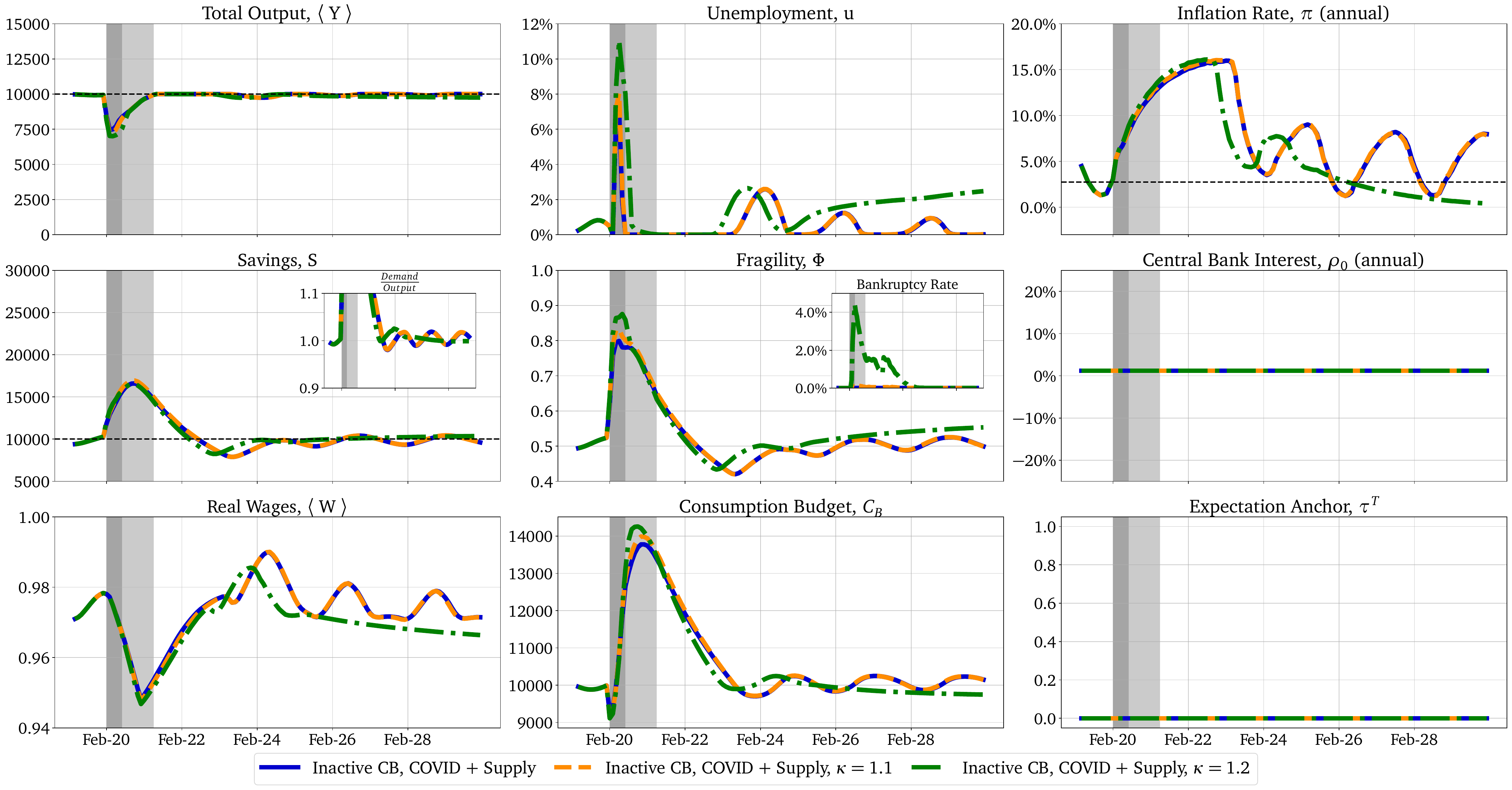}
    \caption{\textbf{Economic dashboard for the Inactive Central Bank scenario without any policy and with a stronger COVID shock:} The dynamics for a COVID shock strength of $\kappa=1.0$ (blue), $\kappa=1.1$ (orange) and  $\kappa=1.2$ (green) in the Inactive Central Bank scenario without any policy. Once the COVID shock gets too strong, the economy collapses in the long run.}
    \label{cf21_covid}
\end{figure}

\subsection{Monetary Policy}\label{appx:dashboards}

\begin{figure}[H]
    \centering
    \includegraphics[width=\textwidth]{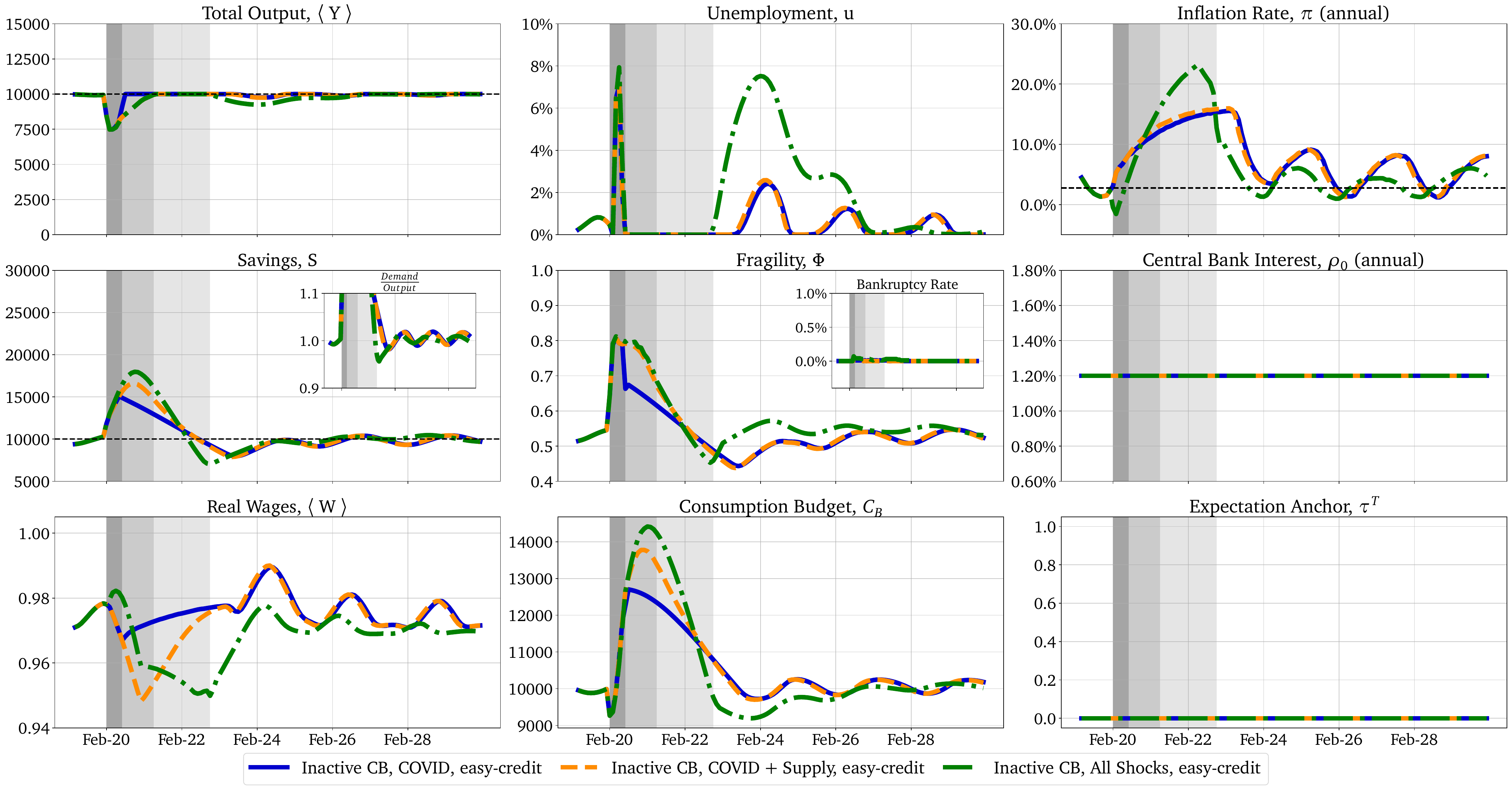}
    \caption{\textbf{Dynamics for the three shocks for the Inactive Central Bank scenario:} The dynamics for the three shocks, COVID only (blue), COVID and Supply Chain shock (orange) and all shocks (green) for the Inactive Central Bank scenario with Easy-Credit policy. The areas shaded in grey indicate the duration of the three shocks: the COVID shock lasting until the end of the dark grey area, the supply chain shock until the end of the grey area, and the price shock until the end of the light grey area.}
    \label{baseline_ap}
\end{figure}

\begin{figure}[H]
    \centering
    \includegraphics[width=\textwidth]{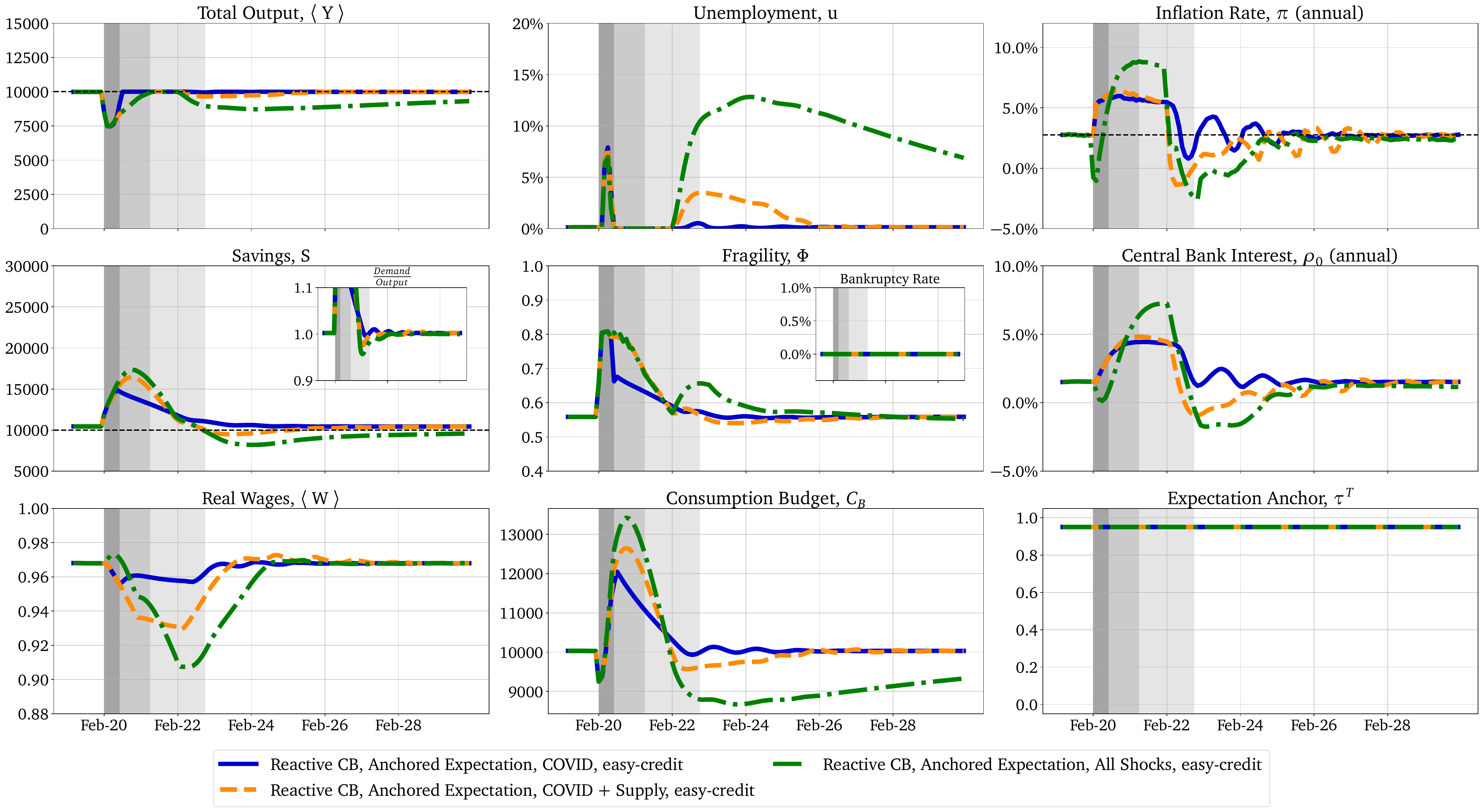}
    \caption{\textbf{Dynamics for the three shocks for the Reactive Central Bank with Anchored Expectation scenario:} The dynamics for the three shocks, COVID only (blue), COVID and Supply Chain shock (orange) and all shocks (green) to the scenario with reactive Central Bank and Anchored Expectation of economic agents with an Easy-Credit policy. The areas shaded in grey indicate the duration of the three shocks: the COVID shock lasting until the end of the dark grey area, the supply chain shock until the end of the grey area, and the price shock until the end of the light grey area.}
    \label{activeCB_ap}
\end{figure}

\begin{figure}[H]
    \centering
    \includegraphics[width=\textwidth]{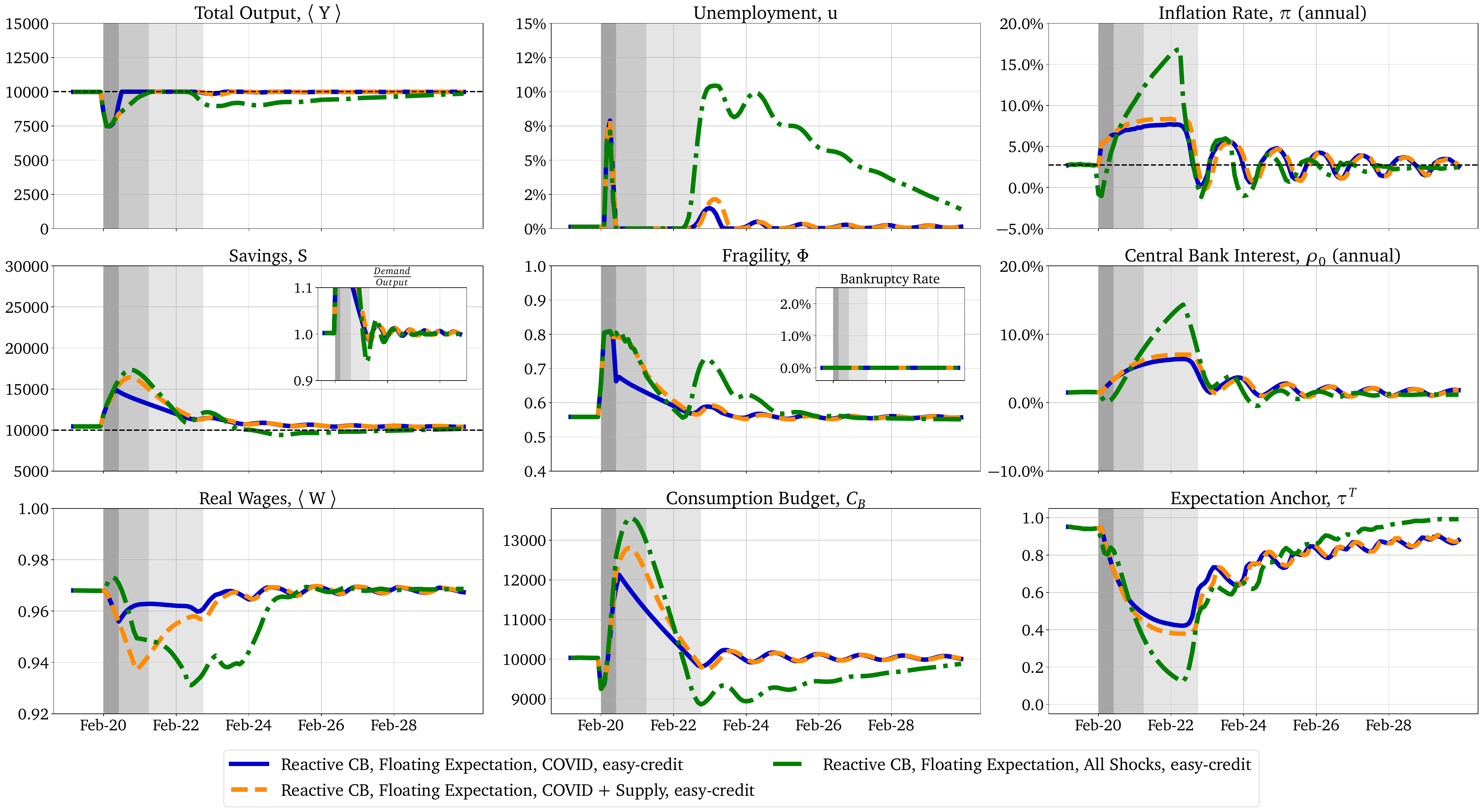}
    \caption{\textbf{Dynamics for the three shocks for the Reactive Central Bank with Floating Expectation scenario:} The dynamics for the three shocks, COVID only (blue), COVID and Supply Chain shock (orange) and all shocks (green) to the scenario with reactive Central Bank and Floating Expectation of economic agents with an Easy-Credit policy. The areas shaded in grey indicate the duration of the three shocks: the COVID shock lasting until the end of the dark grey area, the supply chain shock until the end of the grey area, and the price shock until the end of the light grey area.}
    \label{activeCB_dyn_ap}
\end{figure}

\begin{figure}[H]
    \centering
    \includegraphics[width=\textwidth]{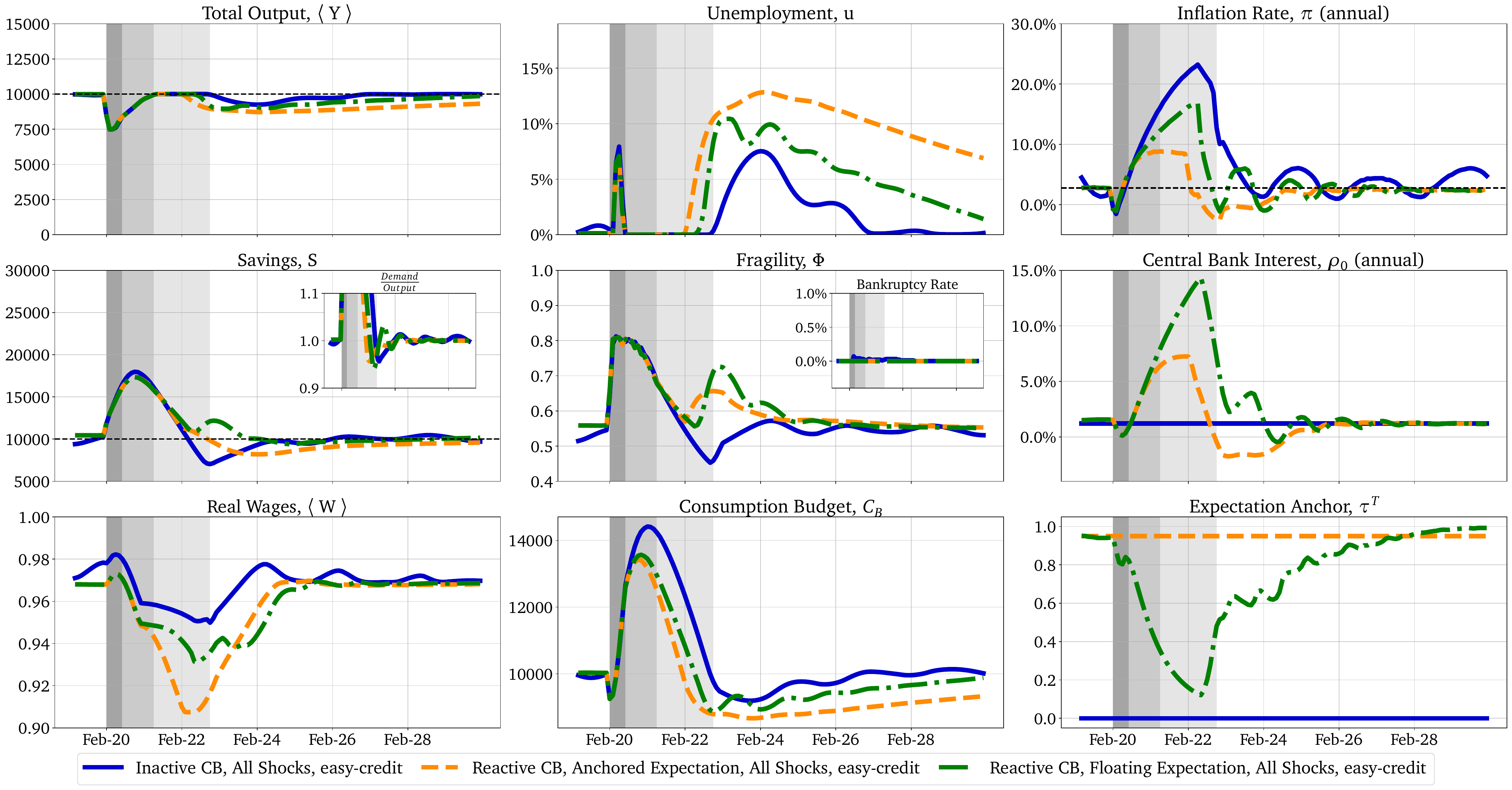}
    \caption{\textbf{Dynamics for all scenarios and all shocks:} The full dashboard for Figure \ref{fig:activeCB_ap_short} All dynamics are with Easy-Credit policy in the Inactive Central Bank scenario (blue), Reactive Central Bank with Anchored Expectation scenario (orange) and Reactive Central Bank with Floating Expectation scenario (green). The areas shaded in grey indicate the duration of the three shocks: the COVID shock lasting until the end of the dark grey area, the supply chain shock until the end of the grey area, and the price shock until the end of the light grey area.}
    \label{all_scenarios}
\end{figure}

\subsection{Importance of Expectation Anchoring}
\label{appsec:anchoring}

\begin{figure}[H]
    \centering
    \includegraphics[width=\textwidth]{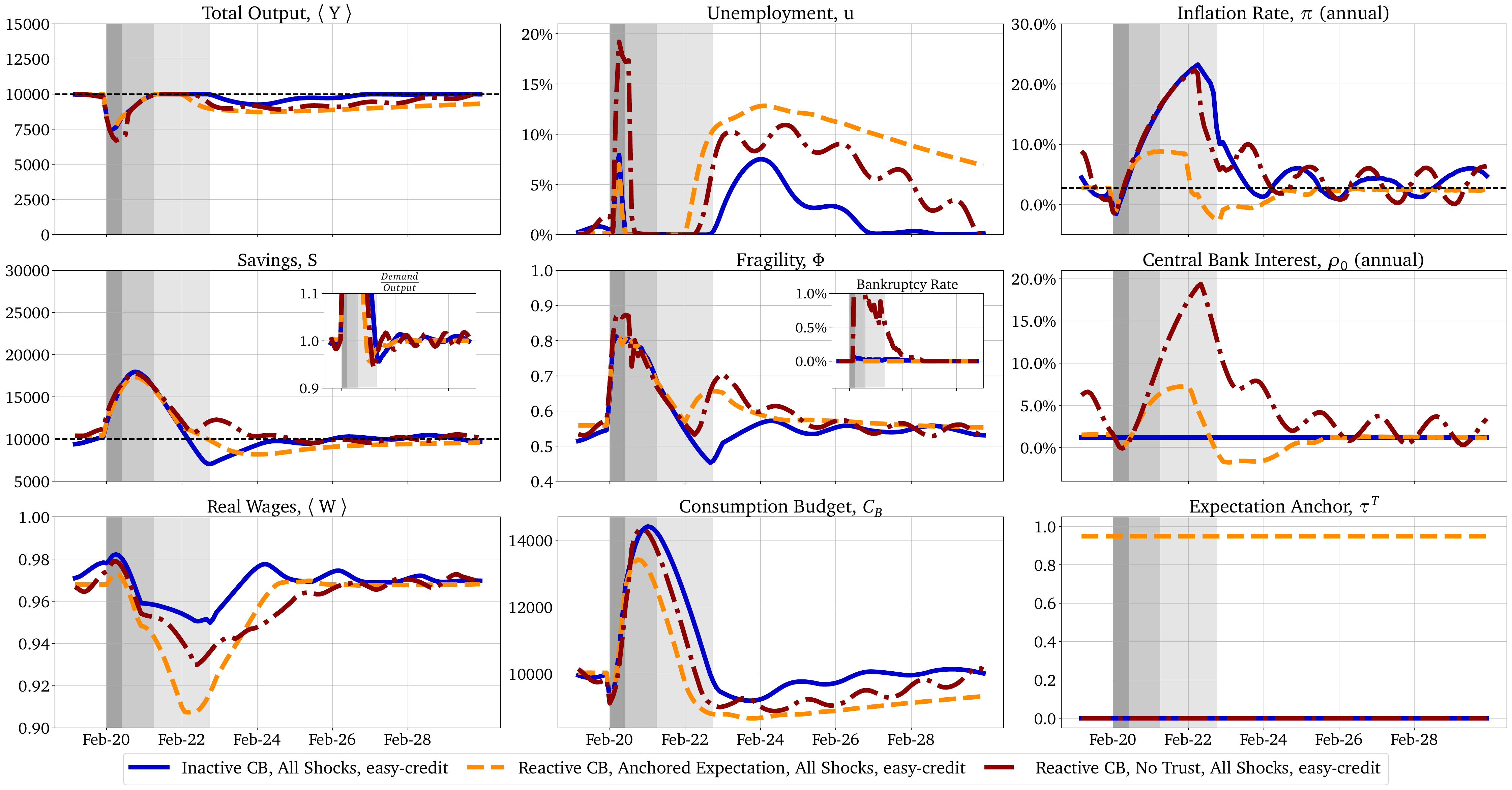}
    \caption{\textbf{Dynamics with de-Anchored Expectation} ($\tau^T=0$): All dynamics are with Easy-Credit policy and all shocks, including the Inactive Central Bank scenario (blue), Reactive Central Bank with Anchored Expectation scenario (orange dashed) and Reactive Central Bank with no trust scenario $\tau^T=0$ (red dash-dotted). The areas shaded in grey indicate the duration of the three shocks: the COVID shock lasting until the end of the dark grey area, the supply chain shock until the end of the grey area, and the price shock until the end of the light grey area. Note that without expectation anchoring, the inflation path is very similar with (red dashed-dotted) and without (blue) monetary policy.}
    \label{app:anchoring}
\end{figure}

\subsection{Monetary Policy with Stronger Central Bank Reaction}\label{appx:strongerCB}
\begin{figure}[H]
    \centering
    \includegraphics[width=\textwidth]{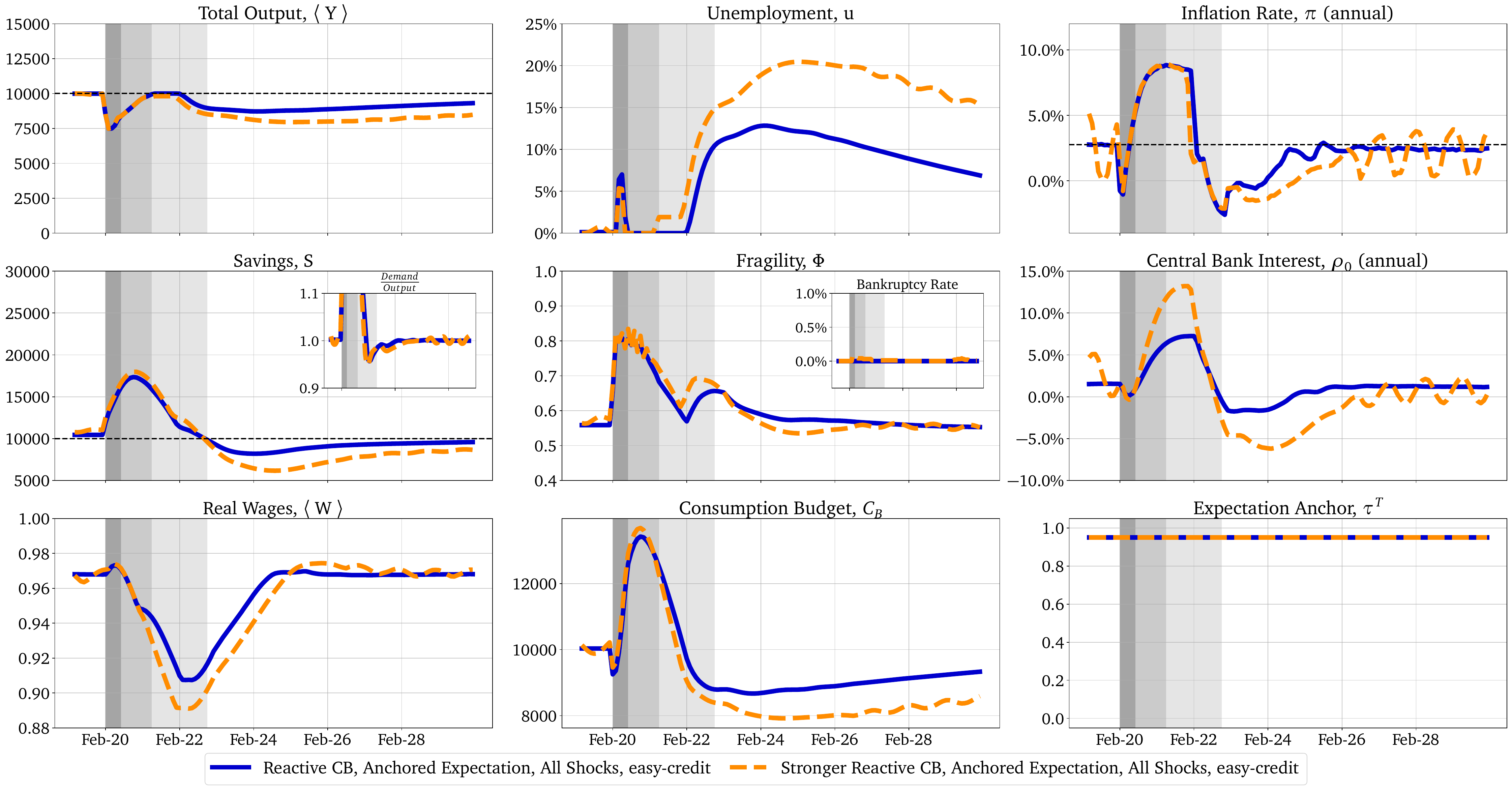}
    \caption{\textbf{Economic dashboard for the stronger Reactive Central Bank with Anchored Expectation scenario and all shocks.} The dynamics for a Central Bank strength of $\phi_\pi=1.0$ (blue) and $\phi_\pi=2.0$ (orange). A stronger central bank is not able to decrease inflation due to an external price shock, but with stronger policies, the consumption of households reduces, which leads to an increase in unemployment.}
    \label{activeCB_stronger_ap}
\end{figure}

\begin{figure}[H]
    \centering
    \includegraphics[width=\textwidth]{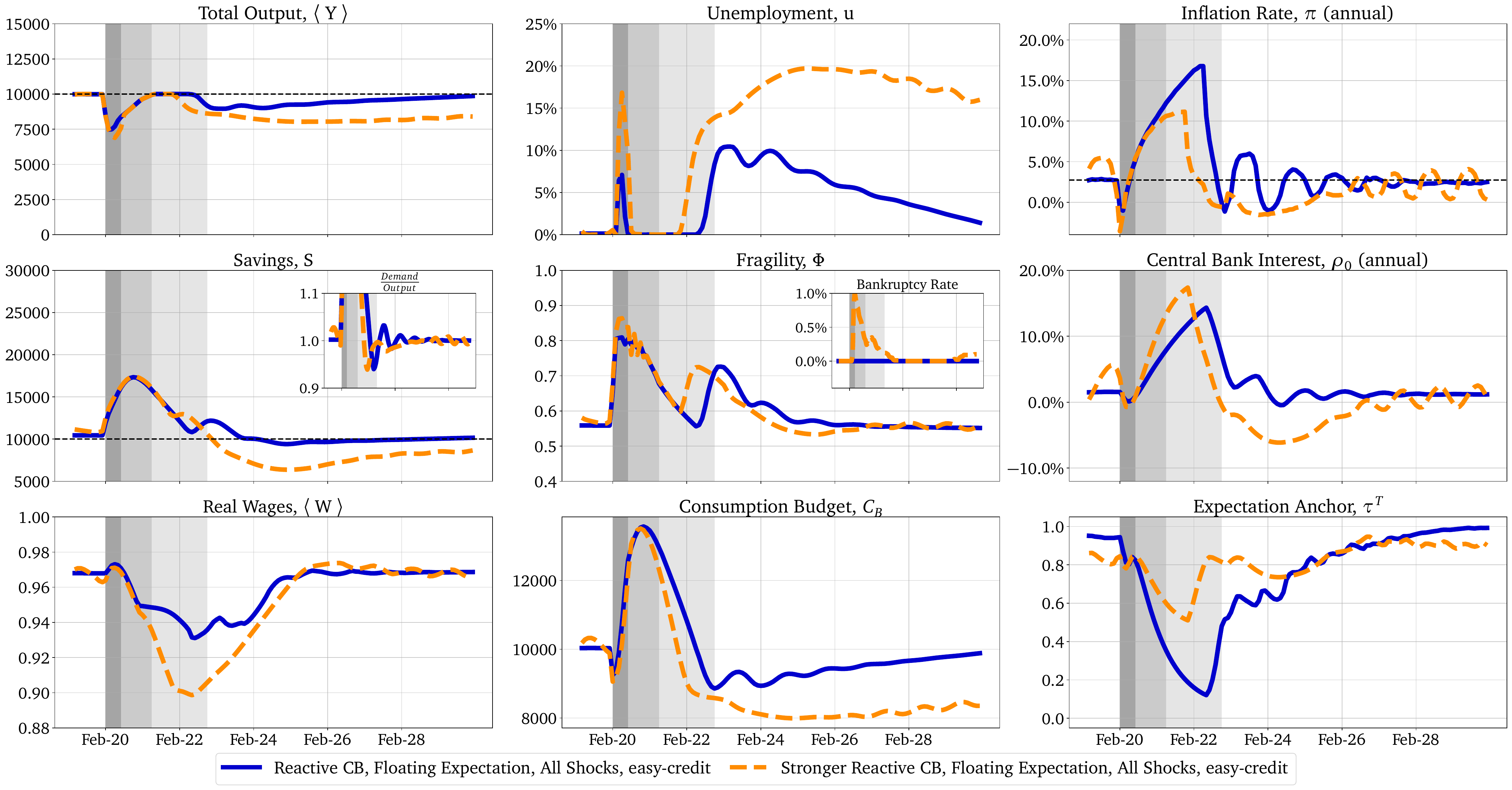}
    \caption{\textbf{Economic dashboard for the stronger Reactive Central Bank with Floating Expectation scenario and all shocks.} The dynamics for a Central Bank strength of $\phi_\pi=1.0$ (blue) and $\phi_\pi=2.0$ (orange) with Floating Expectation of economic agents. A stronger Central bank can reduce excess inflation at the cost of increased unemployment. The central bank must maintain a balance between falling inflation and rising unemployment.}
    \label{activeCB_dyn_stronger_ap}
\end{figure}

\subsection{Sensitivity of Monetary Policy to \texorpdfstring{$\alpha_c$}{Lg} and \texorpdfstring{$\alpha_\Gamma$}{Lg}}\label{appx:alphacgamma}
\begin{figure}[H]
    \centering
    \includegraphics[width=\textwidth]{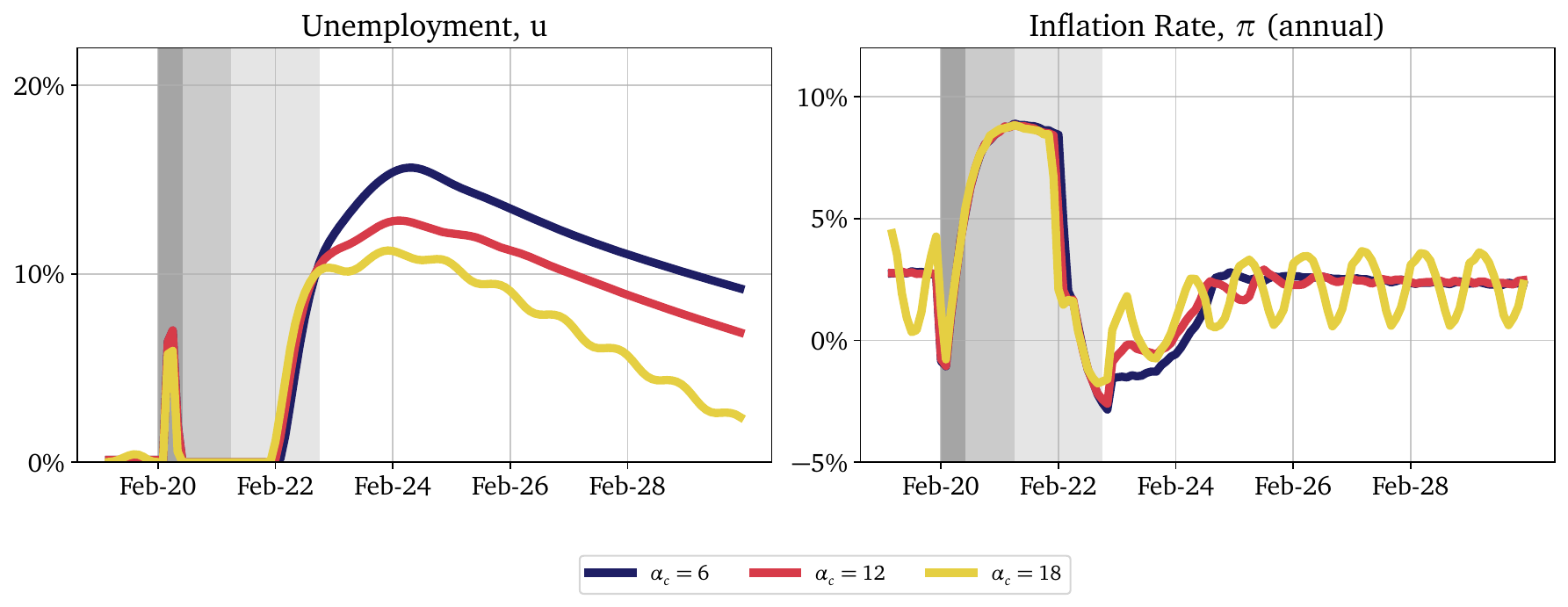}
    \caption{\textbf{Sensitivity of $\alpha_c$ for the Reactive Central Bank with Anchored Expectation scenario and all shocks.} As already described in \cite{GualdiEtAl2017MonetaryPolicyDarka}, larger $\alpha_c$ lead to a greater magnification of price trends (cet. par.). This amplification, in turn, causes the fluctuations in unemployment and inflation. Furthermore, larger $\alpha_c$ increases consumption which reduces unemployment due to higher demand.
    }
    \label{sensitivity_alpha_c_const}
\end{figure}
\begin{figure}[H]
    \centering
    \includegraphics[width=\textwidth]{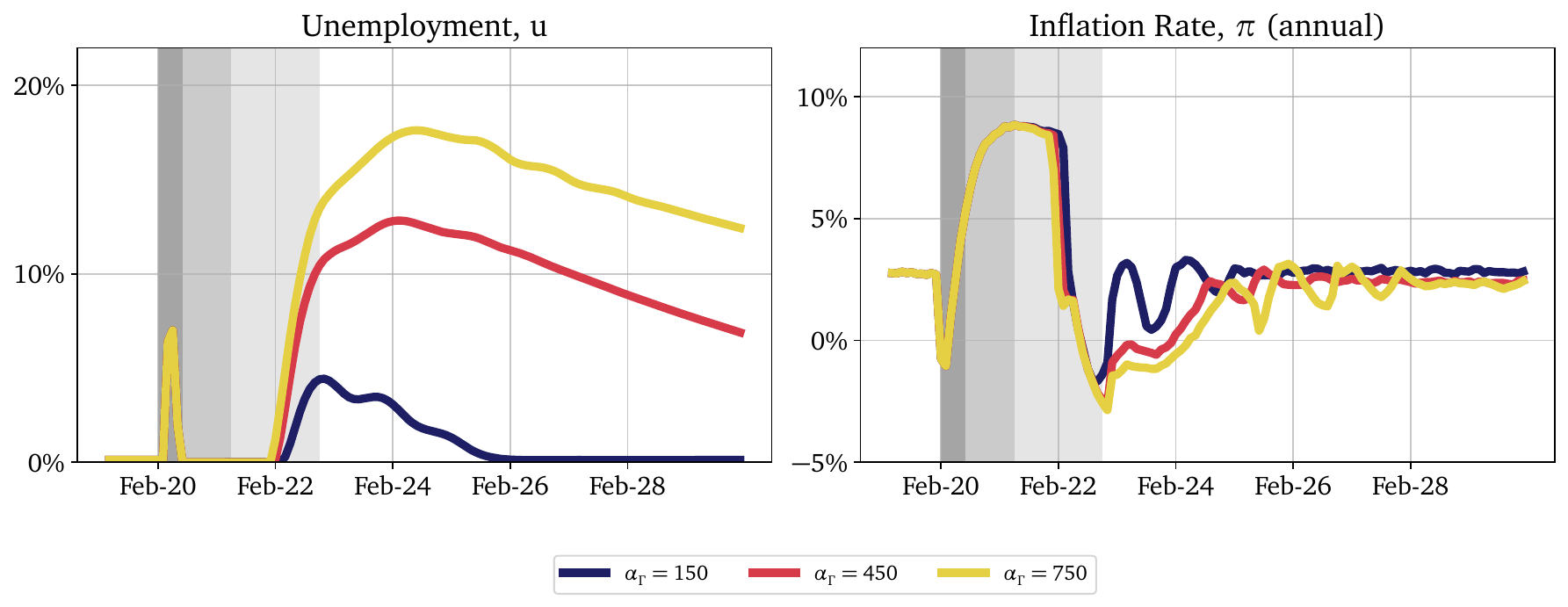}
    \caption{\textbf{Sensitivity of $\alpha_\Gamma$ for the Reactive Central Bank with Anchored Expectation scenario and all shocks.} Larger values of $\alpha_\Gamma$ increases the influence of financial fragility on the hiring/firing policy of firms (cet. par.). This in turn leads to a larger downward adjustment of the workforce which increases unemployment.}
    \label{sensitivity_alpha_gamma_const}
\end{figure}

\subsection{Helicopter Money}\label{appx:helicopter}

\begin{figure}[H]
    \centering
    \includegraphics[width=\textwidth]{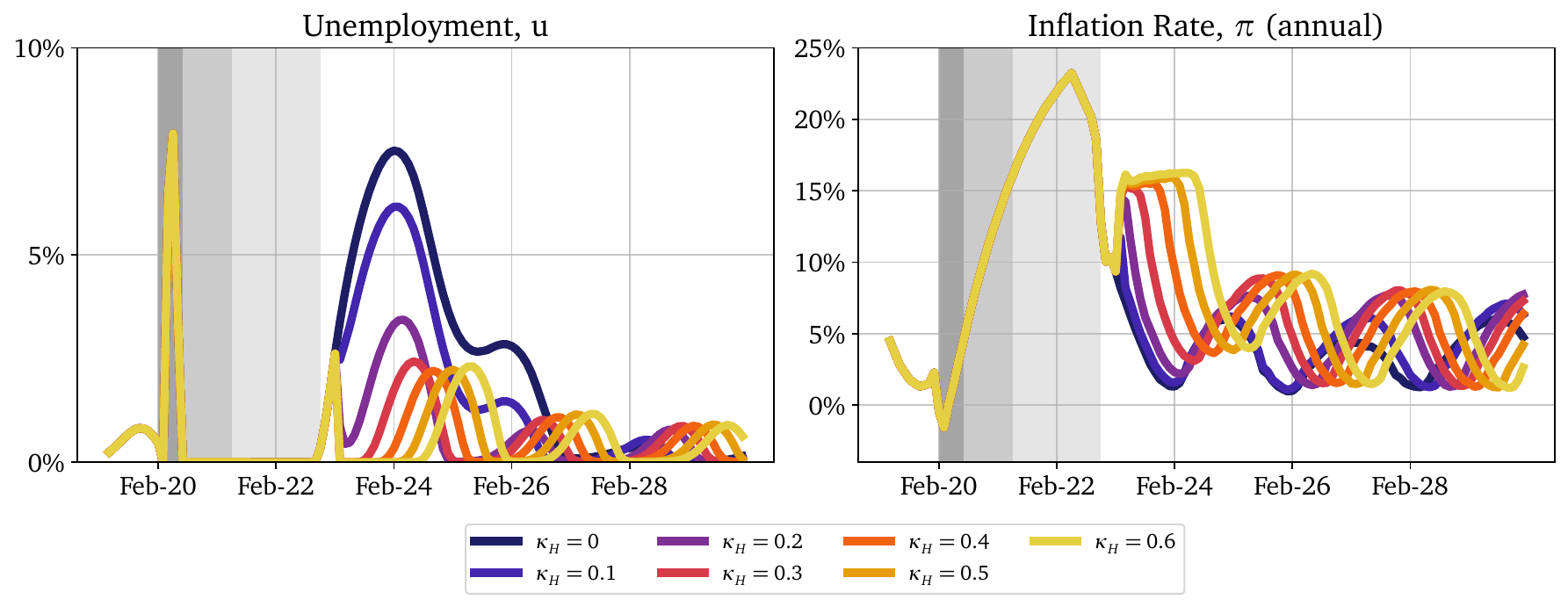}
    \caption{\textbf{Helicopter Money in the Inactive Central Bank scenario and all shocks.} Unemployment (left) and inflation (right) for a helicopter drop of size $\kappa_H S$ one month after the price shock. Already with $\kappa_H\ge 0.2$ unemployment is reduced almost to zero. A further increase of Helicopter Money only increases inflation without reducing unemployment more.}
    \label{baseline_ap_helicopter_spectrum}
\end{figure}

\begin{figure}[H]
    \centering
    \includegraphics[width=\textwidth]{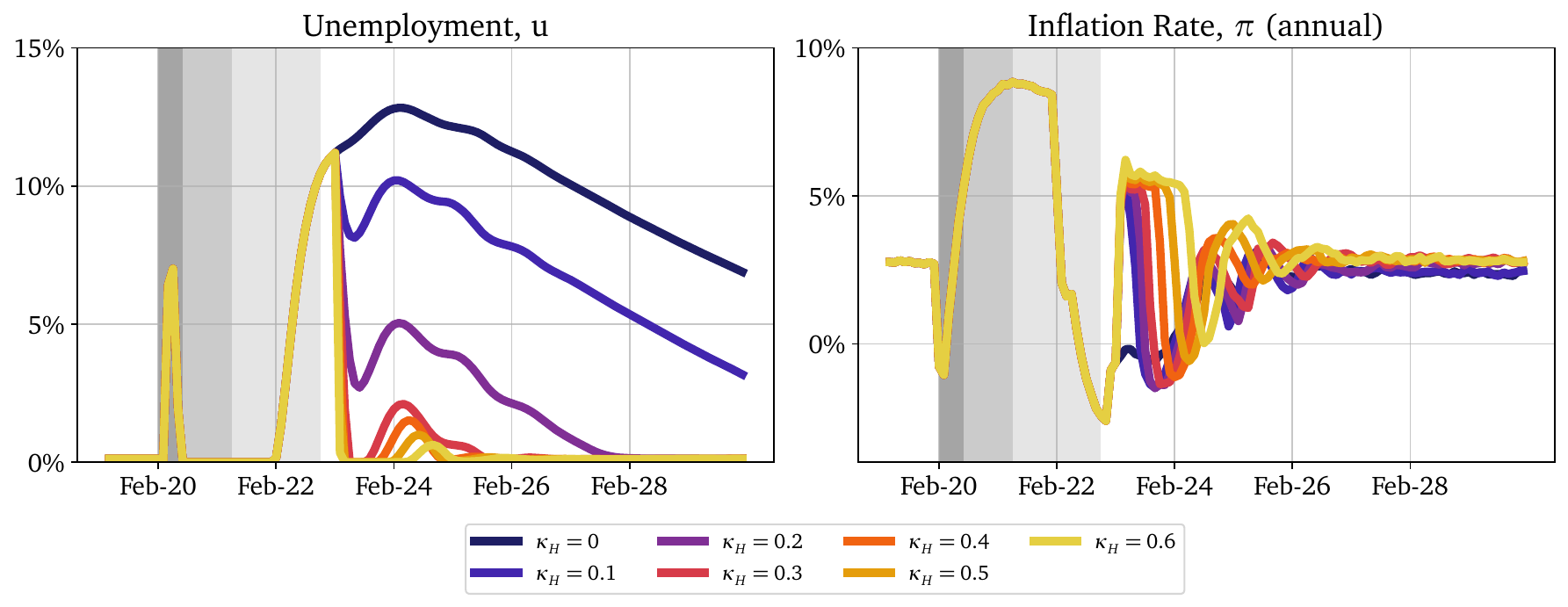}
    \caption{\textbf{Helicopter Money in the Reactive Central Bank with Anchored Expectation scenario.} Unemployment (left) and inflation (right) for a helicopter drop of size $\kappa_H S$ one month after the price shock. With $\kappa_H\ge 0.3$ unemployment is reduced almost to zero. A further increase of Helicopter Money only increases inflation without reducing unemployment further. However, the Central Bank policy manages to keep inflation under control.}
    \label{activeCB_ap_helicopter_spectrum}
\end{figure}

\begin{figure}[H]
    \centering
    \includegraphics[width=\textwidth]{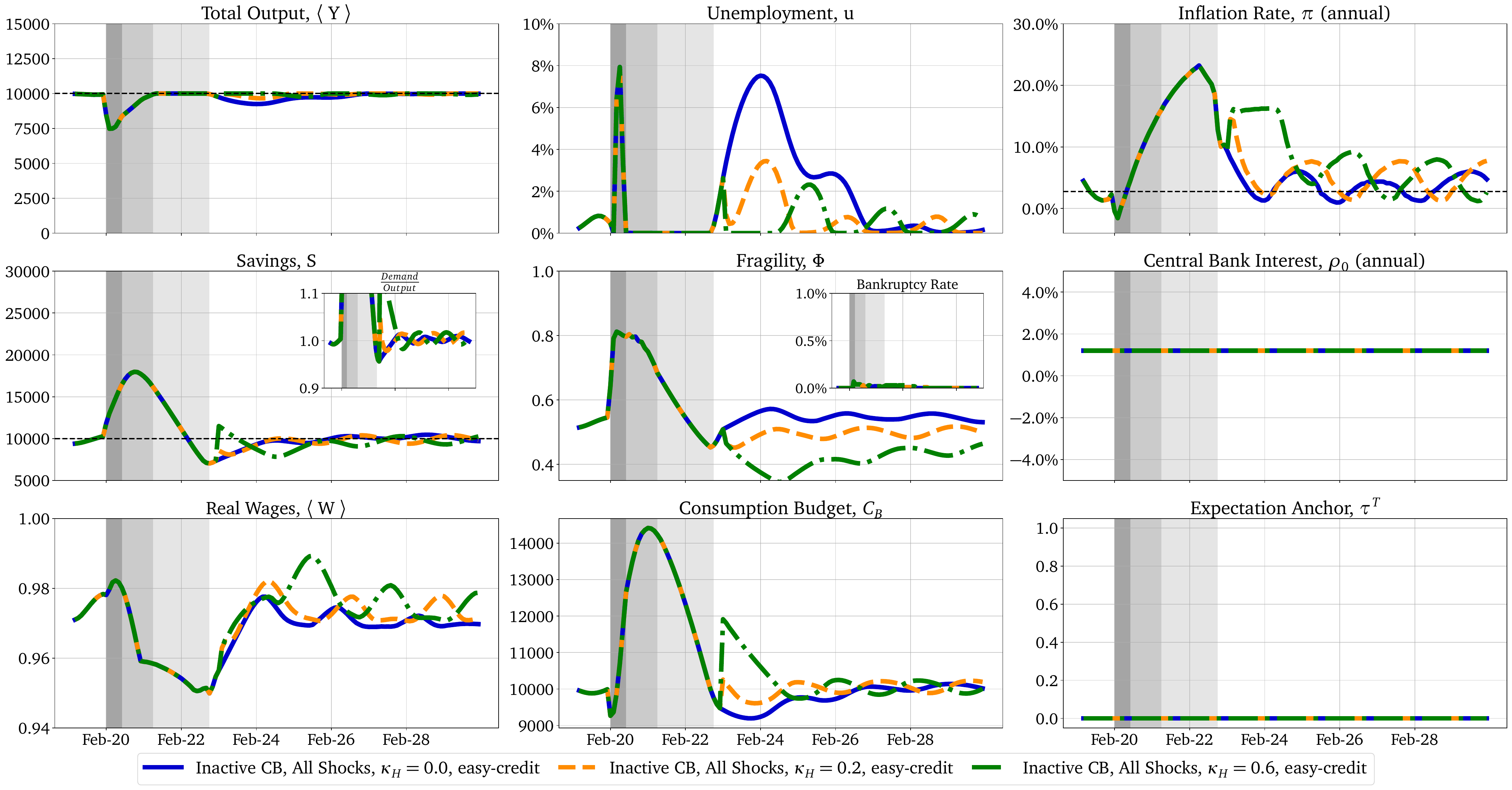}
    \caption{\textbf{Economic dashboard for Helicopter Money in the Inactive Central Bank scenario and all shocks.} Full dynamics for scenario without Helicopter Money (blue), for Helicopter Money with $\kappa_H=0.2$ (orange) and $\kappa_H=0.6$ (green) one month after the price shock and with Easy-Credit policy.}
    \label{baseline_ap_helicopter}
\end{figure}

\begin{figure}[H]
    \centering
    \includegraphics[width=\textwidth]{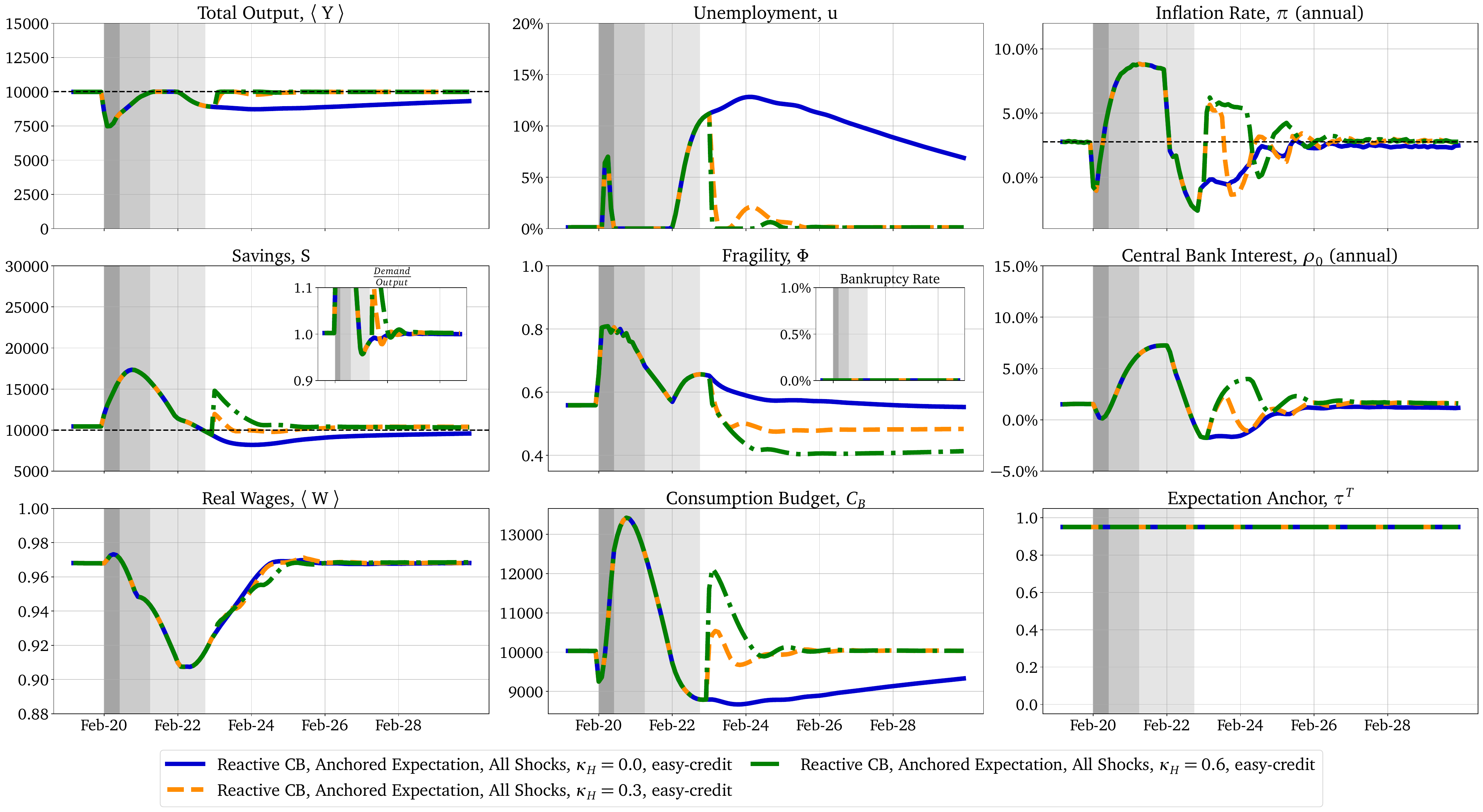}
    \caption{\textbf{Economic dashboard for Helicopter Money in the Reactive Central Bank with Anchored Expectation scenario and all shocks.} Full dynamics for scenario without Helicopter Money (blue), for Helicopter Money with $\kappa_H=0.3$ (orange) and $\kappa_H=0.6$ (green) one month after the price shock and with Easy-Credit policy.}
    \label{activeCB_ap_helicopter}
\end{figure}

\begin{figure}[H]
    \centering
    \includegraphics[width=\textwidth]{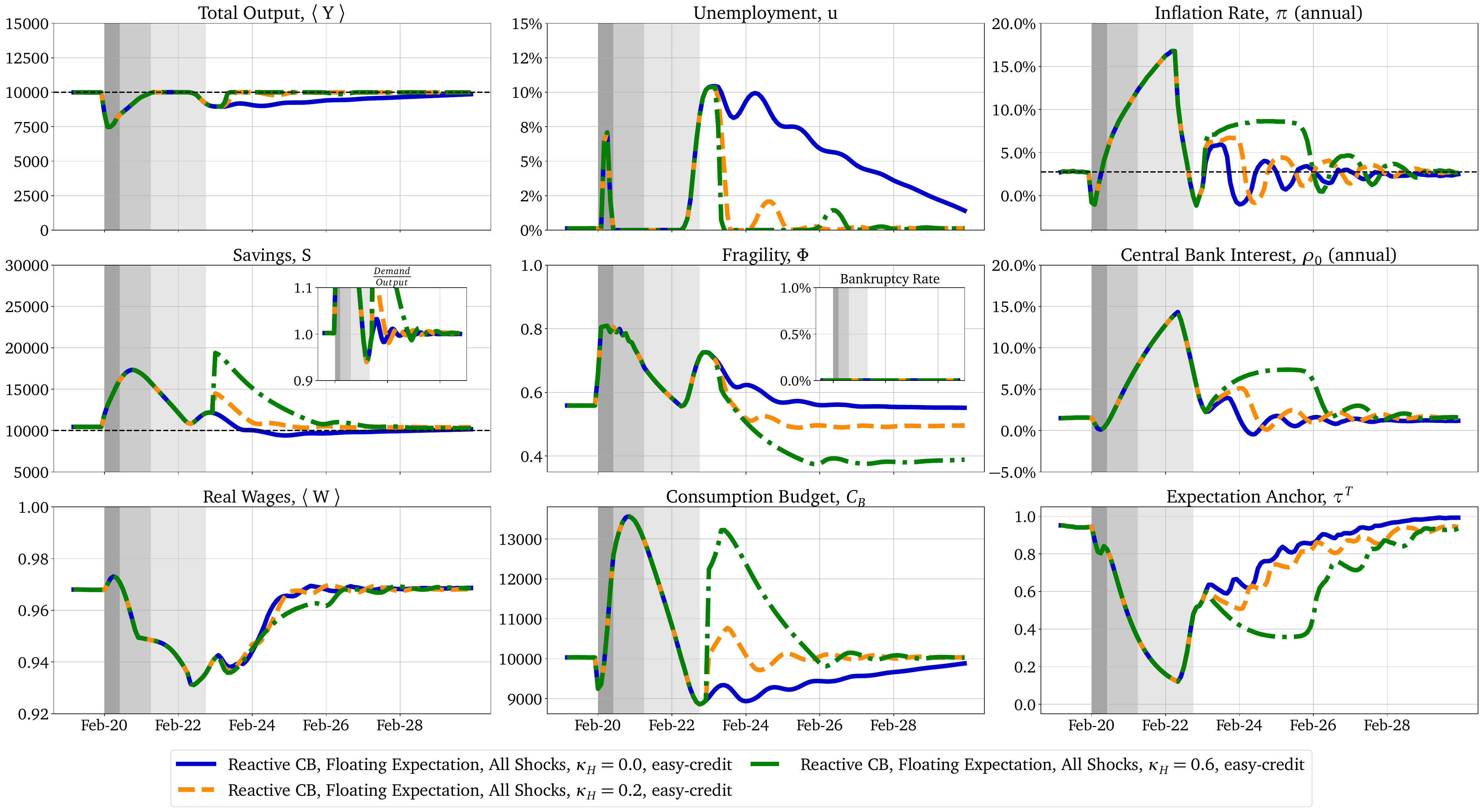}
    \caption{\textbf{Economic dashboard for Helicopter Money in the Reactive Central Bank with Floating Expectation and all shocks.} Full dynamics for scenario without Helicopter Money (blue), for Helicopter Money with $\kappa_H=0.2$ (orange) and $\kappa_H=0.6$ (green) one month after the price shock and with Easy-Credit policy.}
    \label{activeCB_dyn_ap_helicopter}
\end{figure}

\subsection{Windfall Tax}\label{appx:windfall}

\begin{figure}[H]
    \centering
    \includegraphics[width=\textwidth]{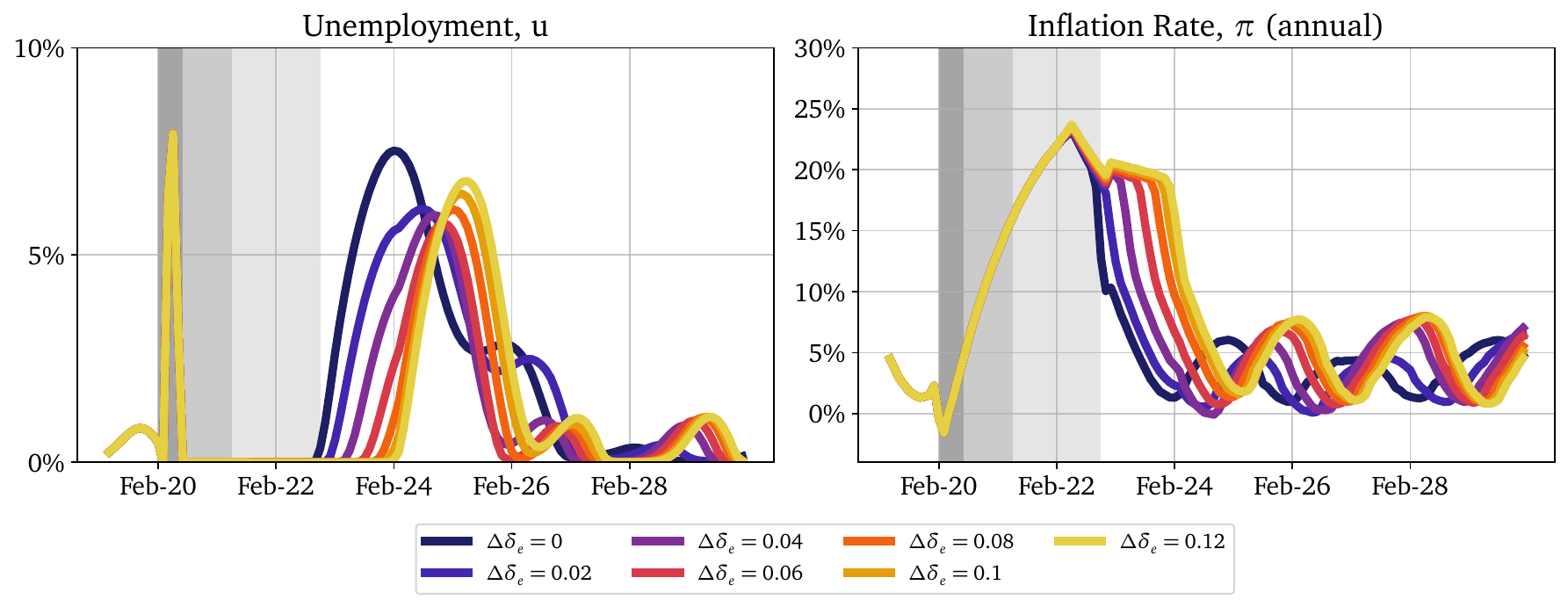}
    \caption{\textbf{Windfall Tax in the Inactive Central Bank scenario.} Unemployment (left) and inflation (right) for Windfall Tax of $\delta_e + \Delta \delta_e$ one year before the end of the price shock with a duration of two years. With $\Delta \delta_e \approx 4\%$ unemployment is reduced strongly. A further increase of tax does only increase unemployment again. For larger dividends, inflation increases because increased demand due to increased savings.}
    \label{baseline_ap_windfall_spectrum}
\end{figure}

\begin{figure}[H]
    \centering
    \includegraphics[width=\textwidth]{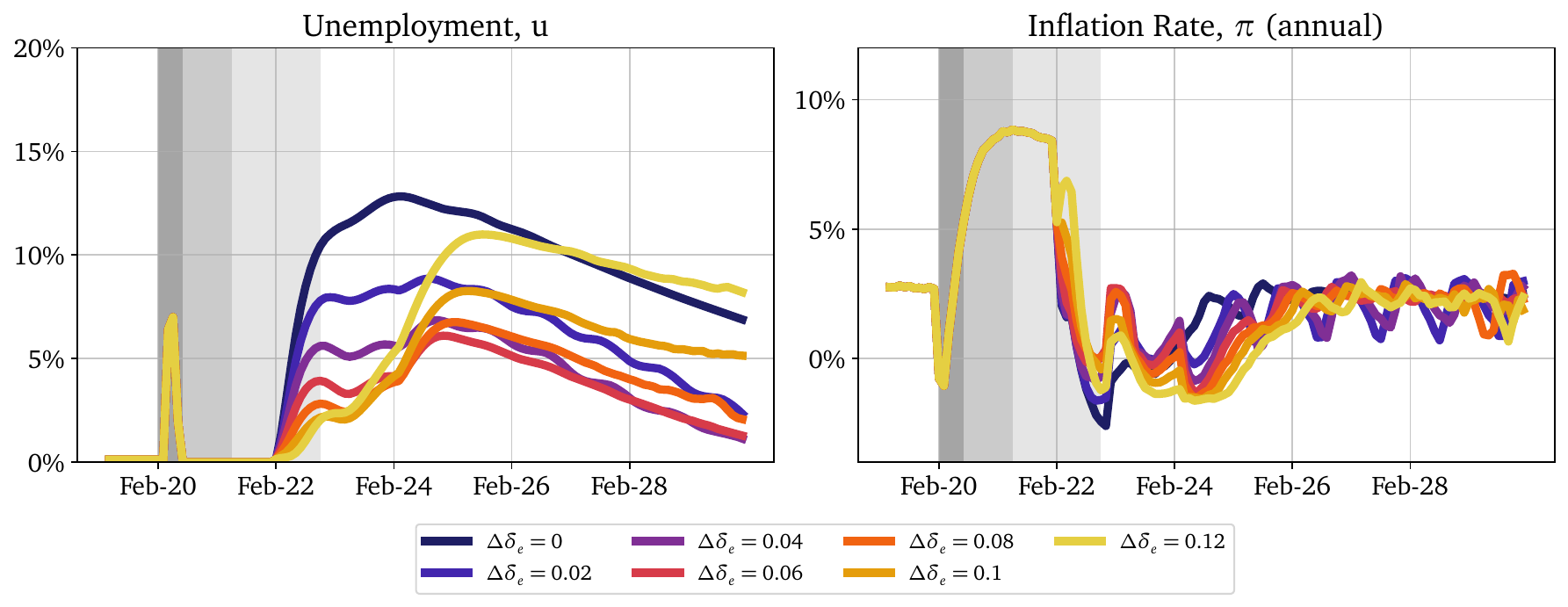}
    \caption{\textbf{Windfall Tax in the Reactive CB with Anchored Expectation scenario.} Unemployment (left) and inflation (right) for Windfall Tax of $\Delta \delta_e$ one year before the end of the price shock with a duration of two years. With $\Delta \delta_e \approx 6\%$ unemployment is reduced strongly. A further increase of $\Delta \delta_e$ leads to a significant resurgence of unemployment. 
    } 
    \label{activeCB_ap_windfall_spectrum}
\end{figure}

\begin{figure}[H]
    \centering
    \includegraphics[width=\textwidth]{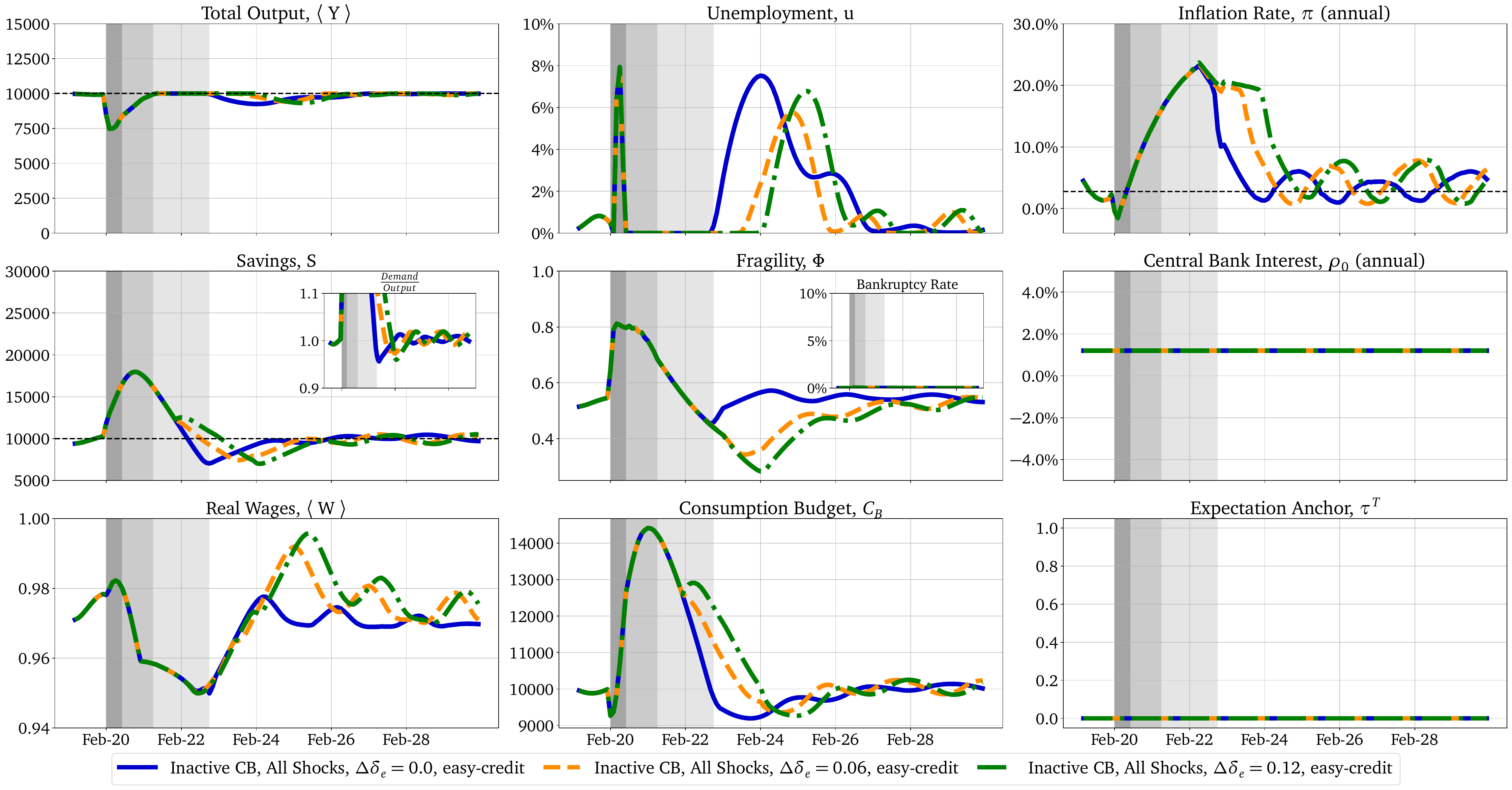}
    \caption{\textbf{Economic dashboard for Windfall Tax in the Inactive Central Bank scenario, all shocks.} Full dynamics for the Inactive Central Bank scenario without Windfall Tax (blue), for tax of $\Delta \delta_e = 6\%$ (orange) and $\Delta \delta_e = 12\%$ (green) one year before the end of the price shock with a duration of two years.}
    \label{baseline_ap_windfall}
\end{figure}

\begin{figure}[H]
    \centering
    \includegraphics[width=\textwidth]{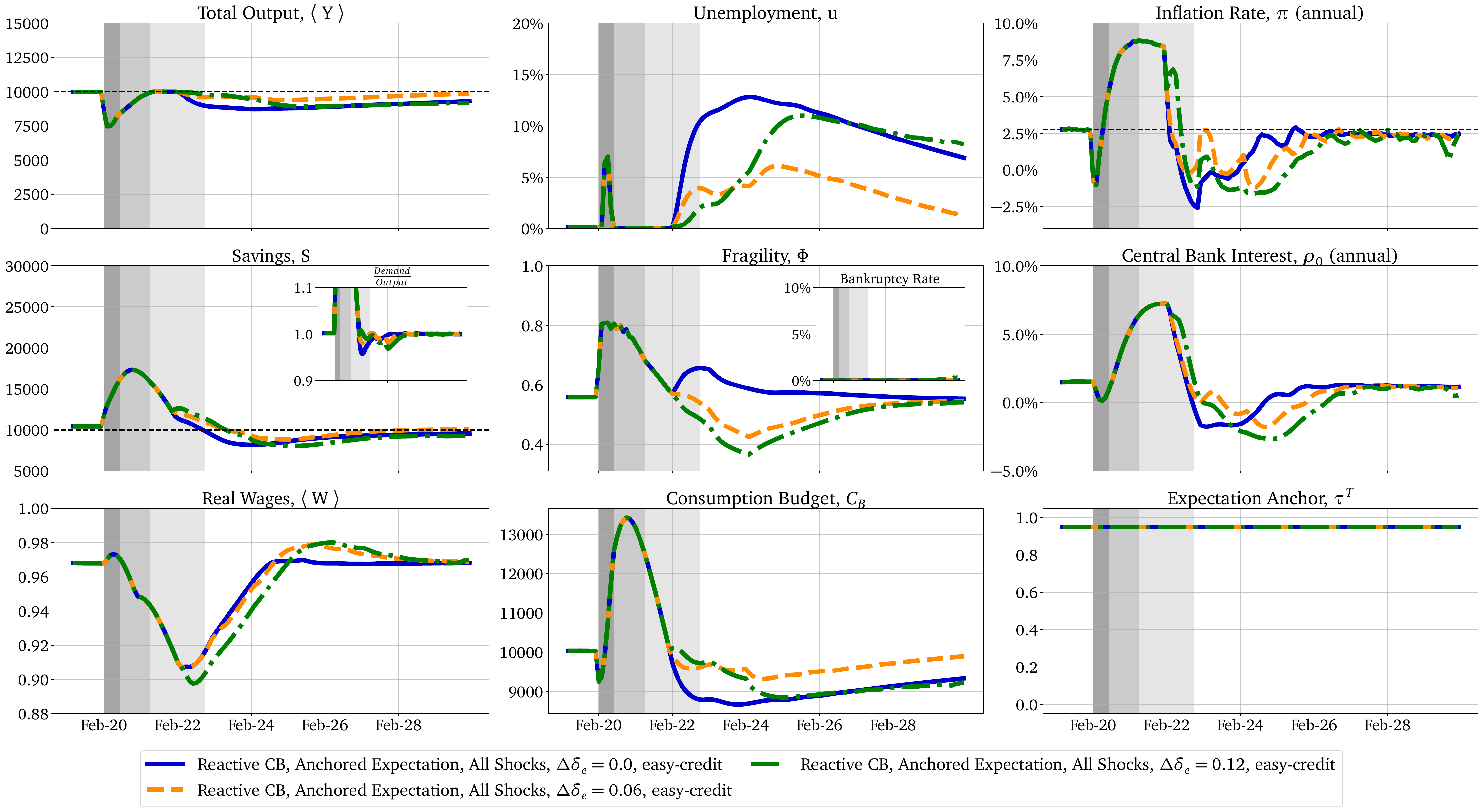}
    \caption{\textbf{Economic dashboard for Windfall Tax in the Reactive Central Bank with Anchored Expectation scenario and all shocks.} Full dynamics for scenario without Windfall Tax (blue), for tax of $\Delta \delta_e = 6\%$ (orange) and $\Delta \delta_e = 12\%$ (green) one year before the end of the price shock with a duration of two years.}
    \label{activeCB_ap_windfall}
\end{figure}
\begin{figure}[H]
    \centering
    \includegraphics[width=\textwidth]{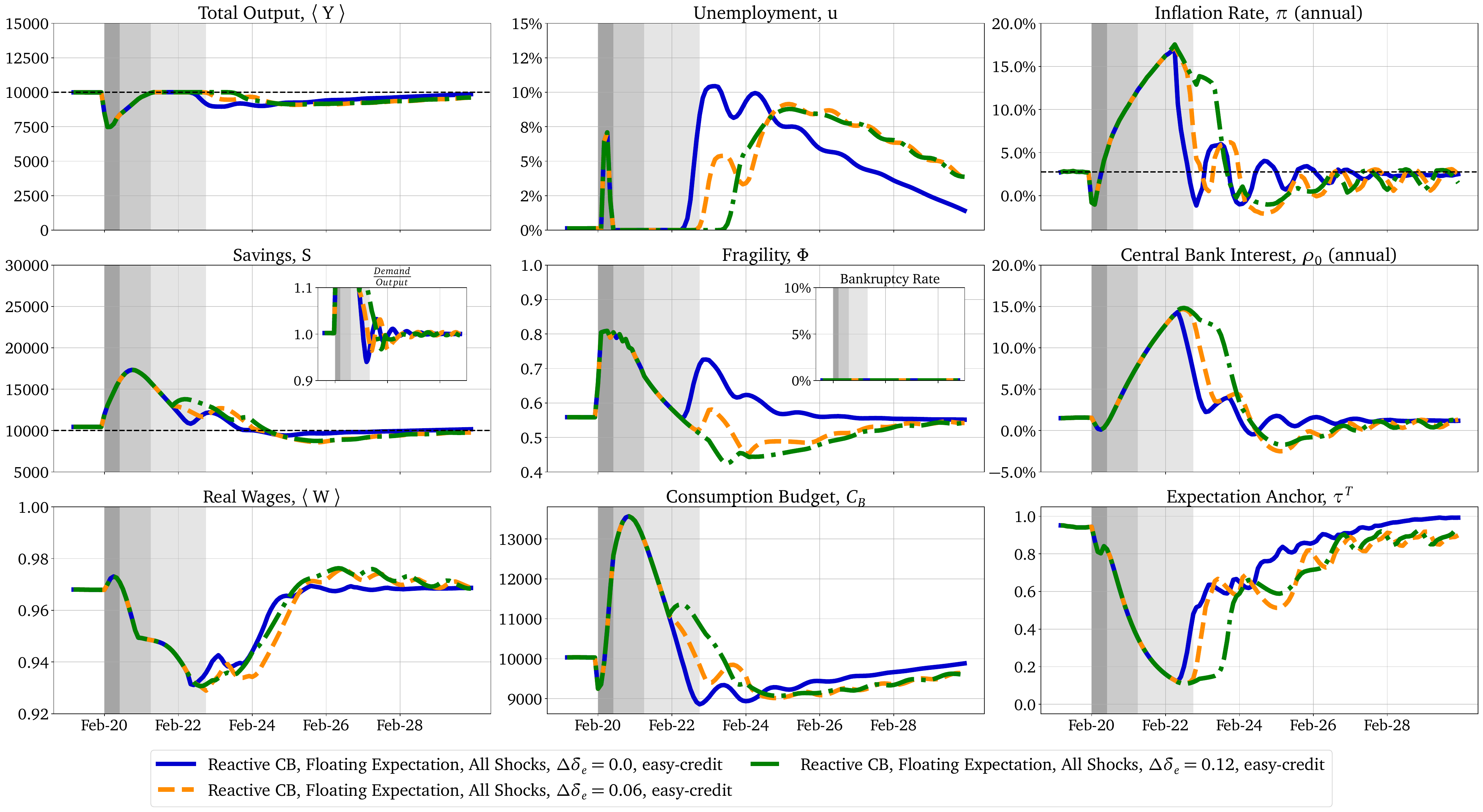}
    \caption{\textbf{Economic dashboard for Windfall Tax in the Reactive Central Bank with Floating Expectation scenario and all shocks.} Full dynamics for scenario without Windfall Tax (blue), for tax of $\Delta \delta_e = 6\%$ (orange) and $\Delta \delta_e = 12\%$ (green) one year before the end of the price shock with a duration of two years.}
    \label{activeCB_dyn_ap_windfall}
\end{figure}

\subsection{Sloppiness}\label{appx:sloppy}
For the sloppiness analysis we follow the same approach as in \cite{NaumannWoleskeEtAl2023ExplorationParameterSpace}. We define a mean-square loss error which is dependent on the set of parameters $\Phi$ and a small deviation $\delta$. The loss function can be written as
\begin{equation}\label{eq:square_loss}
    \mathcal{L}(\Phi, \delta) = \frac{1}{2 S K T}\sum_{s}\sum_{k}\sum_{t}\left(\frac{y_{s, k, t}(\Phi+\delta) - y_{s, k, t}(\Phi)}{\norm{y_{s,k}(\Phi)}}\right)^2,
\end{equation}
where $y_{s, k, t}(\Phi)$ is the realisation of output variable $k\in\{1,\dots,K\}$ at time $t\in\{1,\dots,T\}$ for random realisation $s\in\{1,\dots,S\}$. In this case, we use the output variables inflation and unemployment with 50 random realisations and we average over a period of 11 years that stars 1 year before the COVID shock in February 2020.

The Hessian matrix for the mean-squared loss at point $\Phi$ is then defined as
\begin{equation}\label{eq:hessian_def}
    H_{i,j}^{\mathcal{L}}(\Phi) := \left. \frac{d^2\mathcal{L}(\Phi, \delta)}{d\log\Phi_i d\log\Phi_j} \right|_{\delta = 0},
\end{equation}

\begin{equation}\label{eq:hessian}
    H_{i,j}^{\mathcal{L}}(\Phi) = \frac{1}{S K T}\sum_{s}\sum_{k}\sum_{t}\frac{1}{\norm{y_{s, k, t}(\Phi)}^2} \frac{dy_{s, k, t}(\Phi)}{d\log\Phi_i}\frac{dy_{s, k, t}(\Phi)}{d\log\Phi_j},
\end{equation}

Since the parameters in general have different order of magnitudes, we take the derivative with respect to the log parameters to only consider relative parameter changes. Note, that in Eq. \ref{eq:hessian} the second derivative term vanishes as we evaluate $H_{i,j}^{\mathcal{L}}(\Phi)$ at $\delta=0$. 

The Hessian matrix can be decomposed into its eigenvalues and eigenvectors where the eigenvalues indicates the relative importance of the corresponding eigenvectors that represent the linear combination of parameters in parameter space. We are interested in the stiff directions, so the eigenvectors with largest eigenvalues that bring the biggest change to the dynamics of the system. 

\begin{figure}[H]
    \centering
    \includegraphics[width=\textwidth]{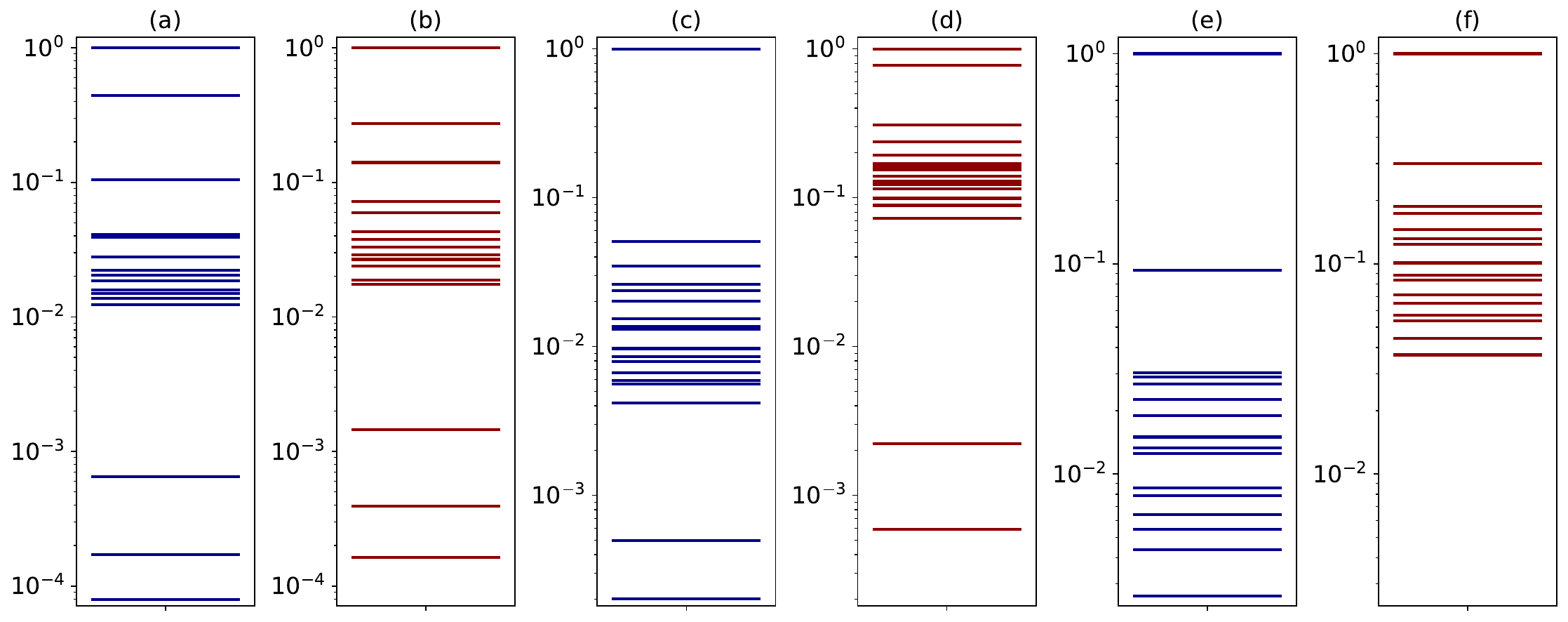}
    \caption{Eigenvalue Spectrum for three scenarios and all shocks. Sloppiness parameter $\epsilon = 0.01$, $n_{seeds} = 50$ target variable inflation (red) unemployment (blue), (a,b) Inactive Central Bank, (c,d) Reactive Central Bank with Anchored Expectation, (e,f) Reactive Central Bank with Floating Expectation.}
    \label{sloppy_eigenvalues}
\end{figure}

\begin{figure}[H]
    \centering
    \includegraphics[width=\textwidth]{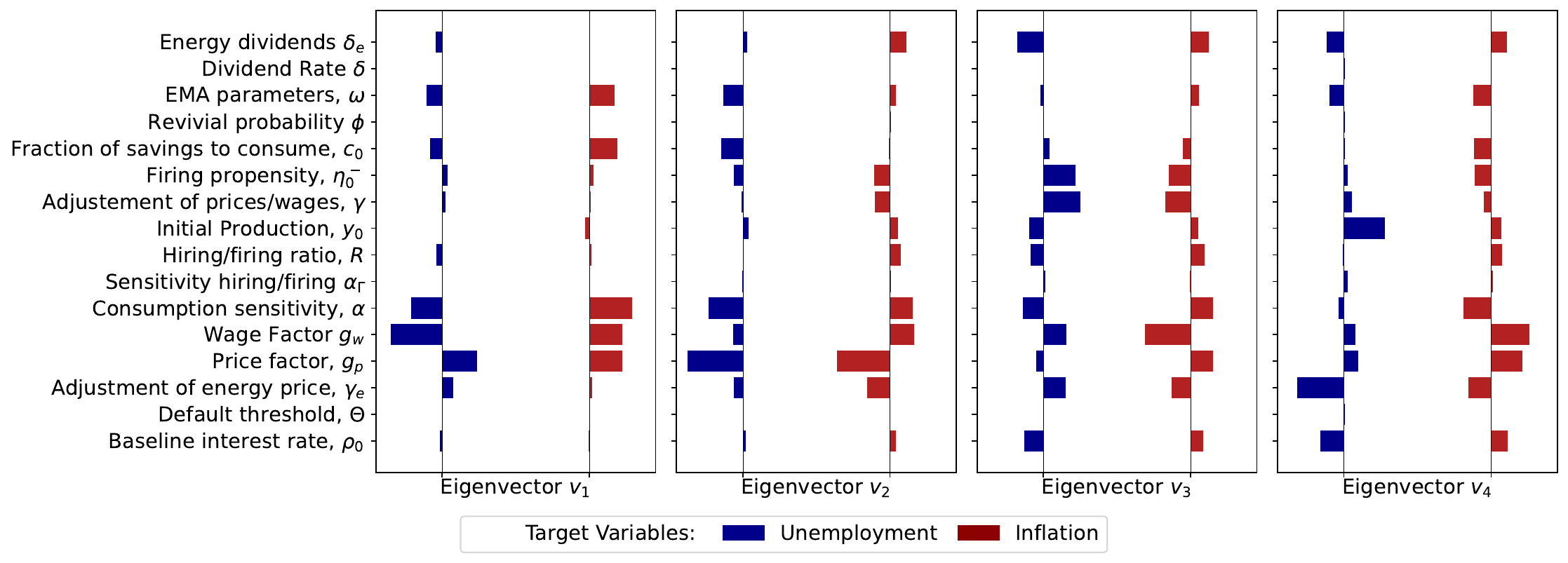}
    \caption{\textbf{Eigenvectors of the Hessian matrix for the Inactive Central Bank Scenario and all shocks.} The length of the eigenvectors is normed to 1 and the colors indicate the target variables, unemployment (blue) and inflation (red). }
    \label{sloppy_evecs_baseline}
\end{figure}

\begin{figure}[H]
    \centering
    \includegraphics[width=\textwidth]{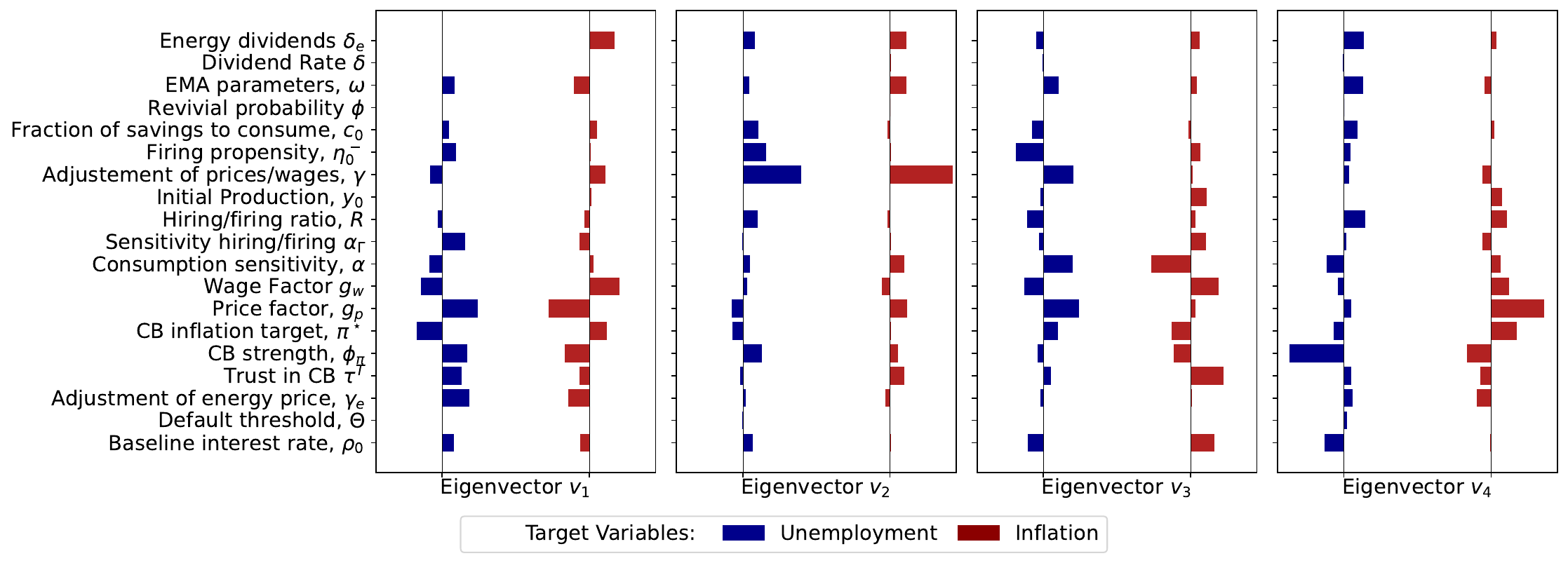}
    \caption{\textbf{Eigenvectors of the Hessian matrix for the Reactive Central Bank with Anchored Expectation scenario and all shocks.} The length of the eigenvectors is normed to 1 and the colors indicate the target variables, unemployment (blue) and inflation (red). }
    \label{sloppy_evecs_activecb}
\end{figure}

\begin{figure}[H]
    \centering
    \includegraphics[width=\textwidth]{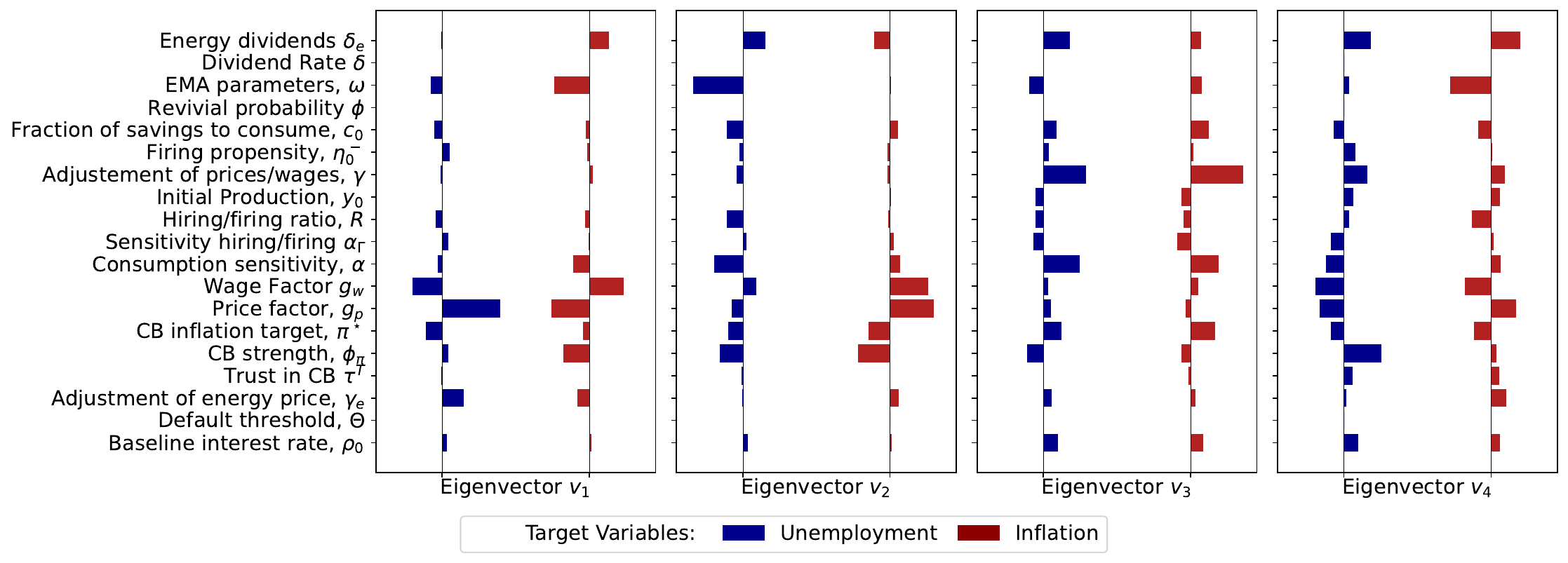}
    \caption{\textbf{Eigenvectors of the Hessian matrix for the Reactive Central Bank with Floating Expectation scenario and all shocks.} The length of the eigenvectors is normed to 1 and the colors indicate the target variables, unemployment (blue) and inflation (red). }
    \label{sloppy_evecs_activecb_dyn}
\end{figure}

The eigenvalue spectrum spans several decades (Figure \ref{sloppy_eigenvalues}), which indicates the sloppiness of the model at that set of parameters. Based on the spectrum, we select the four largest eigenvectors to consider for our analysis (see Figures \ref{sloppy_evecs_baseline}, \ref{sloppy_evecs_activecb}, \ref{sloppy_evecs_activecb_dyn}).

\subsection{Sensitivity to the Indexation Parameters $g_p$ and $g_w$}\label{appx:gpgw}

\begin{figure}[H]
    \centering
    \includegraphics[width=\textwidth]{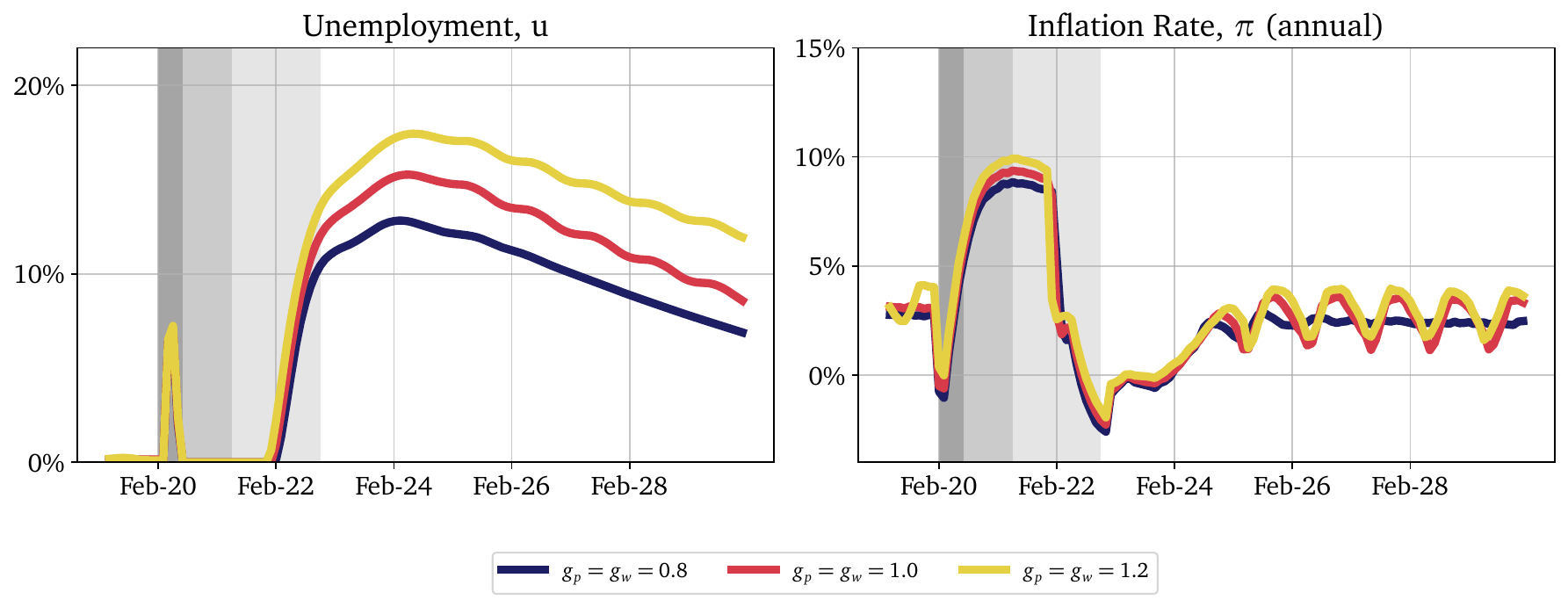}
    \caption{\textbf{Sensitivity of $g_p = g_w = const$ for the Reactive Central Bank with Anchored Expectation scenario and all shocks.} Increasing $g_p=g_w$ increases slightly inflation (cet. par.), causing slightly higher interest rates which leads to a reduction of consumption budget that reduces demand and therefore increases unemployment.}
    \label{sensitivity_gp_gw_const}
\end{figure}

\begin{figure}[H]
    \centering
    \includegraphics[width=\textwidth]{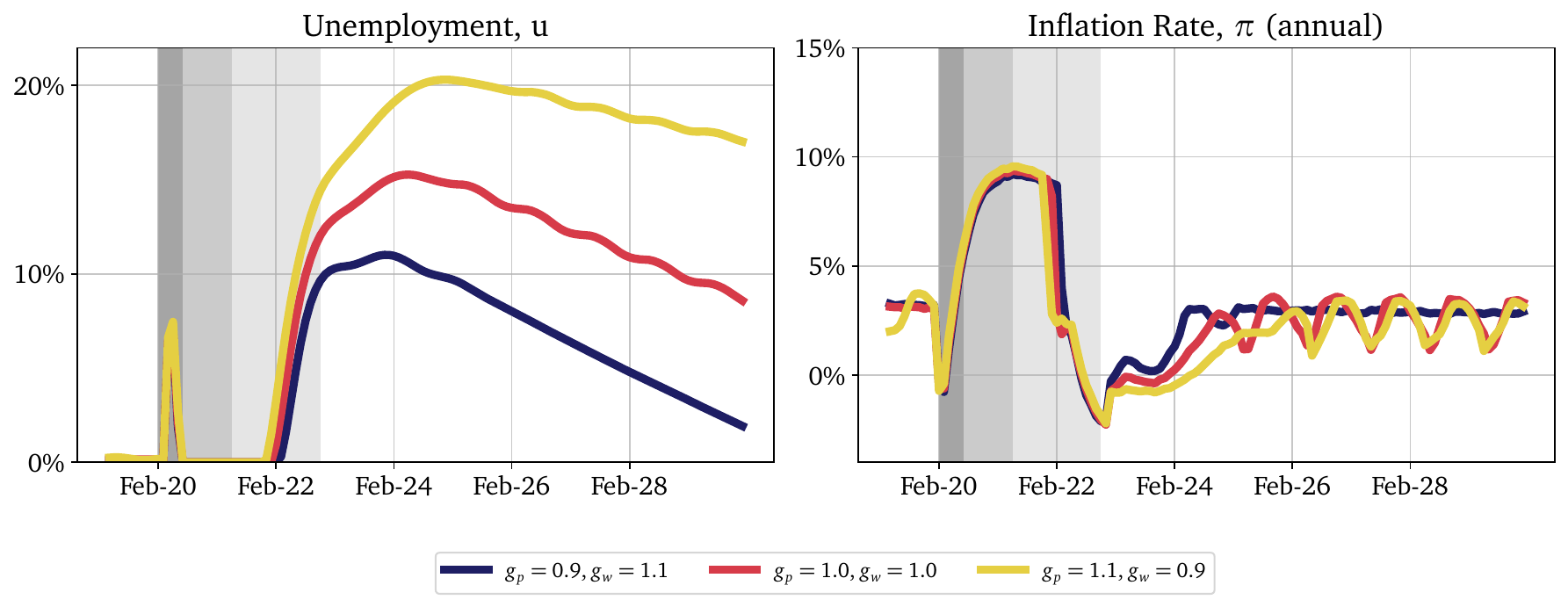}
    \caption{\textbf{Sensitivity of $\frac{g_p + g_w}{2} = const$ for the reactive Central Bank with Anchored Expectation scenario and all shocks.} (blue) When $g_w>g_w$, there is a higher bargaining power of worker which increases wages (cet. par.), therefore there is a higher consumption budget and higher demand which decreases unemployment. (yellow) For larger market power ($g_p>g_w$), there is the reverse effect of decreased wages causing a decrease in consumption budget and demand which therefore increases unemployment.}
    \label{sensitivity_gp_+_gw_const}
\end{figure}

\end{document}